%% file: main.tex
\PassOptionsToPackage{twoside=false}{geometry}
\documentclass[manuscript,screen,nonacm]{acmart}

\setcopyright{none} 
\renewcommand\footnotetextcopyrightpermission{}

\makeatletter
\let\@acmBadgeLTS\@empty
\let\@authorsaddresses\@empty 
\makeatother

\AtBeginDocument{%
  }

\usepackage{tikz}
\usetikzlibrary{
    arrows.meta,   
    positioning,   
    fit,           
    shapes,        
    backgrounds    
}
\usepackage{adjustbox}
\usepackage{multirow}
\usepackage{tabularx}
\usepackage{xspace}
\usetikzlibrary{arrows.meta, positioning}
\usepackage[table]{xcolor}
\usepackage{colortbl}
\usepackage{longtable}
\usepackage{graphicx}
\usepackage{xcolor}
\usepackage{bbm}
\usepackage{booktabs}
\usepackage{makecell}
\usepackage{placeins}
\usepackage{algorithm}
\usepackage{algpseudocode}
\usepackage{enumitem}
\usepackage{wrapfig}
\usepackage{subcaption}

\colorlet{blue}{black}
\colorlet{teal}{black}

\newcommand{\sys}{\texttt{Trace}\xspace}
\begin{document}



\title{Rules, Choice, and Chance: A Welfare Evaluation Framework for Student-School Matching}





\author{Reetahan Mukhopadhyay}
\email{reetahan.m@nyu.edu}
\affiliation{%
  \institution{New York University}
  \city{New York City}
  \state{New York}
  \country{USA}
}

\author{Clara Strasser Ceballos}
\email{Clara.StrasserCeballos@stat.uni-muenchen.de}
\affiliation{%
  \institution{Ludwig-Maximilians-Universität München}
  \city{Munich}
  \state{Bavaria}
  \country{Germany}
}

\author{Augustin Plantureux}
\email{augustin.plantureux@polytechnique.edu}
\affiliation{%
  \institution{École Polytechnique}
  \city{Palaiseau}
  \state{Île-de-France}
  \country{France}
}

\author{Christoph Kern}
\email{christoph.kern@stat.uni-muenchen.de}
\affiliation{%
  \institution{Ludwig-Maximilians-Universität München}
  \city{Munich}
  \state{Bavaria}
  \country{Germany}
}

\author{Julia Stoyanovich}
\email{stoyanovich@nyu.edu}
\affiliation{%
  \institution{New York University}
  \city{New York City}
  \state{New York}
  \country{USA}
}


\begin{abstract}

\textcolor{teal}{Centralized student-school matching systems, adopted in the United States,
Chile, the Netherlands, and the United Kingdom, assign students to schools
based on their submitted preference rankings and schools' priority policies.
Yet welfare in these systems is typically measured through a narrow lens: a
few aggregate statistics published by education authorities, such as the share
of students unmatched or assigned to a top-ranked choice. A single aggregate
cannot reveal how outcomes vary across students, or why. Welfare depends on
three interacting forces: the \emph{rules} policymakers set (algorithm,
tie-breaking, list length limits); the \emph{choices} students make (which and
how many schools to rank); and \emph{chance} (lottery draws and students'
residential location). A \emph{rule} that reduces unmatched rates may
simultaneously lower top-ranked assignment rates for specific groups, and
evaluating such trade-offs in advance is challenging because individual-level
preference data are rarely available. We present \sys (Trade-offs in Rules,
Agency, Chance, and Equity), an open-source framework that synthesizes
individual-level preference profiles from aggregate data using mixtures of
Mallows models and evaluates alternative \emph{rules} across multiple welfare
metrics, disaggregating outcomes by \emph{choice} and \emph{chance}. We apply
\sys to two real-world systems: NYC high school matching, where we study
minimum list length requirements, and Chilean school assignment, where we
compare the current multiple tie-breaking rule with a counterfactual single
tie-breaking rule. Both reveal substantial, heterogeneous welfare trade-offs,
underscoring the value of simulation-based evaluation before implementing
school choice reforms.}

\end{abstract}

\maketitle

\input{sections/01_introduction}

\input{sections/02_related_work}

\input{sections/03_preliminaries}

\input{sections/04_framework}

\input{sections/05_case_studies}

\input{sections/06_conc}

\newpage  

\input{sections/AI_Usage}

\input{sections/07_adverse_impacts}

\input{sections/08_ethical_considerations}

\bibliographystyle{ACM-Reference-Format}
\bibliography{sample-base}

\appendix

\input{sections/01_additional_details_notation}

\newpage
\input{sections/02_additional_details_case_NYC_and_Chile}

\newpage
\input{sections/03_additional_results_chile}

\newpage
\input{sections/03_additional_validation_results_NYC_and_Chile}

\newpage
\input{sections/04_additional_detail_demo}

\end{document}

%% file: sections/01_introduction.tex
\section{Introduction}
\label{sec:intro}

Two-sided matching algorithms are used in many real-world settings, whether to match students to schools~\cite{SchoolChoice2023a}, medical graduates to hospitals~\cite{NRMP2026}, or refugees to locations \cite{strasserceballosLocationMatchingShaky2025}. Particularly in the school setting, the use of these algorithms has gained considerable importance, prompting the replacement of traditional admission practices with centralized matching systems~\cite{cheAllocatingStudentsSchools2025} in the United States~\cite{SchoolChoice2023a}, Chile~\cite{correaSchoolChoiceChile2022}, the Netherlands~\cite{ruijsSchoolChoiceAmsterdam2019}, and the United Kingdom~\cite{terrierImmediateAcceptanceDeferred2026}. These systems promise greater transparency and efficiency, as well as the potential to reduce educational segregation by enabling students to apply to schools beyond their local neighborhoods~\cite{belleiFamiliesResponsesNew2025,abdulkadirogluStrategyProofnessEfficiencyMatching2009}.

Designing such systems, however, is far from straightforward. A natural starting point is the choice of the assignment mechanism itself~\cite{SchoolChoice2023a}. \citet{abdulkadirogluSchoolChoiceMechanism2003a} were among the first to formally study the design of school admission systems using matching theory. They analyzed and compared several mechanisms, including the Immediate Acceptance (IA)~\cite{SchoolChoice2023a}, Top Trading Cycles (TTC)~\cite{shapleyCoresIndivisibility1974}, and the Gale and Shapley Deferred Acceptance (DA) algorithm~\cite{galeCollegeAdmissionsStability1962}. Based on this comparison, the authors identified the DA algorithm as a particularly suitable mechanism for school assignment, largely due to its two key theoretical properties: \textit{stability}, meaning that no student-school pair would prefer to deviate from the assigned outcome, and \textit{strategy-proofness}, meaning that students are best off ranking schools in order of their true preferences \cite{SchoolChoice2023a}. These properties motivated major reforms, with cities such as Boston and New York City (NYC) redesigning their systems in the early 2000s to adopt variants of the student-proposing DA algorithm ~\cite{abdulkadirogluSchoolChoiceMechanism2003a}. 

\smallskip\textbf{Impact of Rules on Welfare.} When reforming school admission systems, policymakers and system designers must consider student welfare under the resulting match. Welfare is commonly measured by the share of students who remain unmatched or are assigned to one of their top-ranked choices. Rules can substantially impact welfare~\cite{marianAlgorithmicTransparencyAccountability2023} and extend beyond the choice of the matching algorithm~\cite{xuDesigningSchoolChoice2024}. They also include constraints on the number of schools students may rank and the mechanism used to resolve ties, i.e., systems may assign each student a single global lottery number (single tie-breaking, STB) or school-specific lottery numbers (multiple tie-breaking, MTB). 

\begin{figure}
    \centering
  \includegraphics[width=0.8\linewidth]{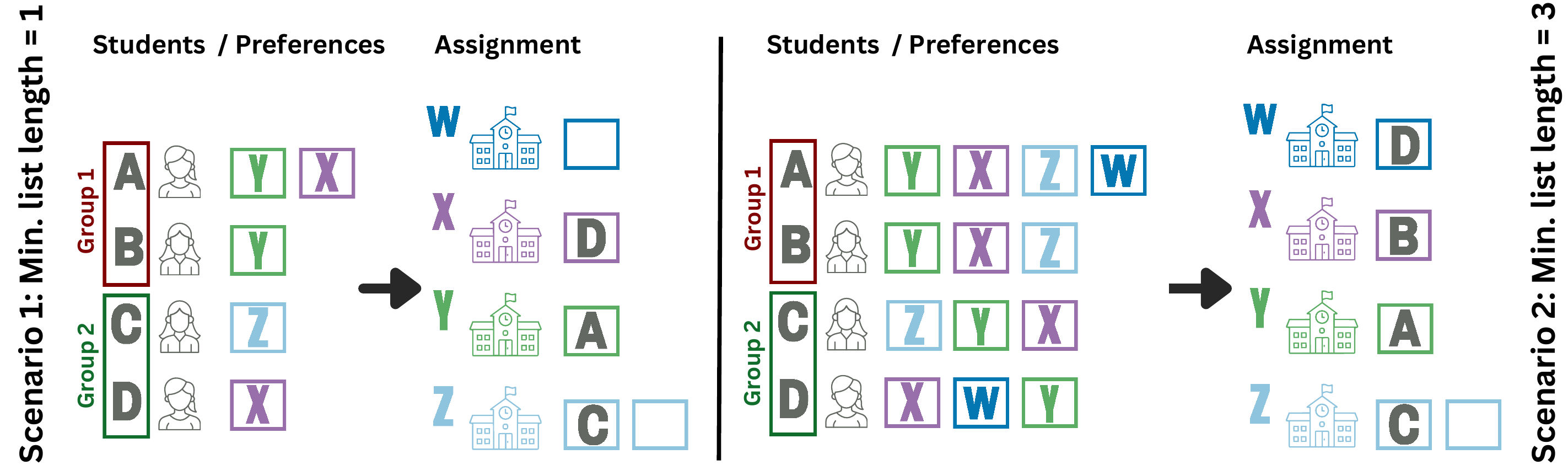}
    \caption{Impact of minimum list length on welfare outcomes. Matching outcomes under student-proposing DA for two scenarios differing only in the minimum list length (1 vs.\ 3). Students are ordered by lottery as $A > B > C > D$ and \textcolor{blue}{grouped by residential location; the grouping does not affect the matching.}}
    \label{fig_motivation}
\end{figure}

Evaluating how these rules affect student welfare is important but complex. Outcomes depend not only on these \emph{rules}, but also on the \emph{choices} students make (which and how many schools to rank) and on \emph{chance} (lottery draws and residential location, which determine with whom a student competes). Because these forces interact, no single welfare metric captures their effects, and choosing among rules involves trade-offs.
For instance, several studies find that the share of students matched to a top-ranked choice is higher under STB, while the share remaining unmatched is lower under MTB \cite{abdulkadirogluStrategyProofnessEfficiencyMatching2009,dehaanPerformanceSchoolAssignment2023, arnostiShortListsCentralized2015a, ashlagiWhatMattersSchool2020}. These trade-offs also differ across students: a rule that performs better overall may help some students and hurt others, depending on their \emph{choices} and on \emph{chance}. Rules should therefore be evaluated ex ante, with outcomes disaggregated
by choice and chance. Figure~\ref{fig_motivation} presents a motivating example.

\begin{example}\label{ex:1}
Suppose four students, $A$, $B$, $C$, and $D$, compete for four schools, \textcolor{blue}{$W$, $X$, $Y$, and $Z$}, under the student-proposing DA algorithm. Each student is assigned a single global lottery number to break ties; students are ordered from best to worst number: $A > B > C > D$. Schools $W$, $X$ and $Y$ each have one seat, while school $Z$ has two  seats. \textcolor{blue}{For evaluation, we group students by residential district: students $A$ and $B$ form Group 1, the others Group 2.} In terms of our framework, the \emph{rules} are DA with STB and a minimum list length. DA and STB are held fixed; we compare \emph{rules} that differ only in the minimum list length, i.e., the min. number of schools each student must rank.
\emph{chance} is represented by the lottery order \textcolor{blue}{and residential location (group membership)}; and \emph{choice} is reflected in how many schools each student decides to rank. We track two welfare metrics: the \textbf{unmatched rate} (share of students who receive no assignment) and the \textbf{top-$1$ match rate} (share of students assigned to their first-ranked school).

\smallskip\noindent\textbf{Scenario 1 (minimum list length $= 1$).} Only student $A$ submits more than one preference. Student $B$ is rejected from their sole-ranked school, exhausts their list, and remains unmatched, leaving one seat at school $Z$ unfilled. Unmatched rate: 25\%. Top-1 match rate: 75\% overall; 100\% for Group~2.

\smallskip\noindent\textbf{Scenario 2 (minimum list length $= 3$).} All other parameters remain unchanged. Because students can continue applying after rejection, all students are ultimately matched. Student $B$ is now assigned to 
their second-ranked school; however, student $D$, previously assigned to their top choice (school $X$), is displaced by a higher-priority student and reassigned to school $W$. Unmatched rate: 0\%. Top-1 match rate: 50\% overall; 50\% for Group~2.
\end{example}

{\textbf{Challenges in Evaluating Welfare Impacts.}} Example~\ref{ex:1} illustrates that a seemingly minor \emph{rule} change, the minimum list length, can substantially affect welfare, and that the resulting trade-off varies by group and is invisible to any single aggregate statistic. For Group~1, the reform improves coverage: the unmatched rate falls from 50\% to 0\%. For Group~2, both students remain matched, but match quality worsens: the top-1 match rate falls from 100\% to 50\%. Thus, the aggregate unmatched rate improves from 25\% to 0\%, while the overall top-1 match rate declines from 75\% to 50\%. A policymaker tracking only the aggregate unmatched rate would conclude that the reform is unambiguously beneficial; one tracking only the top-1 match rate would reach the opposite conclusion. Capturing such trade-offs therefore requires evaluating the welfare impact of alternative \emph{rules} across multiple metrics and groups.

Yet translating this into practice raises two key challenges. \textcolor{blue}{First, individual-level preference data are rarely available. However, simulating outcomes under alternative rules requires students' preference lists, but these data are sensitive and education authorities typically publish only aggregate statistics.} \textcolor{blue}{In the absence of such data, existing studies often assume stylized preferences, e.g., uniformly random or popularity-based~\cite{ashlagiAssigningMoreStudents2019, faenzaLongerListsYield2025}.} \textcolor{teal}{Second, even when individual-level preference data are available, they may be insufficient for evaluating alternative rules. This is because observed preference profiles reflect the rules under which they were submitted and may not capture the preferences relevant under a reform. For instance, submitted lists are censored at the length each student chose under the prevailing rule, so evaluating rules that extend lists, e.g., NYC's removal of its 12-program ranking cap~\cite{nyc2025NYCPUBLIC2025,NYCAdmissionsAre2024}, requires a model of preferences beyond the observed cutoff. Moreover, students may adjust how they form and report preferences when rules change~\cite{dehaanPerformanceSchoolAssignment2023}, e.g., strategizing when the algorithm is not strategy-proof~\cite{calsamigliaStructuralEstimationModel2020}. The observed profile is thus one realization among many that could arise under a reform, and evaluating against it alone can be misleading. A credible evaluation requires generating preference profiles and varying them systematically, even though predicting families' behavioral response remains beyond the scope of existing frameworks, including ours.}

Together, these considerations highlight a gap in the welfare evaluation of student-school matching: the lack of a  framework that can generate realistic individual-level preference profiles, simulate alternative rules across plausible profiles, and evaluate welfare trade-offs across multiple metrics and  groups. \textcolor{blue}{Practitioners echo this: what is missing are tools that show ``in advance, who would gain and who would lose from a proposed change'' \cite{neilsonDecadeCentralizedSchool2026}.}

\smallskip{\textbf{Contributions and Roadmap.}  We present background and related work  in Section~\ref{rel_work}, followed by notation in Section~\ref{prel}. Then, in Section~\ref{framework}, we present our \textbf{first contribution}, \sys (Trade-offs in Rules, Agency, Chance, and Equity), a flexible open-source framework that enables counterfactual welfare analysis in student-school matching \textcolor{blue}{when only aggregate data are available}, examining how the \emph{rules} policymakers set, the \emph{choices} students make, and \emph{chance} jointly shape welfare outcomes.
\sys has two components. The first, a \emph{preference generation module}, fits a Mallows mixture model to aggregate statistics (e.g., applicants per school and group-level match rates) to generate calibrated individual-level preference profiles. The second, a \emph{welfare evaluation module}, simulates the assignment under alternative \emph{rules} and computes group-level welfare metrics to identify trade-offs.

In Section~\ref{sec:eval} we present our \textbf{second contribution}: a validation and application of \sys on the NYC high school (HS) matching process and the Chilean school assignment system, both of which use variants of the student-proposing DA algorithm. In NYC, where only aggregate data are publicly available, we evaluate the welfare impact of imposing a minimum list length, a \emph{rule} that interacts with students' \emph{choices} of how many schools to rank. We find that longer lists improve match rates at the cost of average match quality, with substantial heterogeneity across residential districts. In Chile, where individual-level preferences are available, we use students' observed preference lists to compare the current MTB rule with a counterfactual STB rule. Compared with MTB, STB would amplify the role of \emph{chance}: it steepens welfare gradients across lottery deciles and affects students differently by residential location. \textcolor{blue}{We additionally validate \sys by showing that generated preference profiles reproduce published aggregate statistics, including those not used for calibration.}  \textcolor{blue}{To test fidelity, we repeat the Chilean comparison using only profiles generated by \sys from aggregate statistics and reach the same qualitative conclusions. This suggests that \sys can support meaningful policy analysis when individual-level preferences are unavailable.} \textcolor{teal}{Together, the two modules address both challenges: the preference generation module supplies calibrated profiles where none are published, and re-generating profiles under varied parameters (Section~\ref{framework}, Entry point~2) ensures that no conclusion rests on a single fixed profile.}


We conclude in Section~\ref{conc}.  \sys is available at \url{https://anonymous.4open.science/r/TRACE-5FC3}.

\smallskip\textbf{Scope.} \textcolor{teal}{\sys evaluates design configurations for a fixed, calibrated preference-generating process; it does not predict how families' preferences or reporting would change under a reform. All welfare metrics are computed over \emph{stated} preference lists, which need not coincide with families' true preferences (Section~\ref{prel}). The framework does support sensitivity analysis: profiles can be re-generated under perturbed parameters, e.g., longer list-length distributions or greater preference heterogeneity (Section~\ref{framework}). Equity claims are likewise scoped: we quantify how \emph{rules} redistribute outcomes across lottery position and residence, the dimensions observable in our data (Section~\ref{conc}).}

%% file: sections/02_related_work.tex
\section{Background and Related Work}
\label{rel_work}

\smallskip{\textbf{Design and Welfare.} Our work relates to studies evaluating how design choices affect welfare in student-school matching. Table~\ref{welf_summ} summarizes the design parameters studied in the literature, all of which are \emph{rules} in our framework. Although definitions vary, most studies measure welfare by a common set of match quality metrics.
%

\input{other/welfare_table_pivoted}

There are three key takeaways from the literature summarized in Table \ref{welf_summ}. First, the most prominent \emph{rules} studied are the assignment mechanism and the tie-breaking mechanism \cite{agarwalRevealedPreferenceAnalysis2020}. For the former, many studies compare variants of the Deferred Acceptance (DA) algorithm with the Boston mechanism. For the latter, the literature typically focuses on STB and MTB. An exception is \citet{correaSchoolChoiceChile2022}, which also examines alternative approaches, such as family-level versus student-level lottery numbers. Second, most studies focus on a small number of welfare outcomes, in particular the share of students remaining unmatched and the share being matched to a top-ranked choice. These metrics capture different notions of welfare: match rates reflect assignment coverage; top-ranked rates and average rank capture ordinal welfare; utility and envy/jealousy measures capture cardinal welfare~\cite{xuDesigningSchoolChoice2024}.  Third, the literature has studied design decisions almost exclusively as \emph{rules}: all parameters in Table~\ref{welf_summ} reflect choices made by policymakers. \textcolor{blue}{One exception is \citet{arnostiLotteryDesignSchool2023}, who studies the joint impact of \emph{rules} (tie-breaking mechanism) and \emph{choice} (students' list length). Specifically, comparing STB and MTB conditional on list length, he finds that students with shorter lists have a higher probability of being matched under STB than under MTB, whereas reverse holds for students with longer lists.} 

These studies motivate our evaluation of the welfare impact of \emph{rules} across a comprehensive set of metrics. We focus on assignment coverage and ordinal welfare metrics given their prominence in the literature.

\smallskip{\textbf{Equity and School Choice.}} Several studies critically evaluate the equity promises of centralized school assignment, arguing that expected gains may fail to materialize because modeling assumptions do not reflect real-world conditions \citep{robertsonModelingAssumptionsClash2021}. One key concern is that families'\emph{choices} are shaped by unequal starting conditions, particularly access to information about schools and the assignment process \citep{robertsonModelingAssumptionsClash2021, robertsonNotAnotherSchool2022}. \textcolor{blue}{In NYC, for instance, lottery numbers were first released in 2021, with little accompanying information, so the ability to interpret them likely benefited more privileged families \citep{marianAlgorithmicTransparencyAccountability2023}.} This matters because beliefs about admission chances influence which and how many schools families rank \citep{kaporHeterogeneousBeliefsSchool2020, agteSearchBiasedBeliefs2024}. For example, families who underestimate their risk of remaining unmatched tend to rank too few schools \citep{arteagaSmartMatchingPlatforms2022}, with lower-SES families more likely to underestimate this risk \citep{medinaMakingAlgorithmicTransparency2026}. Families also differ in their preferences, including regarding schools' racial composition \citep{haileyRacialPreferencesSchools2022}. Beyond \emph{choice}, residential location and admission criteria shape which schools students can access in the first place \citep{idouxIntegratingNewYork2022, kutscherCentralizedAdmissionSystems2023}. Together, these factors can lead to unequal assignment outcomes \citep{kloostermanSchoolChoiceAsymmetric2020} and may reinforce existing patterns of segregation \citep{idouxIntegratingNewYork2022, kutscherCentralizedAdmissionSystems2023}. These findings motivate our approach: evaluating how alternative assignment \emph{rules} affect welfare across students who differ in their \emph{choices} and the \emph{chance} they face.

\smallskip{\textbf{Preference Estimation and Mixtures of Mallows Models.} Our work builds on mixtures of Mallows models \cite{mallowsNonNullRankingModels1957a, liuEfficientlyLearningMixtures2018a}. These models describe heterogeneous populations in which each group's rankings vary randomly around a central ranking, with a dispersion parameter controlling the variation \cite{awasthiLearningMixturesRanking2014}.
To estimate these parameters, existing methods rely on individual-level ranking data, which need not be complete: methods handle pairwise preferences \cite{luEffectiveSamplingLearning2014, stoyanovichWorkloaddrivenLearningMallows2016}, partially observed top-\(p\) lists \cite{crispinoEfficientAccurateInference2022, chierichetti2018mallows, haddadanGeneralizedTopkMallows2025}, and rankings under alternative distance metrics \cite{crispinoEfficientAccurateInference2022, fengMallowsTypeModel2022}. 

Estimating individual-level preferences is essential for evaluating welfare under alternative design choices, a goal shared by the studies in Table~\ref{welf_summ}~\cite{awadelkarimRankheterogeneousPreferenceModels2023}. Mixtures of Mallows models suit this task because, once central rankings and dispersion are inferred, individual-level preferences can be generated without being observed.
We depart from convention in two respects: we share dispersion parameters across groups, rather than per group, following the concentric mixture structure of \citet{collasConcentricMixturesMallows2021}, and we calibrate Mallows parameters using only aggregate match statistics rather than individual-level observations. To the best of our knowledge, no prior work has simultaneously used Mallows mixtures for welfare evaluation and calibrated their parameters from aggregate match statistics alone.

%% file: other/welfare_table_pivoted.tex
\begin{table}[b!]
\centering
\caption{\textbf{Design parameters and welfare metrics in school choice literature.}}
\small
\setlength{\tabcolsep}{4pt}
\renewcommand{\arraystretch}{1.15}
\resizebox{\linewidth}{!}{%
\begin{tabular}{l | p{2.4cm} p{2.4cm} p{1.8cm} p{1.8cm} p{1.4cm}}
\toprule
\textbf{Welfare Metric}
  & \makecell[l]{\textbf{Assignment}\\\textbf{mechanism}}
  & \makecell[l]{\textbf{Tie-breaking}\\\textbf{mechanism}}
  & \makecell[l]{\textbf{List cap}}
  & \makecell[l]{\textbf{School}\\\textbf{priorities}}
  & \textbf{Quota} \\
\midrule
Assigned to current school
  &
  &
  &
  &
  & \cite{correaSchoolChoiceChile2022} \\
  Average rank
  &
  & \cite{ashlagiAssigningMoreStudents2019,ashlagiWhatMattersSchool2020}
  & \cite{kanoriaCompetitionPartnersMatching2025}
  &
  & \\
Average utility
  & \cite{xuDesigningSchoolChoice2024}
  & \cite{xuDesigningSchoolChoice2024}
  & \cite{ajayiEmpiricalAnalysisSchool2017, xuDesigningSchoolChoice2024,kanoriaCompetitionPartnersMatching2025}
  &
  & \\
Equality of opportunity
  & \cite{prietoEvaluatingSchoolPriorities2023a}
  &
  &
  & \cite{prietoEvaluatingSchoolPriorities2023a}
  & \\
Expected assignment utility
  & \cite{dehaanPerformanceSchoolAssignment2023,abdulkadirogluWelfareEffectsCoordinated2017,calsamigliaStructuralEstimationModel2020}
  & \cite{dehaanPerformanceSchoolAssignment2023}
  &
  &
  & \\
Rank distribution
  & \cite{dehaanPerformanceSchoolAssignment2023}
  & \cite{ashlagiAssigningMoreStudents2019,ashlagiWhatMattersSchool2020,dehaanPerformanceSchoolAssignment2023}
  &
  &
  & \\
Share matched to top-$p$ choice
  & \cite{prietoEvaluatingSchoolPriorities2023a,dehaanPerformanceSchoolAssignment2023,arnostiShortListsCentralized2015a,agarwalDemandAnalysisUsing2018,xuDesigningSchoolChoice2024,cheEfficiencyStabilityLarge2019}
  & \cite{correaSchoolChoiceChile2022,arnostiLotteryDesignSchool2023,abdulkadirogluStrategyProofnessEfficiencyMatching2009,xuDesigningSchoolChoice2024,dehaanPerformanceSchoolAssignment2023}
  &
  & \cite{prietoEvaluatingSchoolPriorities2023a,casteraCorrelationRankingsMatching2025}
  & \cite{correaSchoolChoiceChile2022} \\
  Share matched/unmatched
  & \cite{cheEfficiencyStabilityLarge2019}
  & \cite{arnostiLotteryDesignSchool2023,correaSchoolChoiceChile2022,ashlagiAssigningMoreStudents2019,abdulkadirogluStrategyProofnessEfficiencyMatching2009}
  & \cite{ajayiEmpiricalAnalysisSchool2017,faenzaLongerListsYield2025}
  & \cite{casteraCorrelationRankingsMatching2025}
  & \cite{correaSchoolChoiceChile2022} \\
Share of envied/jealous students
  & \cite{xuDesigningSchoolChoice2024,cheEfficiencyStabilityLarge2019}
  & \cite{xuDesigningSchoolChoice2024}
  & \cite{xuDesigningSchoolChoice2024}
  &
  & \\
Siblings to same school
  &
  & \cite{correaSchoolChoiceChile2022}
  &
  &
  & \\
  Variance of rank
  &
  & \cite{ashlagiWhatMattersSchool2020}
  &
  &
  & \\
\bottomrule
\end{tabular}}
\label{welf_summ}
\end{table}

%% file: sections/03_preliminaries.tex
\section{Preliminaries}
\label{prel}

\subsection{Student-School Matching}
\label{subsec:matching}

Let $I = \{1,\ldots,n\}$ denote the set of \textbf{applicants} and $\mathcal{J} = \{1,\ldots,m\}$ the set of \textbf{programs}. Schools $s \in \mathcal{S}$ may offer multiple programs $j \in \mathcal{J}_s$, so $\mathcal{J} = \bigcup_{s \in \mathcal{S}} \mathcal{J}_s$; because admissions are organized at the program level, we work with $\mathcal{J}$ throughout. Each program $j \in \mathcal{J}$ has a capacity $c_j \in \mathbb{N}$.  Each applicant $i \in I$ is characterized by an attribute vector $x_i \in \mathcal{X}$ (e.g., gender and residential district), with $X = (x_i)_{i \in I}$ denoting the full attribute profile.

Applicants are partitioned into \textbf{subdivisions} $d \in \mathcal{D}$, with $I_d \subseteq I$ denoting applicants in subdivision~$d$. Subdivisions define the groups for which central preference rankings are estimated (Section~\ref{framework}), and serve as the primary unit for disaggregated welfare reporting. In our two case studies, they correspond to the 32 residential districts of NYC and the 56 provinces of Chile
(Section~\ref{sec:eval}).

Each applicant $i \in I$ submits a \textbf{preference list over programs}, an ordered sequence $R_i = (j_{i1}, j_{i2}, \ldots, j_{iL_i})$ of distinct programs, where $j_{iq} \in \mathcal{J}$ is the program ranked in position~$q$. \textcolor{blue}{We treat $R_i$ as a \emph{stated} preference list, which need not reflect the applicant's true preferences~\citep{fackTruthTellingPreferenceEstimation2019}.} Programs not listed by applicant~$i$ are treated as unacceptable. The \textbf{list length} $L_i$ satisfies $L_{\min} \le L_i \le L_{\max}$, where the minimum and maximum list lengths are \emph{rules} set by the policymaker; typically $L_{\max} \ll m$, so applicants rank only a small fraction of programs. Within these bounds, $L_i$ reflects the applicant's \emph{choice}, which we model as a draw from a distribution $\mathcal{L}(\theta_L)$ with parameters~$\theta_L$. We denote by $P = (R_i)_{i \in I}$ the full preference profile and by $A_j(P) = \{i \in I : j \in R_i\}$ the set of applicants who list program~$j$.


\textbf{Programs rank applicants} by a priority score $\rho_{ij} \in \mathbb{R}$, where lower values indicate higher priority: $i \succ_j i' \iff \rho_{ij} < \rho_{i'j}$. The score combines three components, $\rho_{ij} = \alpha b_{ij} + \beta t_{ij} + \lambda_{ij}$: a \emph{reserve bucket}~$b_{ij}$ that assigns the applicant to a seat pool (e.g., general education or Students with Disabilities); a \emph{priority tier}~$t_{ij}$ that encodes program-specific priorities such as sibling enrollment or continuing student status; and a \emph{lottery number} $\lambda_{ij} \in [0,1)$ that breaks remaining ties. Choosing $\alpha \gg \beta \gg 1$ ensures that buckets dominate tiers, which dominate the lottery. Under STB, $\lambda_{ij} = \lambda_i$ for all programs~$j$; under MTB, each $\lambda_{ij}$ is drawn independently.

Under \textbf{single tie-breaking} (STB), each applicant receives one lottery draw $\lambda_i \in [0,1)$ used at every program ($\lambda_{ij} = \lambda_i$ for all $j$). Under \textbf{multiple tie-breaking} (MTB), $\lambda_{ij}$ is drawn independently for each applicant-program pair, giving each applicant an independent lottery position at every program. Each jurisdiction determines how $b_{ij}$, $t_{ij}$, and $\lambda_{ij}$ are constructed from applicant attributes and admission policies; the matching algorithm requires only the resulting priority matrix $\rho = (\rho_{ij})_{i \in I,\, j \in \mathcal{J}}$.

A \textbf{matching} is a function $\mu : I \to \mathcal{J} \cup \{\emptyset\}$, where $\mu(i) = j$ means applicant~$i$ is assigned to program~$j$, and $\mu(i) = \emptyset$ means $i$ is unmatched. A matching is feasible if $|\{i \in I : \mu(i) = j\}| \le c_j$ for every $j \in \mathcal{J}$. A \textbf{matching mechanism}~$M$ takes as input the preference profile~$P$, the capacity vector $c = (c_j)_{j \in \mathcal{J}}$, and the priority matrix~$\rho$, and returns a feasible matching $\mu = M(P, c, \rho)$.
The \textbf{realized rank} of applicant~$i$ under~$\mu$ is $r_i(\mu) = q$ if $\mu(i) = j_{iq}$, and $r_i(\mu) = +\infty$ if $\mu(i) = \emptyset$.



The primary student-side \textbf{welfare metrics} (summarized in Table~\ref{tab:welfare}, Appendix~\ref{app:notation}) are the unmatched rate $W_{\mathrm{unmatch}}(\mu) = \frac{1}{n}\sum_{i\in \mathcal{I}}\mathbbm{1}\{\mu(i)=\emptyset\}$ (or match rate $W_{\mathrm{match}}(\mu) = 1 - W_{\mathrm{unmatch}}(\mu)$), the average matched rank $\bar{r}(\mu) = \frac{1}{|\{i:\mu(i)\neq\emptyset\}|}\sum_{i:\mu(i)\neq\emptyset} r_i(\mu)$, and the top-$p$ match rate $W_p(\mu) = \frac{1}{n}\sum_{i\in \mathcal{I}}\mathbbm{1}\{r_i(\mu)\leq p\}$. Each metric can be computed for any subgroup $\mathcal{I}_g$, where $g$ may be a subdivision, lottery decile, gender, or other characteristic, which we write as, e.g., $W_p(\mu \mid g)$; \textbf{subdivision-level welfare} $W_p(\mu \mid d)$ is the primary lens of our policy analyses. On the program side, $u_j(\mu)=\frac{1}{c_j}\sum_{i\in \mathcal{I}}\mathbbm{1}\{\mu(i)=j\}\in[0,1]$ measures capacity utilization, and for a subdivision~$d$ we report the capacity-weighted aggregate $\bar u_d(\mu)=\frac{\sum_{j\in \mathcal{J}_d}\sum_{i\in \mathcal{I}}\mathbbm{1}\{\mu(i)=j\}}{\sum_{j\in \mathcal{J}_d}c_j}$, which we also use to compare simulated and observed utilization during calibration.

\subsection{Mixture of Mallows Models}
\label{subsec:mallows}

A \textbf{Mallows distribution} over $\mathcal{J}$ is parameterized by a \emph{central ranking} $\sigma$ (a strict ordering of $\mathcal{J}$) and a \emph{dispersion parameter} $\phi \in (0,1)$. The probability of a complete ranking $\tau$ is $\Pr(\tau \mid \sigma, \phi) \;\propto\; \phi^{\,d_K(\tau,\,\sigma)}$, where $d_K(\tau, \sigma)$ is the Kendall distance: the number of pairwise disagreements between $\tau$ and $\sigma$. Smaller $\phi$ concentrates mass near $\sigma$; as $\phi \to 1$ the distribution approaches uniform over all rankings. The Repeated Insertion Model (RIM)~\cite{Doignon_Pekec_Regenwetter_2004} can be used to efficiently draw top-truncated preferences of length~$L_i$ consistent with $\mathrm{Mallows}(\sigma, \phi)$.


A \textbf{mixture of $K$ Mallows models} for subdivision $d$ is a weighted combination of Mallows distributions sharing the central ranking $\sigma_d$ but differing in dispersion. Formally, a complete ranking $\tau_i$ for applicant $i \in I_d$ is drawn as $\tau_i \sim \sum_{k=1}^{K} w_k \cdot \mathrm{Mallows}(\sigma_d, \phi_k)$, where $w_k \ge 0$, $\sum_k w_k = 1$ are mixture weights, and $\phi_1, \ldots, \phi_K \in (0,1)$ are component-specific dispersion parameters. This structure captures within-subdivision heterogeneity (through the mixture of dispersions $\{\phi_k\}$) while anchoring each component to the same subdivision-level preference ordering $\sigma_d$.



%% file: sections/04_framework.tex
\section{Trace (Trade-offs in Rules, Agency, Chance, and Equity)}
\label{framework}

\input{other/compact_pseudocode}

\sys operates in two phases: \textbf{Phase~1} (Lines~1--11 of Algorithm~\ref{alg:trace}) calibrates a mixture of Mallows models to aggregate welfare statistics; \textbf{Phase~2} (Lines~13--14) samples a synthetic preference profile from the calibrated model, runs a matching mechanism, and computes welfare metrics.  Three entry points determine where a user begins. \textbf{Entry~point~1} (\emph{realized profile available}): the user supplies $P$ directly, skips Phase~1, and starts at the matching step by choosing a mechanism $M$ and priority configuration $\rho$. \textbf{Entry~point~2} (\emph{mixture parameters available}): the user supplies $\{\sigma_d,\phi_k\}$ directly and enters at the sampling step; this mode also supports robustness analysis by varying $\sigma$ and $\phi$ explicitly. \textbf{Entry~point~3} (\emph{aggregate statistics only}): the user
supplies observed or target welfare statistics $\{o_d\}_{d\in\mathcal{D}}$ and the framework executes the full pipeline. The remainder of this section describes Entry~point~3, which subsumes the others.

\smallskip\textbf{Inputs.} Entry~point~3 requires four classes of input: (i)~\emph{student-side data}: subdivision sizes $\{|I_d|\}$ 
and the empirical list-length distribution $\mathcal{L}(\theta_L)$; (ii)~\emph{program-side data}: identifiers $\mathcal{J}$, capacities $c_j$, and utilization counts; (iii)~\emph{aggregate welfare statistics} $\{o_d\}_{d\in\mathcal{D}}$, which serve as calibration targets; and (iv)~a \emph{priority configuration} specifying reserve-bucket fractions, priority tiers, and tie-breaking rules used to construct $\rho_{ij}$. 

\smallskip\textbf{Phase~1: Parameter Estimation.} Mixture weights are initialized uniformly: $w_k = 1/K$; dispersion parameters are initialized as $\phi_k \sim \mathrm{Beta}(a,b)$. For each $d\in\mathcal{D}$, the central ranking $\sigma_d$ is initialized by sorting programs in descending order of a competitiveness measure $\gamma_d(j)$. For example, the ratio $T_{d,j}^2/A_{d,j}$ of true squared to total applicants from subdivision $d$ at program $j$ (see Section~\ref{sec:eval}).  Next, during EM-like optimization loop, the algorithm iterates for up to $T_{iter}$ rounds. In each round, for component $k$ in sequence and candidate value $\varphi$, 
$N_{\mathrm{sim}}$ synthetic preference profiles are simulated under the current parameter vector, 
with $\phi_k$ temporarily replaced by $\varphi$: for each $i \in I_d$, a component $k_i$ is drawn from $\{w_k\}$, a full ranking $\tau_i \sim \mathrm{Mallows}(\sigma_d, \phi_{k_i})$ is sampled via RIM~\citep{Doignon_Pekec_Regenwetter_2004} and truncated to length $L_i \sim \mathcal{L}(\theta_L)$ to obtain $R_i$~\citep{Chakraborty2021}. Each profile is then matched using $M(P^{(m)}, c, \rho)$, and the welfare vector $\tilde{o}_{d,k}^{(m)}(\varphi) \in \mathbb{R}^v$ is computed per subdivision, with v denoting the number of welfare statistics included in the calibration target. A shared lottery draw $(\lambda_{ij})_{i \in I,\; j \in \mathcal J}$ is held fixed across all $N_{\mathrm{sim}}$ simulations and $\phi$-evaluations within a round to reduce variance in the stochastic objective. 

The empirical mean and covariance of the simulated welfare vectors, for a candidate value $\varphi$ replacing $\phi_k$, are
\begin{equation}\label{eq:gauss}
    \bar{o}_{d,k}(\varphi) :=
    \frac{1}{N_{\mathrm{sim}}}
    \sum_{m=1}^{N_{\mathrm{sim}}} \tilde{o}_{d,k}^{(m)}(\varphi),
    \qquad
    \Sigma_{d,k}(\varphi) :=
    \frac{1}{N_{\mathrm{sim}}-1}
    \sum_{m=1}^{N_{\mathrm{sim}}}
    \left(\tilde{o}_{d,k}^{(m)}(\varphi)-\bar{o}_{d,k}(\varphi)\right)
    \left(\tilde{o}_{d,k}^{(m)}(\varphi)-\bar{o}_{d,k}(\varphi)\right)^\top
    + \varepsilon I_v,
\end{equation}
where $\varepsilon I_v$ is a small regularization term. The Gaussian log-likelihood of the observed statistics $o_d$ is
\begin{equation}\label{eq:lik}
    \ell_{d,k}(\varphi) :=
    -\tfrac{1}{2}
    \left(o_d-\bar{o}_{d,k}(\varphi)\right)^\top
    \Sigma_{d,k}(\varphi)^{-1}
    \left(o_d-\bar{o}_{d,k}(\varphi)\right)
    -\tfrac{1}{2}\log\left|\Sigma_{d,k}(\varphi)\right|,
\end{equation}
and the total objective is
\begin{equation}\label{eq:obj}
    \mathcal{Q}_k(\varphi) :=
    \sum_{d\in\mathcal{D}} \ell_{d,k}(\varphi).
\end{equation}
$\mathcal{Q}_k(\varphi)$ is maximized over $\varphi\in(0,1)$ using Brent's method~\citep{brent1973algorithms}, initialized with a bracket near the current $\phi_k$; the updated value is held fixed before optimizing $\phi_{k+1}$.
After all $\{\phi_k\}$ are updated, the central rankings are revised via a utilization-based nudge. For each program $j$, define the filled-count error $\delta_j = \hat{f}_j - \tilde{f}_j$, where $\hat{f}_j$ and $\tilde{f}_j$ are the observed and mean-simulated filled counts. Popularity scores are updated as $\gamma_d(j) \leftarrow \gamma_d(j) + \eta\cdot\delta_j$, and $\sigma_d$ is re-derived by re-sorting programs by the updated scores. Programs consistently overfilled in simulation are demoted; programs consistently underfilled are promoted. The nudge is applied after recording the best-so-far parameters, serving as a warm start for the next round.
The loop terminates early when $\max_k|\Delta\phi_k|<\epsilon_{tol}$. The configuration $(\{\phi_k^\ast\},\{\sigma_d^\ast\},\{w_k\})$ achieving the highest recorded objective value $Q^\star$ across all rounds is returned.

\smallskip\textbf{Phase~2: Welfare Evaluation.}  Given $(\{\phi_k^*\},\{\sigma_d^*\},\{w_k\})$, or parameters supplied directly at Entry~point~2, a synthetic profile $P = \{R_i\}_{i\in I}$ and attribute profile $X$ are generated: for each $i\in I_d$, draw $k_i \sim \mathrm{Categorical}(w_1,\ldots,w_K)$, sample $\tau_i \sim \mathrm{Mallows}(\sigma_d^*,\phi_{k_i}^*)$ via RIM, and truncate to $L_i \sim \mathcal{L}(\theta_L)$.  The resulting lists reflect subdivision-level preference structure through $\sigma_d$ and within-subdivision heterogeneity through the mixture of dispersions $\{\phi_k\}$.
The profile $P$ (or, at Entry point~1, an observed profile) is passed to the matching mechanism, yielding $\mu = M(P,c,\rho)$. The framework then computes the welfare metrics defined in Section~\ref{prel}, both globally and conditional on applicant subgroups, such as subdivisions $I_d$, groups defined by applicant attributes $x_i$, list length $L_i$, priority status, or lottery position.

%% file: other/compact_pseudocode.tex
\begin{algorithm}[b!]
\caption{\sys$(\{I_d\},\,\mathcal{J},\,c,\,\rho,\ M,\{o_d\},\,\mathcal{L}(\theta_L);\;
               K,\,T_{iter},\,N_{\mathrm{sim}},\,\eta,\,\epsilon_{tol})$}
\label{alg:trace}
\small 
\begin{algorithmic}[1]
\Statex \hrulefill\;\textit{Phase~1}\;\hrulefill
\State \textbf{Init:}\;
  $w_k \leftarrow 1/K$;\;
  $\phi_k \sim \mathrm{Beta}(a,b)$;\;
  $\sigma_d \leftarrow \mathrm{argsort}_j\,\gamma_d(j)$; $Q^\star \leftarrow -\infty$
\For{$t = 1,\ldots,T_{iter}$}
  \For{$k = 1,\ldots,K$}
    \State \parbox[t] {0.88\linewidth}{Simulate $N_{\mathrm{sim}}$ profiles from 
$k_i \sim Categorical(w_1,…,w_K)$, 
$\tau_i \sim \mathrm{Mallows}(\sigma_{d(i)}, \phi_{k_i})$, 
replacing $\phi_k$ by its candidate value
           via RIM + truncation to $L_i \sim \mathcal{L}(\theta_L)$}
    \State Run $M(P^{(m)},c,\rho)$ each simulation;
           compute $\tilde{o}_d^{(m)}$ per $d$
    \State $\phi_k \leftarrow \arg\max_{\varphi \in (0,1)} Q_k(\varphi)$
           \hfill\Comment{Brent's method, Eqs.~\ref{eq:gauss}--\ref{eq:obj}}
  \EndFor
  \State \textbf{if} $Q^{(t)} > Q^\star$:  
          $Q^\star \leftarrow Q^{(t)}$ \textbf{and} Record $(\{\phi_k\},\{\sigma_d\})$ 
  \State \textbf{break} if $\max_k|\Delta\phi_k|<\epsilon_{tol}$
  \State \textbf{for all } $d \in D$ \textbf{ do }
        $\sigma_d \leftarrow \operatorname{argsort}_{j \in \mathcal{J}} \gamma_d(j)$
        where $\gamma_d(j) \leftarrow \gamma_d(j) + \eta\delta_j$
        \hfill\Comment{Utilization nudge, warm start}
\EndFor
\State \Return $\{\phi_k^*\},\,\{\sigma_d^*\},\,\{w_k\}$
       \hfill\Comment{$w_k=1/K$ fixed; Phase~1 output}
\Statex \hrulefill\;\textit{Phase~2}\;\hrulefill
\State Sample $P{=}\{R_i\}_{i\in I}$ from $\prod_{d\in D}\prod_{i\in I_d} \left[\sum_{k=1}^{K} w_k\, \operatorname{Trunc}_{L_i}(\mathrm{Mallows}(\sigma_d^*,\phi_k^*))\right]$
via RIM + truncation.
\State $\mu \leftarrow M(P,c,\rho)$;\;
       compute and report $W(\mu)$ per Section~\ref{prel}
\end{algorithmic}
\end{algorithm}

%% file: sections/05_case_studies.tex
\section{Case Studies}
\label{sec:eval}

We apply \sys to two real-world school matching systems. In \textbf{NYC}, where only aggregate welfare statistics are available (Entry Point 3, Section~\ref{framework}), we use \sys to generate synthetic preference profiles and study the welfare effects of a minimum list length requirement. \textcolor{blue}{In \textbf{Chile}, we use both entry points. For the welfare analysis, we compare MTB with STB on the observed individual-level profiles $P$ (Entry Point 1). For validation, we also generate profiles from aggregate data alone (Entry Point 3) and check that they yield the same conclusions, showing that \sys is useful when individual-level data are unavailable.}  Both analyses disaggregate results by two dimensions of \emph{chance}: lottery position and residential location (boroughs in NYC, regions in Chile). \emph{Choice} enters through how many schools students rank and is revisited in Section~\ref{sec:eval:rules}. 

\smallskip\textbf{NYC HS students} are matched via the student-proposing DA algorithm \cite{abdulkadirogluNewYorkCity2005a, nyc2025NYCPUBLIC2025}, submitting rank-ordered lists during October--November. Programs allocate seats across reserve pools (e.g., general education and Students with Disabilities) and rank applicants by priority tiers such as district residence and sibling enrollment within each pool. Ties are broken by a single global lottery draw (STB) $\lambda_i \in [0,1)$ shared across all programs, 
which is also the primary admission criterion in \textit{Unscreened} programs \cite{marianDecodingNYCSchool2025}. Recent reforms include \textcolor{blue}{visible lottery numbers for families \cite{NYCFamiliesApplying2023}}, 
the removal of the 12-program ranking cap 
~\cite{marianResults2025NYC2025a, NYCAdmissionsAre2024}, and an official admission prediction tool~\cite{marianResults2025NYC2025a, OfferChances}. 

We use open-source data from the 2024--2025 admission cycle: application and offer statistics~\cite{StudentApplicationsAdmissions}, directory data~\cite{DirectoryData}, and admission outcomes~\cite{AdmissionsOutcomes}. \textcolor{blue}{Importantly, applicants in our study period already had access to their lottery numbers, reflecting the earlier transparency reform.} 
Together, these data sources provide the inputs to Algorithm~\ref{alg:trace}: (1) district-level welfare moments $W_p(\mu \mid d)$ and $W_{\mathrm{unmatch}}(\mu \mid d)$ as calibration targets; (2) program-level capacities $c_j$; and (3) the statistic $\mathrm{Ratio}_{d,j} = T_{d,j}^2 / A_{d,j}$ to initialize central rankings $\sigma_d$. Here, $A_{d,j} = |A_j(P) \cap I_d|$ denotes the number of \emph{total applicants}, i.e., students from district $d$ who listed program $j$. In contrast, $T_{d,j} = |T_j(P,\mu) \cap I_d|$ denotes the number of \emph{true applicants}, i.e., students who listed $j$ but were not matched to any higher-ranked program, so that their application
to $j$ was genuinely competitive. The ratio upweights programs with concentrated competitive demand relative to their total applicant pool. Our dataset covers 70{,}807 HS applicants across the 32  residential districts of NYC defining subgroups $I_d$. \textcolor{blue}{We calibrate at the district level and aggregate results to the five boroughs for reporting.}

\smallskip\textbf{Chilean HS students} are matched via the 
student-proposing DA~\cite{correaSchoolChoiceChile2022}, submitting rank-ordered lists during August--October. Programs rank applicants by nationally defined priority tiers: enrolled siblings, priority students (up to a 15\% quota), children of school employees, and former students. Unlike NYC, ties are broken by application-specific lottery draws (MTB) $\lambda_{ij} \in [0,1)$: each draw is independent across school-program pairs, so a student's lottery luck varies by application. A 2026 reform replaces random draws with deterministic numbers obtained by hashing student IDs and program identifiers~\cite{lydSAECAMBIANDOTODO2025, vialDIARIOOFICIALREPUBLICA}, leaving the assignment logic intact. We study the welfare effects of implementing STB (a single $\lambda_i$ shared across all programs) versus the existing MTB. We use open-source data from the 2024--2025 admission cycle \cite{ministeriodeeducacionSistemaAdmisionEscolar}. Our dataset covers 124,492 HS applicants across the 56 provinces  of Chile defining subgroups $I_d$. \textcolor{blue}{We calibrate at the province level and aggregate results to the regions for reporting.}

\subsection{Validation}
\label{sec:eval:val}

\begin{figure}[t]
\centering
\begin{minipage}[t]{0.48\linewidth}
  \centering
  \includegraphics[width=0.85\linewidth]{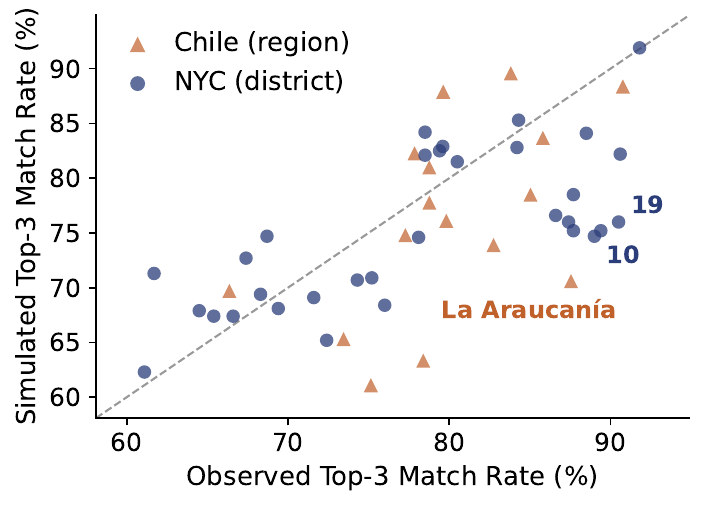}
  \caption*{(a) Top-3 match rate $W_3(\mu \mid d)$.}
\end{minipage}
\hfill
\begin{minipage}[t]{0.48\linewidth}
  \centering
  \includegraphics[width=0.85\linewidth]{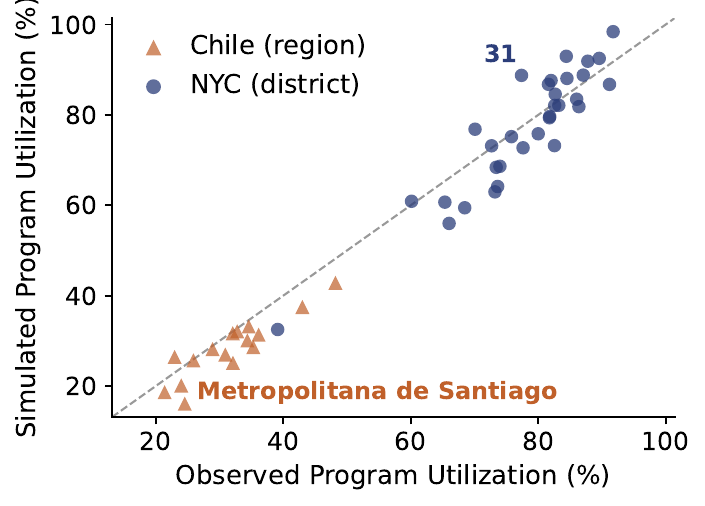}
  \caption*{(b) Program utilization $\bar u_d(\mu)$.}
\end{minipage}
\caption{\textbf{Observed vs.\ simulated validation metrics for NYC districts (circles) and Chilean regions (triangles).} Points near the 45° dashed line indicate good fit. We report the mean absolute error (MAE) across units, in percentage points (pp). For the top-3 match rate, NYC fits more tightly than Chile (MAE 5.55pp vs.\ 6.58pp); for program utilization, Chile fits more tightly than NYC (3.71pp vs.\ 4.90pp). \textcolor{blue}{Two simpler baselines fit much worse. (i) \emph{No dispersion:} keeping each district's/region's central ranking $\sigma$ but setting all $\phi_i=0$ (every student submits $\sigma$) yields MAEs of 68.05pp (top-3) and 58.60pp (utilization) for NYC, and 54.33pp and 24.35pp for Chile. (ii) \emph{No local ranking:} keeping the fitted $\phi$ values but replacing the local rankings with a single system-wide $\sigma$ that orders programs by total applicants yields 45.44pp and 48.71pp for NYC, and 18.44pp and 15.68pp for Chile.}}

\label{fig:val_scatter}
\vspace{-0.4cm}
\end{figure}

\textcolor{blue}{We validate our framework in three ways. First, we compare observed and simulated welfare metrics across NYC districts and Chilean regions (Figure~\ref{fig:val_scatter}; further metrics in Appendix~\ref{app:obs_vs_sim}), discussed below. Second, in Chile, we calibrate on a subset of aggregate statistics and evaluate fit on the remaining held-out ones (Appendix~\ref{app:held_out}).}\footnote{\textcolor{blue}{``Held-out'' does not refer to a machine-learning train/test split. These are aggregate statistics we deliberately exclude when calibrating model parameters, so that they provide an independent check of fit, reflecting whether the fitted model generalizes beyond statistics that it was explicitly optimized to match.}} \textcolor{blue}{Third, also in Chile, we rerun the counterfactual analysis on profiles generated by \sys from aggregate statistics alone, and check whether the main conclusions drawn from the observed individual-level profiles still hold (Appendix~\ref{app:fidelity}).} 

\textbf{NYC.} The synthetic preference model recovers the cross-district pattern reasonably well. The simulated share of students matched to a top-3 choice, $W_3(\mu \mid d)$, deviates from the observed share by 5.55pp on average across the 32 districts (MAE); 27 of 32 districts lie within 10pp. This error is smaller than in Chile (6.58pp), where comparable fit preserves the policy conclusions (Appendix~\ref{app:fidelity}). For program utilization $\bar u_d(\mu)$, the MAE is 4.90pp. 


\smallskip\textbf{Chile.} Since full preference profiles $P$ are observed, we aggregate them as for NYC (Entry point 3), apply our framework to the aggregates alone, and compare the simulated welfare statistics to the observed ones.\footnote{\textcolor{blue}{These simulated profiles are reused in Appendix~\ref{app:fidelity}.}} We run the model at the province level (Santiago is split into sub-provinces) and report results for the 16 regions. Across regions, the MAE is 6.58pp on $W_3(\mu \mid d)$ and of 3.71pp on $\bar u_d(\mu)$. Several provinces are recovered with near-zero error: Capitán Prat (0.00pp), Ranco (0.10pp), Tamarugal (0.20pp), and Copiapó (0.40pp).
The largest $W_3$ deviation occurs in La Araucan\'{i}a. Its province Cautín, which accounts for 77.5\% of the region's applicants, has unusually dispersed preferences: only 13.4\% of students' top-3 choices overlap with the top 3 of $\sigma_d$, compared with 32.9\% nationally.


\subsection{Chance under Current Rules}
\label{sec:eval:chance}

Before evaluating any \emph{rule} change, we ask how much lottery \emph{chance} already shapes welfare under each system's current rules. One such rule is the tie-breaking mechanism used: NYC uses a single global draw $\lambda_i$ (STB), while Chile uses per-school draws $\lambda_{ij}$ (MTB). 

\begin{figure}[t]
\centering
\begin{minipage}[t]{0.48\textwidth}
  \centering
  \includegraphics[width=\textwidth]
{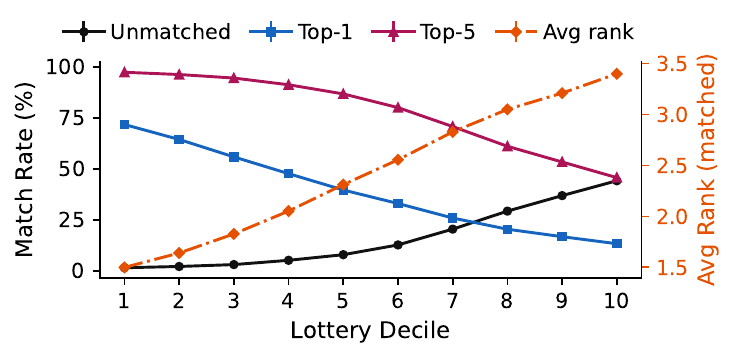}
 \caption{\textbf{NYC:} $W_1$, $W_5$, $W_{\mathrm{unmatch}}$, and $\bar{r}$ by lottery decile. The 1st decile achieves a $W_5$ of 97.3\% and an $\bar{r}$ of 1.50, compared with 45.7\% and 3.41 for the 10th decile. The resulting 51.6 pp $W_5$ gap highlights the welfare gradient induced by lottery chance. \textcolor{blue}{95\%  CIs are smaller than the markers.}}
  \label{fig:lotto-dist}
\end{minipage}
\hfill
\begin{minipage}[t]{0.48\textwidth}
  \centering 
\includegraphics[width=\textwidth]
{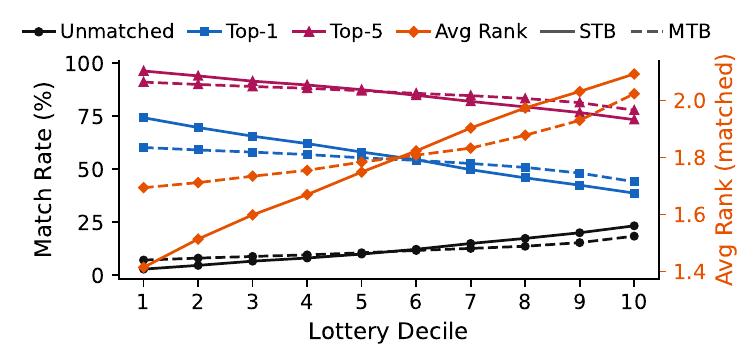}
  \caption{\textbf{Chile:} $W_1$, $W_5$, $W_{\mathrm{unmatch}}$, and $\bar{r}$ by lottery decile under STB and MTB, averaged over 10 simulations. MTB deciles rank students by their best relative lottery position across top-5 choices; STB deciles rank by the single lottery draw. \textcolor{blue}{ 95\% CIs are smaller than the markers.}}
  \label{fig:chilestbmtblottery}
\end{minipage}
\end{figure}

In \textbf{NYC}, the single draw $\lambda_i$ acts as a global admission priority across every program on a student's list. Figure~\ref{fig:lotto-dist} shows the consequences of moving from the luckiest lottery decile to the unluckiest, $W_{\mathrm{unmatch}}$ rises from 1.5\% to 44.1\%, $W_1$ falls from 71.8\% to 13.3\%, and $\bar{r}$ jumps from 1.50 to 3.41. A gap of 51.6pp in $W_5$ separates the best and worst deciles, driven entirely by chance.

 In \textbf{Chile}, the MTB baseline looks different. Figure~\ref{fig:chilestbmtblottery} shows that outcomes vary much less across lottery deciles under MTB: $W_1$ falls by 16pp from the luckiest to the unluckiest decile, compared with 58.5pp in NYC.
  Because each $\lambda_{ij}$ is drawn independently, a poor draw at one program does not carry over to others.

Together, these comparisons show that \emph{chance} is not a fixed background condition: its influence on welfare is itself shaped by the prevailing tie-breaking \emph{rule}. The following sections examine what happens when those \emph{rules} change.

\subsection{Rules and Choice: Coverage--Quality Trade-offs}
\label{sec:eval:rules}

A recurring pattern across both systems is that \emph{rule} changes do not uniformly improve welfare: gains in match coverage come at the cost of match quality, and vice versa.

\begin{figure}[t!]
\centering
\includegraphics[width=\linewidth]
{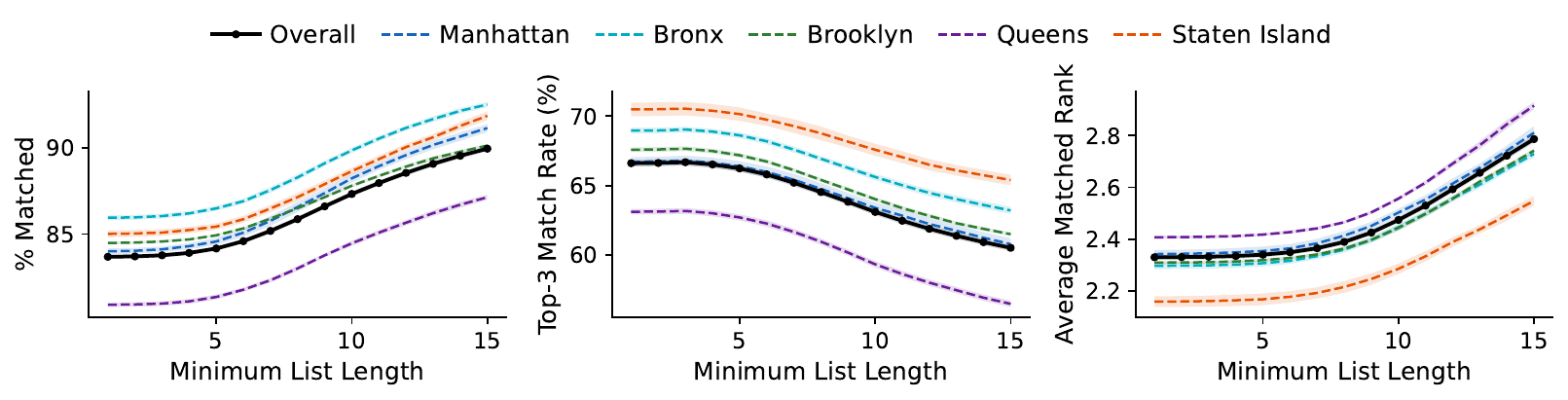}
\caption{\textbf{NYC:} Impact of minimum list length on $W_{\mathrm{match}}$, $W_3$, and $\bar{r}$ by borough. \textcolor{blue}{The minimum acts as a floor on submitted list length, so it binds only for students who would otherwise rank fewer than $L_{\min}$ programs}. \textcolor{blue}{$\bar{r}$ is computed only over students who are matched at the baseline $L_{\min}=1$. Longer lists improve overall coverage but worsen rank-based welfare, with heterogeneous effects across boroughs. The shading represents a 95\% CI.}}
\label{fig:nyc_min_impact_main}
\end{figure}

\smallskip\textbf{NYC: minimum list length.} Figure~\ref{fig:nyc_min_impact_main} shows the effect of imposing a minimum list length, disaggregated by borough. As the minimum increases from 1 to 15, $W_{\mathrm{match}}$ rises in all boroughs, from 83.7\%  to 90.0\% on average.  At the same time, $W_3$ falls from 66.6\% to 60.5\%, and $\bar{r}$ worsens from 2.33 to 2.79. \textcolor{blue}{ Importantly, $\bar{r}$ is computed only over students already matched at the baseline $L_{\min}=1$, so its increase does not reflect newly matched students at low ranks, but shows that \emph{previously matched students are pushed to less preferred programs}.} All boroughs follow this pattern, but at different levels: Queens has the lowest match and top-3 rates throughout, and Staten Island the highest top-3 rate (see Section \ref{sec:eval:geo}).

\textcolor{blue}{These results show how a \emph{rule} interacts with student \emph{choice}. In our simulation, students who would rank fewer than $L_{\min}$ programs are extended using the next programs in their own preference ranking. This is a best case: the added programs reflect students' true preferences. In practice, ranking programs in an informed way requires time and resources that are unequally distributed across families \citep{robertsonModelingAssumptionsClash2021, kloostermanSchoolChoiceAsymmetric2020}, so students forced to lengthen their list may fill the added slots with guesses. Minimum list-length rules may therefore need to be paired with accessible school information for longer lists to translate into better matches across the socioeconomic spectrum.}

\begin{wrapfigure}{r}{0.56\textwidth}
  \centering
\includegraphics[width=0.5\textwidth]{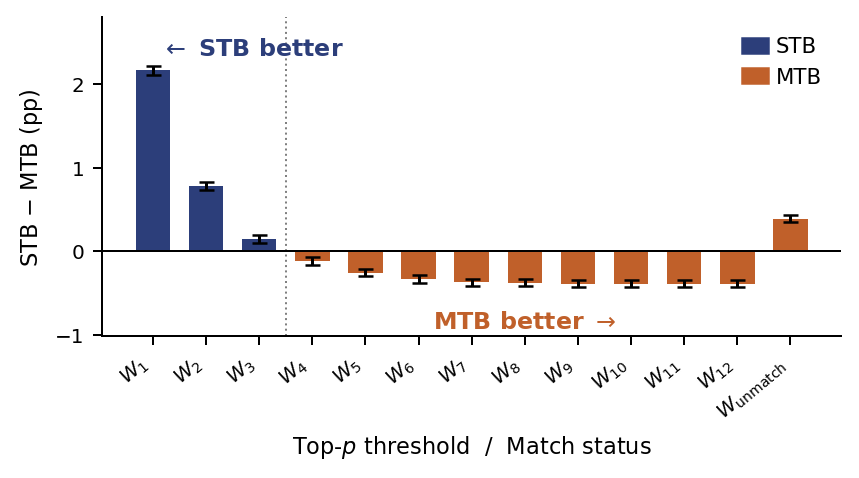}
  \caption{\textbf{Chile:} STB$-$MTB difference in $W_p$ and $W_{\mathrm{unmatch}}$ (pp). Blue bars favor STB, orange bars favor MTB. STB gains 2.18pp at $p=1$, but MTB holds a small, stable edge from $p=4$ and leaves 0.39pp fewer students unmatched. Error bars show 95\% CIs.}
  \label{fig:stb_mtb_diff}
\end{wrapfigure}

\smallskip\textbf{Chile: STB vs.\ MTB.} 
Figure~\ref{fig:stb_mtb_diff} shows an analogous trade-off but for the tie-breaking rule. STB raises $W_1$ by 2.18pp (from 53.94\% under MTB to 56.12\%) and outperforms MTB up to $p=3$. From $p = 4$ the ordering reverses: MTB is marginally better, and $W_{\mathrm{unmatch}}$ is 0.39pp higher under STB. The same trade-off holds for female and non-female students, with nearly identical STB$-$MTB differences (Table~\ref{tab:mtb_stb_summary}, Appendix~\ref{app:mtb_stb}). \textcolor{blue}{This crossover aligns with the theoretical results of \citet{arnostiLotteryDesignSchool2023} (Appendix~\ref{app:stb_mtb_crossover_theory}): a favorable draw under STB applies at every program and favors top choices, while MTB's independent draws give students new chances further down their lists.}

Taken together, these results illustrate a central finding of our framework: A \emph{rule} that improves one welfare outcome may worsen another, and the size of this trade-off varies across groups. Evaluating \emph{rules} through a single metric, such as $W_{\mathrm{match}}$ or $W_1$ alone, would systematically obscure these tensions.  More broadly, the results across both settings illustrate all three components of the framework: the \emph{rule} reshapes the opportunity set, \emph{chance} determines how that opportunity is realized, and \emph{choice} - which schools to rank and how many - mediates how much of any potential gain students can capture.

\subsection{Rules and Chance: Lottery Position}
\label{sec:eval:lottery}

The interaction between \emph{rules} and \emph{chance} is not symmetric across lottery positions: a \emph{rule} change can help students with unfavorable draws while leaving those with favorable draws largely unaffected, or amplify the advantage of a good draw.

\begin{figure}[t!]
\centering
\includegraphics[width=\linewidth]
{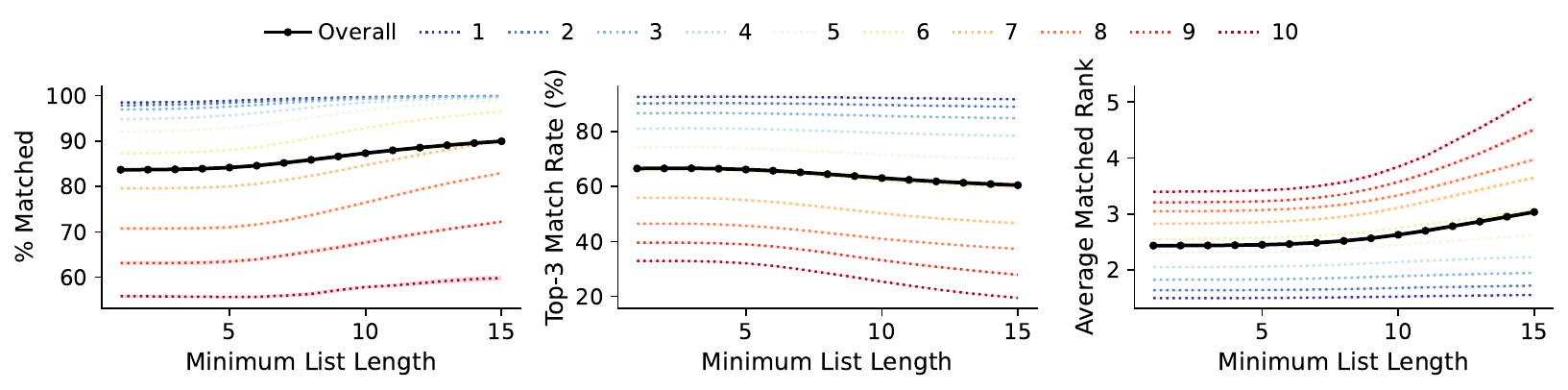}
\caption{\textbf{NYC:} Impact of minimum list length on $W_{\mathrm{match}}$, $W_3$, and $\bar{r}$ by lottery decile. $\bar{r}$ is computed only over students matched at the baseline $L_{\min}=1$. Longer lists disproportionately improve $W_{\mathrm{match}}$ for worse-draw students, at the cost of  lower $W_3$ and higher $\bar{r}$.}
\label{fig:nyc_min_impact_lotto}
\end{figure}

\smallskip\textbf{NYC: minimum list length.} Figure~\ref{fig:nyc_min_impact_lotto} disaggregates the impact of the minimum list length by lottery decile (1-10). Students in the most favorable deciles (1-4) experience relatively stable outcomes as the minimum increases: $W_{\mathrm{match}}$ remains high, and $\bar{r}$ worsens only modestly once the minimum exceeds roughly ten programs. Students with poor draws gain coverage but lose match quality: $W_{\mathrm{match}}$ rises by 12.2 pp for the 8th decile, while for the 10th decile $\bar{r}$ worsens from 3.40 to 6.42 and $W_3$ falls from 32.9\% to 19.4\%. The rule thus reduces the $W_{\mathrm{match}}$ gap between lucky and unlucky students  but widens the quality gap, at the expense of the very students it is meant to help. \textcolor{blue}{In absolute terms, a 15-program minimum would match about 4,450 additional students, while about 4,300 fewer students would receive a top-3 match.} 

\smallskip\textbf{Chile: STB vs.\ MTB.} Figure~\ref{fig:chilestbmtblottery} shows that the contrast between STB and MTB along the lottery dimension is stark. Under MTB, $W_{\mathrm{match}}$ and top-choice rates vary much less across deciles, since per-school draws spread lottery luck across applications. Under STB, the common tie-breaker $\lambda_i$ creates a steep gradient: students in the best deciles are substantially more likely to receive their first or top-5 choice, while outcomes for the worst deciles deteriorate sharply. Thus, compared with the current MTB rule, a counterfactual STB rule would amplify the role of chance, making a student's single lottery draw matter far more for match quality. Because MTB assigns no single draw per student, we sort MTB students by their best draw across their top-5 programs. The steeper STB gradient in $W_5$ and $W_{\mathrm{unmatch}}$ holds for all alternative definitions (see  Figure \ref{fig:fig4_alternates}).

Together, these analyses show that the lottery-luck dimension of \emph{chance} is sensitive to \emph{rule} design. \emph{Rules} that appear neutral, e.g., requiring longer lists, or standardizing tie-breaking, can substantially reshape who bears the cost of an unfavorable lottery draw.

\subsection{Rules and Chance: Students' Residential Location}
\label{sec:eval:geo}

\begin{figure}[b]
\centering
\includegraphics[width=\linewidth]%
{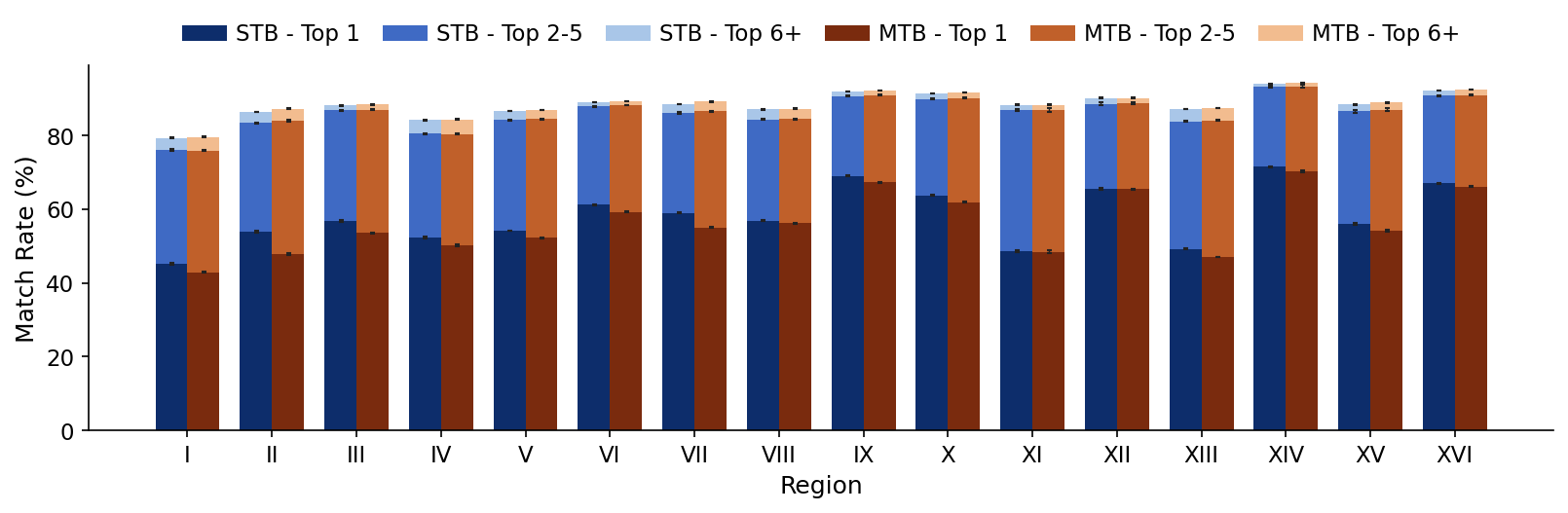}
\caption{\textbf{Chile:} Assignment outcomes under STB (blue) and MTB (orange) by region. Stacked segments show the share of students matched to their first choice, to a choice ranked 2--5, and to a lower-ranked choice; total bar height is $W_{\mathrm{match}}$. Error bars show 95\% CIs. Region numerals are listed in Table~\ref{tab:chilean_regions_roman}.}
\label{fig:chilestbmtbregion}
\end{figure}

We now turn to residence as the second dimension of \emph{chance}.  The borough-level results in Figure~\ref{fig:nyc_min_impact_main} already show that \emph{rule} effects are not uniform across residential areas. \textcolor{teal}{Queens students face structurally lower $W_{\text{match}}$ and higher $\bar{r}$ at every minimum list length, and bear the steepest rank degradation as the minimum increases. This pattern is present in the observed data as well (Appendix~\ref{app:obs_vs_sim}); we note that district-level fit is weak for Queens, so the magnitude, though not the direction, of this gradient carries uncertainty.}
Manhattan benefits the most in overall $W_{\mathrm{match}}$ ($+$7.2 pp) while Staten Island suffers the mildest $W_3$ decline ($-$5.1 pp). Chile's data allows a richer analysis of this dimension. Figure~\ref{fig:chilestbmtbregion} reports assignment outcomes under MTB and STB for each Chilean region. Two patterns stand out. First, $W_{\mathrm{match}}$, given by the total height of each bar, varies substantially across regions regardless of the tie-breaking rule: from lowest in Tarapacá (I) to highest in Los Ríos (XIV). Second, within each region the two rules yield nearly identical total bar heights but differ in $W_1$, given by the darkest segment: STB matches more students to their first choice, whereas MTB matches more to choices ranked 2--5. The size of this STB--MTB gap in $W_1$ varies across regions, from about 6.1 pp in Antofagasta (II) to nearly zero in Aysén  (XI) and Magallanes (XII). 

This regional heterogeneity reinforces the central message of our framework: the interaction between \emph{rules} and \emph{chance} is not uniform. Changing the tie-breaking rule reshapes the assignment chances associated with where students happen to live, making some regions more exposed than others to gains or losses in match quality.

%% file: sections/06_conc.tex
\section{Conclusion, Limitations, and Future Work}
\label{conc}

We presented \sys, an open-source framework for (1) generating individual-level preference profiles from aggregate statistics via mixtures of Mallows models and (2) evaluating welfare trade-offs in student-school matching along three interacting forces: the \emph{rules} policymakers set, the \emph{choices} students make, and \emph{chance}. Our analyses of NYC and Chile show that these forces must be considered jointly: \emph{rules} that improve one welfare metric come at the cost of another, with impacts strongly shaped by \emph{chance} and \emph{choice}. In NYC, increasing the minimum list length raises overall match rates while lowering top-$p$ rates for students with unfavorable lottery draws. In Chile, moving from MTB to STB improves top-$1$ match rates but widens welfare differences across lottery deciles and residential location. We hope \sys contributes to making simulation-based, ex ante welfare evaluation a standard step in school choice reform.

\smallskip\textbf{Limitations.} \textcolor{blue}{Although \sys can re-sample preference profiles under alternative design parameters, it does not predict how families' preferences or reporting behavior would actually respond to a reform. Moreover, welfare is evaluated relative to stated preference lists, which may differ from families' underlying preferences because of strategic reporting, incomplete information, or unobserved outside options. In particular, the framework cannot determine whether unranked programs are genuinely unacceptable or whether families would accept programs added under a counterfactual minimum list length.}
Further, our welfare metrics are defined from a planner's perspective and may not reflect how families themselves perceive welfare. We also operationalize each force through a limited set of observable dimensions: for \emph{choice}, we consider only the number of programs students rank, leaving aside, e.g.,  strategic behavior; for \emph{chance}, we consider lottery deciles and residential context, leaving aside more granular or legally protected group characteristics. Finally, \sys relies on Mallows mixtures and does not benchmark against alternatives such as Plackett-Luce, MNL~\cite{awadelkarimRankheterogeneousPreferenceModels2023,agarwalRevealedPreferenceAnalysis2020,cheAllocatingStudentsSchools2025}, or random preference generation~\cite{ashlagiAssigningMoreStudents2019}.

\smallskip\textbf{Future Work.} A natural direction is to use \sys to identify and test interventions, rather than only evaluate outcomes. One example is a post-assignment step that reassigns matched students to lower-ranked but still available programs whenever an unmatched student shares their priority status, trading lottery-based advantages for a reduction in unmatched rates. 
\textcolor{teal}{Future work should also expand the welfare and force dimensions identified above: richer, family-centered welfare measures, a broader operationalization of \emph{choice} and \emph{chance}, and, central to equity, an analysis of whether rule changes compound or mitigate pre-existing disparities, e.g., whether the STB--MTB trade-off differs for Chile's socioeconomically vulnerable priority students, whose status our data already records (Table~\ref{tab:demo_chile}).}
Finally, \sys can be adapted beyond school matching to settings such as preference-based refugee-location matching~\cite{strasserceballosLocationMatchingShaky2025}, where welfare considerations are equally critical and individual-level preference data are scarce.

%% file: sections/AI_Usage.tex
\section*{Generative AI Usage Statement}

\textbf{Writing:} The authors confirm that no generative AI tool was used to write the original manuscript. Generative AI (GPT-6 and Claude Opus 5.5) was used to correct grammar, fix typos, and improve clarity of existing sentences. \textbf{Method \& Analysis:} The authors confirm that no generative AI tool was used to design the framework, generate the data, run the analysis, or interpret the results. Generative AI (Claude Sonnet 5 via Claude Code) was used to adjust plotting scripts, aid in debugging errors as they occurred, format figures and create \LaTeX{} tables.

%% file: sections/07_adverse_impacts.tex
\section*{Adverse Impact Statement}
Our work may have the following unintended impacts.

\begin{enumerate}
    \item \textbf{Incentivizing synthetic over real data.} Our framework may inadvertently reward ``lazy'' research practices by making it convenient to synthetically generate preference profiles from aggregate data rather than seeking or requesting access to existing individual-level data. However, such data, as demonstrated in the Chilean case study, not only enables stronger validation of the framework, but also yields richer insights at more granular levels than those accessible through aggregate statistics alone.
    \item \textbf{Over-trust in simulation outputs.} Policymakers may interpret \sys's simulated welfare outcomes as precise predictions rather than as scenario analyses subject to modeling assumptions. This could lend unwarranted confidence to reform decisions, particularly when the aggregate data is sparse or noisy. 
    \item \textbf{Enabling welfare-washing.} 
    Given that different welfare metrics (e.g., match rate vs. top-1 assignment rate) can point in opposite directions, a policymaker or institution could selectively report only the metric that portrays a chosen policy design most favorably, while concealing trade-offs that harm specific groups. Our framework may make such selective reporting easier by supplying a rich menu of metrics to choose from.
\end{enumerate}

%% file: sections/08_ethical_considerations.tex
\section*{Ethical Consideration  Statement}
Our research raises three ethical challenges.

\begin{enumerate}
    \item \textbf{Participatory welfare evaluation.} Our framework is purely computational in nature. We acknowledge the importance of complementing it with a participatory approach that includes families directly in the evaluation process. Understanding the welfare impact of specific design decisions, and the resulting trade-offs, may be best achieved by engaging the families who are most affected by these choices, rather than relying solely on algorithmic analysis.
    \item \textbf{Unequal participation conditions.} Our framework models student choice as driven purely by preferences, but in practice families face heterogeneous conditions when navigating the matching process that shape their choices and ultimately their outcomes, in ways our framework does not currently capture.
    \item \textbf{Limitations of group stratification.} Our welfare analysis is stratified along a small, carefully chosen set of group dimensions: residential district or borough (NYC), gender, and region or province (Chile), driven by the attributes available in our public datasets. We fully acknowledge that this is insufficient for a comprehensive equity evaluation, which would also require examining outcomes by race, ethnicity, socioeconomic status, and other protected attributes. 
\end{enumerate}

%% file: sections/01_additional_details_notation.tex
\section{Notation and Welfare Metrics}
\label{app:notation}

We summarize the notation used throughout the paper in Table \ref{tab:notation_guide} and the welfare metrics in Table \ref{tab:welfare}.

\input{other/notation_table}

\input{other/metrics_table}

%% file: other/notation_table.tex
\providecommand{\chg}[1]{#1}
\begingroup
\setlength{\tabcolsep}{4pt}

\begin{longtable}{@{}
>{\raggedright\arraybackslash}p{0.17\textwidth}
>{\raggedright\arraybackslash}p{0.22\textwidth}
>{\raggedright\arraybackslash}p{0.57\textwidth}
@{}}
\caption{Notation guide.}
\label{tab:notation_guide}\\

\toprule
\textbf{\chg{Symbol}} & \textbf{Type or range} & \textbf{Notes} \\
\midrule
\endfirsthead

\toprule
\textbf{\chg{Symbol}} & \textbf{Type or range} & \textbf{Notes} \\
\midrule
\endhead

\endfoot

\multicolumn{3}{l}{\textit{Applicants, schools, programs, and attributes}}\\
\midrule
$I$ & Set & Set of applicants, $I=\{1,\ldots,n\}$. \\
$n$ & $\mathbb{N}$ & Total number of applicants. \\
$i$ & Element of $I$ & Individual applicant. \\
$\mathcal{S}$ & Set & Set of schools. \\
$s$ & Element of $\mathcal{S}$ & Individual school. \\
$\mathcal{J}_s$ & Set & Set of programs offered by school $s$. \\
$\mathcal{J}$ & Set & Set of programs, \chg{$\mathcal{J}=\bigcup_{s\in\mathcal{S}}\mathcal{J}_s$}$=\{1,\ldots,m\}$. Admissions and matching are modeled at the program level. \\
$m$ & $\mathbb{N}$ & Total number of programs. \\
$j$ & Element of $\mathcal{J}$ & Individual program. \\
$c_j$ & $\mathbb{N}$ & Capacity of program $j$. \\
\chg{$c$} & \chg{Vector} & \chg{Capacity vector, $c=(c_j)_{j\in\mathcal{J}}$.} \\
$\mathcal{X}$ & Set & \chg{Attribute space (e.g., gender, home language, socioeconomic status, disability status, residential district).} \\
$x_i$ & $\mathcal{X}$ & Observed attribute vector of applicant $i$. \\
$X$ & Profile & Full attribute profile, $X=(x_i)_{i\in I}$. \\
\chg{$\mathcal{D}$} & Set & Set of subdivisions used for preference estimation and disaggregated welfare reporting. \\
$d$ & \chg{Element of $\mathcal{D}$} & \chg{Individual subdivision: a residential district in NYC or a province in Chile (Chilean results are aggregated to regions for reporting).} \\
$I_d$ & Set & Set of applicants in subdivision $d$, with $I_d\subseteq I$. \\
\chg{$d(i)$} & \chg{Element of $\mathcal{D}$} & \chg{Subdivision of applicant $i$, i.e., the $d$ with $i\in I_d$.} \\
\chg{$I_g$} & \chg{Set} & \chg{Subgroup of applicants on which welfare metrics are conditioned; $g$ may denote a subdivision $d$, a lottery decile, gender, or another applicant characteristic.} \\

\midrule
\multicolumn{3}{l}{\textit{Preferences, priorities, and matching}}\\
\midrule
$R_i$ & Ordered sequence & Strict preference list submitted by applicant $i$ over programs. \\
$j_{iq}$ & Element of $\mathcal{J}$ & Program ranked in position $q$ by applicant $i$. \\
$L_i$ & $\{1,\ldots,L_{\max}\}$ & Length of applicant $i$'s submitted preference list, with $L_{\max}\ll m$. \\
$L_{\min}$ & $\mathbb{N}$ & Minimum allowed preference-list length when such a policy constraint is imposed. \\
$\mathcal{L}(\theta_L)$ & Distribution & Distribution of list length $L_i$, parameterized by $\theta_L$. \\
$P$ & Profile & Full preference profile, $P=(R_i)_{i\in I}$. \\
$A_j(P)$ & Set & Set of applicants who list program $j$, $A_j(P)=\{i\in I:j\in R_i\}$. \\
$\rho_{ij}$ & $\mathbb{R}$ & Priority score of applicant $i$ at program $j$; lower values indicate higher priority. \\
$\succ_j$ & Order relation & Program-side priority order at program $j$. For $i,i'\in A_j(P)$, $i\succ_j i'$ iff $\rho_{ij}<\rho_{i'j}$. \\
$b_{ij}$ & \chg{$\mathbb{N}$} & Reserve bucket of applicant $i$ at program $j$; lower values indicate higher priority. \\
$t_{ij}$ & \chg{$\mathbb{N}$} & Priority tier of applicant $i$ at program $j$; lower values indicate higher priority. \\
$\lambda_{ij}$ & $[0,1)$ & Stochastic tie-breaker of applicant $i$ at program $j$. Lower values correspond to higher priority when all other priority components are equal. \\
\chg{$\lambda_i$} & \chg{$[0,1)$} & \chg{Single lottery draw of applicant $i$ under STB.} \\
STB & Lottery regime & Single tie-breaking: the same draw $\lambda_i$ is used for applicant $i$ at every program, so that $\lambda_{ij}=\lambda_i$ for all $j\in \mathcal{J}$. \\
MTB & Lottery regime & Multiple tie-breaking: an independent draw $\lambda_{ij}$ is generated for each applicant-program pair $(i,j)$. \\
$\alpha,\beta$ & $\mathbb{R}$ & Weights used in the composite priority score $\rho_{ij}=\alpha b_{ij}+\beta t_{ij}+\lambda_{ij}$, with $\alpha\gg\beta\gg 1$. \\
$\rho$ & Matrix & Priority score matrix, $\rho=(\rho_{ij})_{i\in I,j\in \mathcal{J}}$. \\
$M$ & Mechanism & Matching mechanism that maps preferences, capacities, and priorities to a feasible matching. \\
$\mu$ & Function & Matching, \chg{$\mu:I\to\mathcal{J}\cup\{\emptyset\}$}. \\
$\mu(i)$ & $\mathcal{J}\cup\{\emptyset\}$ & Assignment of applicant $i$. \\
$\emptyset$ & Symbol & Outcome corresponding to an unmatched applicant. \\
$M(P,c,\rho)$ & Matching & Outcome of the mechanism applied to preferences, program capacities, and priorities. \\
$r_i(\mu)$ & $\mathbb{N}\cup\{+\infty\}$ & Realized rank of applicant $i$ under matching $\mu$; equals $q$ if $\mu(i)=j_{iq}$ and $+\infty$ if $\mu(i)=\emptyset$. \\
$p$ & \chg{$\mathbb{N}$} & Rank cutoff used for top-$p$ welfare measures, typically \chg{$p\in\{1,3,5\}$}. \\

\midrule
\multicolumn{3}{l}{\textit{Empirical application statistics}}\\
\midrule
$T_j(P,\mu)$ & Set & Set of true applicants to program $j$, i.e., applicants who reach program $j$ in the assignment process. \\
$A_{d,j}$ & $\mathbb{N}$ & Number of applicants from subdivision $d$ who list program $j$. \\
$T_{d,j}$ & $\mathbb{N}$ & Number of true applicants from subdivision $d$ to program $j$. \\
\chg{$\mathcal{J}_d$} & Set & Set of programs located in subdivision $d$, when $d$ corresponds to a geographic area. \\
\chg{$u_j(\mu)$} & \chg{$[0,1]$} & \chg{Capacity utilization of program $j$, $u_j(\mu)=\frac{1}{c_j}\sum_{i\in I}\mathbbm{1}\{\mu(i)=j\}$.} \\
$\bar u_d(\mu)$ & $[0,1]$ & Capacity-weighted utilization of programs in subdivision $d$, defined as \chg{$\frac{\sum_{j\in\mathcal{J}_d}\sum_{i\in I}\mathbbm{1}\{\mu(i)=j\}}{\sum_{j\in\mathcal{J}_d}c_j}$}. \\

\midrule
\multicolumn{3}{l}{\textit{Mallows simulation and calibration}}\\
\midrule
$K$ & $\mathbb{N}$ & Number of Mallows mixture components. \\
$k_i$ & $\{1,\ldots,K\}$ & Mixture component drawn for applicant $i$. \\
$w_k$ & $[0,1]$ & Weight of mixture component $k$, with $\sum_{k=1}^K w_k=1$. \\
$\sigma_d$ & Ranking of $\mathcal{J}$ & Central ranking of programs for subdivision $d$. \\
$\phi_k$ & $(0,1)$ & Dispersion parameter of mixture component $k$. \\
\chg{$a,b$} & \chg{$\mathbb{R}_{+}$} & \chg{Parameters of the $\mathrm{Beta}(a,b)$ distribution used to initialize $\phi_k$.} \\
$\tau_i$ & Complete ranking & Complete ranking sampled for applicant $i$ before truncation. \\
\chg{$d_K(\tau,\sigma)$} & $\mathbb{N}$ & Kendall distance between ranking $\tau$ and central ranking $\sigma$. \\
$\mathrm{Mallows}(\sigma_d,\phi_k)$ & Distribution & Mallows distribution centered at $\sigma_d$ with dispersion $\phi_k$. \\
\chg{$R_i=\operatorname{Trunc}_{L_i}(\tau_i)$} & Truncation & Submitted preference list obtained by keeping the top $L_i$ entries of $\tau_i$. \\
$\gamma_d(j)$ & $\mathbb{R}$ & \chg{Competitiveness score used to construct $\sigma_d$ by sorting programs in descending order; initialized as $T_{d,j}^2/A_{d,j}$.} \\
$N_{\mathrm{sim}}$ & $\mathbb{N}$ & Number of simulated preference profiles used during calibration. \\
\chg{$T_{\mathrm{iter}}$} & $\mathbb{N}$ & Maximum number of calibration rounds. \\
\chg{$\epsilon_{\mathrm{tol}}$} & \chg{$\mathbb{R}_{+}$} & \chg{Stopping tolerance: calibration ends early when $\max_k|\Delta\phi_k|<\epsilon_{\mathrm{tol}}$.} \\
\chg{$v$} & \chg{$\mathbb{N}$} & \chg{Number of welfare statistics included in the calibration target.} \\
$o_d$ & $\mathbb{R}^v$ & Observed or target welfare statistics for subdivision $d$. \\
\chg{$\varphi$} & \chg{$(0,1)$} & \chg{Candidate value for $\phi_k$ during calibration.} \\
\chg{$\tilde{o}_{d,k}^{(r)}(\varphi)$} & $\mathbb{R}^v$ & Simulated welfare statistics for subdivision $d$ in simulation \chg{$r$}, when candidate value $\varphi$ replaces $\phi_k$. \\
$\bar{o}_{d,k}(\varphi)$ & $\mathbb{R}^v$ & Mean simulated welfare vector for subdivision $d$ under candidate value $\varphi$ for component $k$. \\
$\Sigma_{d,k}(\varphi)$ & $\mathbb{R}^{v\times v}$ & Empirical covariance matrix of simulated welfare vectors for subdivision $d$ under candidate value $\varphi$ for component $k$. \\
\chg{$\varepsilon$} & \chg{$\mathbb{R}_{+}$} & \chg{Small regularization constant; $\varepsilon I_v$ is added to $\Sigma_{d,k}(\varphi)$.} \\
$\ell_{d,k}(\varphi)$ & $\mathbb{R}$ & Gaussian log-likelihood contribution of subdivision $d$ when updating component $k$ with candidate value $\varphi$. \\
$Q_k(\varphi)$ & $\mathbb{R}$ & Calibration objective for updating component $k$, aggregated across subdivisions. \\
$Q^\star$ & $\mathbb{R}$ & Best recorded calibration objective value. \\
\chg{$\hat f_j$} & \chg{$\mathbb{N}$} & \chg{Observed number of filled seats at program $j$.} \\
\chg{$\tilde f_j$} & \chg{$\mathbb{R}_{+}$} & \chg{Mean simulated number of filled seats at program $j$.} \\
$\delta_j$ & $\mathbb{R}$ & Filled-count error for program $j$, \chg{$\delta_j=\hat f_j-\tilde f_j$}, used to update $\gamma_d(j)$. \\
$\eta$ & $\mathbb{R}_{+}$ & Step size for the utilization-based update of $\gamma_d(j)$. \\

\bottomrule
\end{longtable}
\endgroup

%% file: other/metrics_table.tex
\providecommand{\chg}[1]{{\boldmath\bfseries #1}}

\begin{table}[ht]
\centering
\caption{Welfare metrics.}
\setlength{\tabcolsep}{4pt}
\renewcommand{\arraystretch}{1.5}
\begin{tabular}{lll}
\toprule
\textbf{Welfare Metric} & \textbf{Formula} & \textbf{Description} \\
\midrule
\multicolumn{3}{l}{\chg{\textit{Student-side}}}\\
Unmatched rate
  & $W_{\mathrm{unmatch}}(\mu) = \frac{1}{n}\sum_{i\in I}
     \mathbbm{1}\{\mu(i)=\emptyset\}$
  & Share remaining unmatched. \\
Matched rate
  & $W_{\mathrm{match}}(\mu) =$ \chg{$1-W_{\mathrm{unmatch}}(\mu)$}
  & Share receiving an assignment. \\
Avg.\ matched rank
  & $\bar{r}(\mu) = \frac{1}{|\{i:\mu(i)\neq\emptyset\}|}
     \sum_{i:\mu(i)\neq\emptyset} r_i(\mu)$
  & Mean rank among matched applicants. \\
Top-$p$ match rate
  & $W_p(\mu) = \frac{1}{n}\sum_{i\in I}\mathbbm{1}\{r_i(\mu)\le p\}$
  & Share assigned to a top-$p$ choice. \\
\chg{Subgroup metrics}
  & \chg{$W_p(\mu\mid g) = \frac{1}{|I_{g}|}
     \sum_{i\in I_{g}}\mathbbm{1}\{r_i(\mu)\le p\}$}
  & \chg{Any metric above restricted to $I_g$ (e.g., $g=d$).} \\
\midrule
\multicolumn{3}{l}{\chg{\textit{Program-side}}}\\
\chg{Program utilization}
  & \chg{$u_j(\mu) = \frac{1}{c_j}\sum_{i\in I}\mathbbm{1}\{\mu(i)=j\}$}
  & \chg{Share of program $j$'s seats filled.} \\
\chg{Subdivision utilization}
  & \chg{$\bar u_d(\mu) = \frac{\sum_{j\in\mathcal{J}_d}\sum_{i\in I}\mathbbm{1}\{\mu(i)=j\}}{\sum_{j\in\mathcal{J}_d}c_j}$}
  & \chg{Capacity-weighted utilization in $d$.} \\
\bottomrule
\end{tabular}
\label{tab:welfare}
\end{table}

%% file: sections/02_additional_details_case_NYC_and_Chile.tex
 \section{Additional Details about the Case Studies}
 \label{app:data}
This section provides additional details on the two case studies, serving two purposes. First, it facilitates the interpretation of figures in the main text by mapping NYC districts to boroughs and neighborhoods (Table~\ref{tab:nyc_districts}) and Chilean regions to the Roman numerals used as labels (Table~\ref{tab:chilean_regions_roman}). Second, it supports a modeling choice from Section~\ref{framework}. \sys gives each subdivision $d$ its own central ranking $\sigma_d$ (residential districts in NYC, provinces in Chile). This choice only makes sense if students in different locations rank different schools. Figure~\ref{validation_central_ranking_NYC} (NYC) and Table~\ref{validation_central_ranking_Chile} (Chile) show that they do. For Chile, where individual-level data are available, we also describe who applies  (Table~\ref{tab:demo_chile}) and how many schools they rank (Table~\ref{tab:demo_chile_second}).

\subsection{NYC}\label{app:details_nyc}

\paragraph{\textbf{Subdivisions.}} NYC consists of 32 residential districts across five boroughs (Table~\ref{tab:nyc_districts}). We calibrate \sys at the district level and validate the fit for each district
 (Appendix~\ref{app:obs_vs_sim}), but report the counterfactual results by borough (Section~\ref{sec:eval:rules}).

\input{other/tab_cat_residential_district_NYC}

\begin{figure}[ht]
\centering
\includegraphics[width=0.6\linewidth]{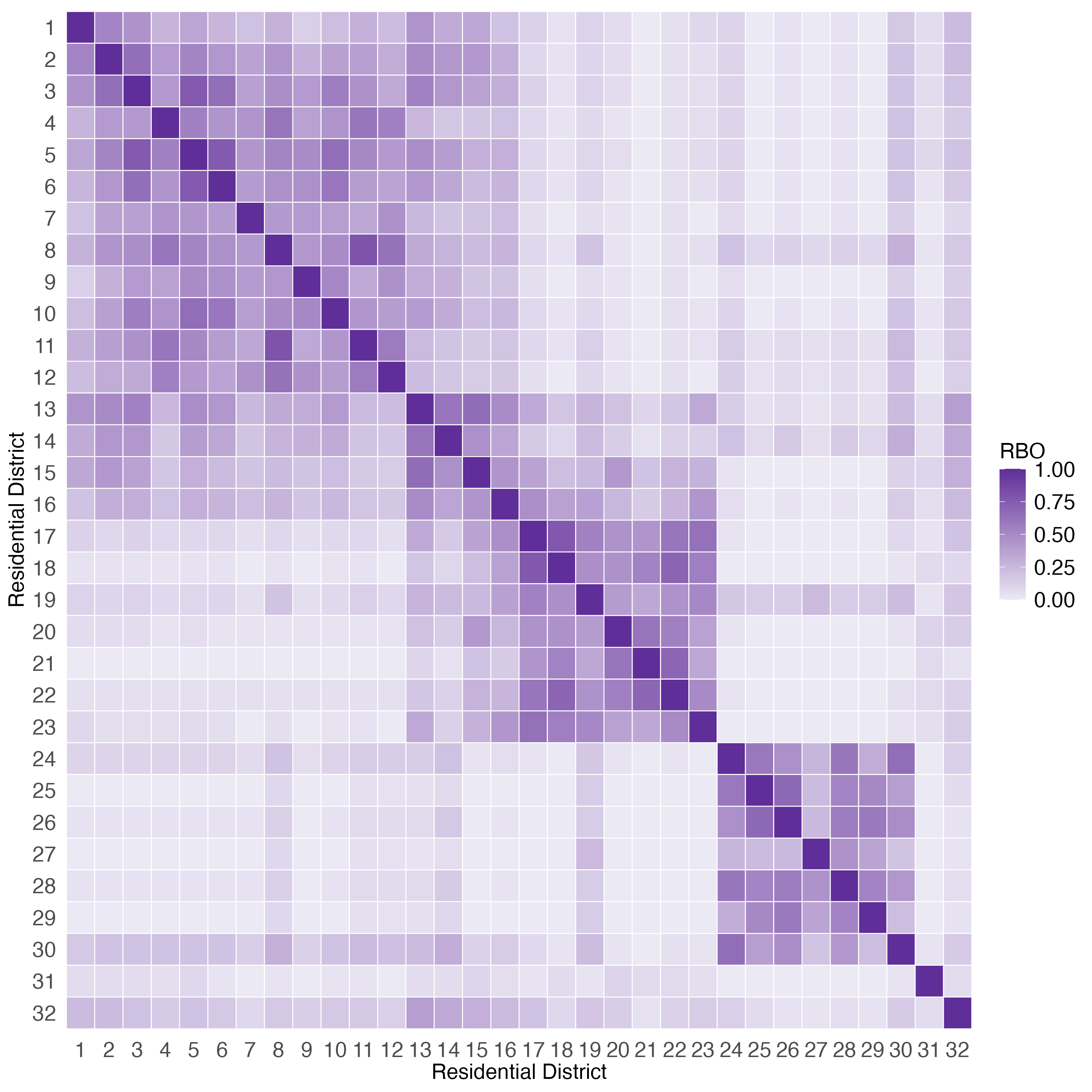}
\caption{\textbf{NYC:} Similarity of top-12 program rankings between residential districts, measured by Rank-Biased Overlap (RBO; persistence parameter $0.9$). Programs are ordered within each district by $\mathrm{Ratio}_{d,j}$. RBO ranges from $0$ (no shared programs) to $1$ (identical lists) and gives more weight to agreement at the top of the lists. Dark blocks along the diagonal largely follow borough boundaries (Table~\ref{tab:nyc_districts}), while districts in distant boroughs share almost no top programs.}
\label{validation_central_ranking_NYC}
\end{figure}

\paragraph{\textbf{Are rankings localized?}} Figure~\ref{validation_central_ranking_NYC} compares the school rankings across all pairs of districts. For each district $d$, we order programs by $\mathrm{Ratio}_{d,j}$, the statistic used to initialize $\sigma_d$ (Section~\ref{sec:eval}), and keep the top 12. Each cell reports the Rank-Biased Overlap (RBO) between two districts' rankings, ranging from $0$ (no shared programs) to $1$ (identical rankings), with greater weight given to agreement near the top. A dark cell away from the diagonal indicates that two districts favor similar programs, and a light cell indicates that they have little in common.

The darkest blocks follow borough boundaries (e.g., Brooklyn 17--23), Manhattan and the Bronx (1--12) form a looser shared block, and distant boroughs (e.g., Staten Island 31) share almost no top programs. These patterns support the use of district-specific central rankings, which capture geographic variation in school rankings that a single citywide ranking would miss.

\FloatBarrier
\subsection{Chile}
\label{app:details_chile}

\paragraph{\textbf{Subdivisions.}} Chile has 16 regions and 56 provinces. We calibrate \sys at the province level and aggregate the results to regions for reporting (Section~\ref{sec:eval:val}). Figures in the main text use Roman numerals to label regions, with their full names listed in  Table~\ref{tab:chilean_regions_roman}. 

\input{other/tab_cat_regions_Chile}

\paragraph{\textbf{Who applies.}} Table~\ref{tab:demo_chile} describes the 124,492 high-school (HS) applicants in the 2024--2025 admission cycle. Gender is roughly balanced (52\% male, 48\% female). Priority students, who are classified as socioeconomically vulnerable, make up a large majority (69\%), and 23\% are high-performing. Keep in mind that being a priority student is common, while the seats reserved for priority students are capped at 15\% of a program's capacity (Section~\ref{sec:eval}). Priority status alone therefore does not guarantee admission. Among school-specific priorities, which may overlap, sibling priority (15\%) and already-enrolled priority (14\%) are the most common.

\input{other/tab_demographics_chile}

\paragraph{\textbf{How many schools they rank.}} Table~\ref{tab:demo_chile_second} reports the length of submitted preference lists. The median applicant ranks 3 schools in almost every group (4 for non-priority students and former students). Group means range from $3.21$ to $4.25$. The largest differences are by priority: applicants with a sibling priority ($3.29$ vs.\ $3.97$), a school staff priority ($3.21$ vs.\ $3.87$), or priority-student status ($3.70$ vs.\ $4.25$) submit shorter lists
than their counterparts, consistent with higher admission chances reducing the need to rank additional schools. Gaps by gender and academic performance are small (at most $0.21$), with high-performing students submitting slightly longer lists. Short lists mean that most of a student's outcome is decided among the first few choices.

\input{other/tab_ranking_behaviour_chile}

\paragraph{\textbf{Are rankings localized?}}  $98.3\%$ of applicants rank only schools in their own region (Table~\ref{validation_central_ranking_Chile}); the share is at least $97\%$ everywhere except Los Lagos (X, $95.8\%$) and Los Ríos (XIV, $92.9\%$). We observe the same pattern at the province level but omit the full breakdown for brevity. School rankings are thus highly localized, supporting the use of province-specific rather than a single national central ranking.

\input{other/tab_validation_central_ranking_chile}

%% file: other/tab_cat_residential_district_NYC.tex
\begin{table}[ht]
\caption{\textbf{NYC:} Districts and their constituent neighborhoods.}
\label{tab:nyc_districts}
\centering
\small
\begin{tabular}{lll}
\toprule
\textbf{District} & \textbf{Borough} & \textbf{Neighborhoods} \\
\midrule
1  & Manhattan & East Village, Lower East Side \\
2  & Manhattan & Financial District, Tribeca, West Village, Midtown, Gramercy, Upper East Side \\
3  & Manhattan & Lincoln Square, Upper West Side \\
4  & Manhattan & East Harlem, Randall's Island \\
5  & Manhattan & Central Harlem, Morningside Heights \\
6  & Manhattan & Inwood, Washington Heights \\
7  & Bronx     & Mott Haven, Port Morris \\
8  & Bronx     & Country Club, Edgewater Park, Soundview, Hunts Point \\
9  & Bronx     & Morris Heights, Mount Eden \\
10 & Bronx     & Riverdale, Bedford Park, Norwood \\
11 & Bronx     & Wakefield, Co-op City, Pelham Parkway \\
12 & Bronx     & East Tremont, Claremont Village \\
13 & Brooklyn  & Brooklyn Heights, Ft.\ Greene, Clinton Hill \\
14 & Brooklyn  & Greenpoint, Williamsburg \\
15 & Brooklyn  & Sunset Park, Cobble Hill \\
16 & Brooklyn  & Bedford Stuyvesant, Weeksville \\
17 & Brooklyn  & Prospect Park, Wingate \\
18 & Brooklyn  & Canarsie, East Flatbush \\
19 & Brooklyn  & Cypress Hills, East New York, Starrett City \\
20 & Brooklyn  & Bay Ridge, Fort Hamilton, Dyker Heights \\
21 & Brooklyn  & Coney Island, Sheepshead Bay, Gravesend, Ocean Parkway \\
22 & Brooklyn  & Marine Park, Georgetown, Flatlands \\
23 & Brooklyn  & Brownsville, Ocean Hill \\
24 & Queens    & Glendale, Ridgewood, Maspeth, Jackson Heights, Sunnyside \\
25 & Queens    & College Point, Whitestone, Hillcrest \\
26 & Queens    & Floral Park, Little Neck, Bayside, Fresh Meadows \\
27 & Queens    & Richmond Hill, Woodhaven, Howard Beach, South Ozone Park \\
28 & Queens    & Rego Park, Forest Hills, Kew Gardens \\
29 & Queens    & Rosedale, Saint Albans, Cambria Heights, Queens Village \\
30 & Queens    & Hunters Point, Long Island City, Astoria, Steinway \\
31 & Staten Island & All of Staten Island \\
32 & Brooklyn  & Bushwick \\
\bottomrule
\end{tabular}
\end{table}

%% file: other/tab_cat_regions_Chile.tex
\begin{longtable}{lc}
\caption{\textbf{Chile:} Regions and their roman numeral equivalents. Note: These numerals are no longer part of the official regional nomenclature, they are retained here for simplicity in presenting the results.} \label{tab:chilean_regions_roman} \\

\hline
\textbf{Region} & \textbf{Roman Numeral} \\
\hline
\endfirsthead

\multicolumn{2}{c}{\tablename\ \thetable{} -- continued from previous page} \\
\hline
\textbf{Region} & \textbf{Roman Numeral} \\
\hline
\endhead

\hline
\multicolumn{2}{r}{\textit{Continued on next page}} \\
\endfoot

\hline
\endlastfoot

Tarapacá Region & I \\
Antofagasta Region & II \\
Atacama Region & III \\
Coquimbo Region & IV \\
Valparaíso Region & V \\
Libertador General Bernardo O'Higgins Region & VI \\
Maule Region & VII \\
Biobío Region & VIII \\
La Araucanía Region & IX \\
Los Lagos Region & X \\
Aysén del General Carlos Ibáñez del Campo Region & XI \\
Magallanes and Chilean Antarctica Region & XII \\
Metropolitan Region of Santiago & XIII \\
Los Ríos Region & XIV \\
Arica and Parinacota Region & XV \\
Ñuble Region & XVI \\

\end{longtable}

%% file: other/tab_demographics_chile.tex
  \begin{longtable}{lrr}
  \caption{\textbf{Chile:} Number and share of high-school applicants in the 2024--2025 admission cycle by gender, priority status, high-performing status, and school-specific priorities. A school-specific priority means that the applicant holds that priority at one or more of the programs they rank; these priorities can overlap. Percentages are rounded.}
  \label{tab:demo_chile} \\
  \hline
  \textbf{Variable} & \textbf{N} & \textbf{\%} \\
  \hline
  \endfirsthead
  \hline
  \textbf{Variable} & \textbf{N} & \textbf{\%} \\
  \hline
  \endhead
  \hline
  \endfoot
    Total & 124492 & 100 \\
    Gender &  &  \\
    \quad... Male & 64147 & 52 \\
    \quad... Female & 60345 & 48 \\
    Priority Student &  &  \\
    \quad... No & 38733 & 31 \\
    \quad... Yes & 85759 & 69 \\
    High-Performing Student &  &  \\
    \quad... No & 96196 & 77 \\
    \quad... Yes & 28296 & 23 \\
    Sibling Priority &  &  \\
    \quad... No & 105329 & 85 \\
    \quad... Yes & 19163 & 15 \\
    Already Enrolled Priority &  &  \\
    \quad... No & 107091 & 86 \\
    \quad... Yes & 17401 & 14 \\
    Former Student Priority &  &  \\
    \quad... No & 118009 & 95 \\
    \quad... Yes & 6483 & 5 \\
    School Staff Priority &  &  \\
    \quad... No & 123667 & 99 \\
    \quad... Yes & 825 & 1 \\
  \end{longtable}

%% file: other/tab_ranking_behaviour_chile.tex
  \begin{longtable}{lrrrrr}
  \caption{\textbf{Chile:} Distribution of applicants' preference list lengths by gender, priority status, high-performing status, and school-specific priorities. Q1 and Q3 are the 25th and 75th percentiles. A school-specific priority means that the applicant holds that priority at one or more of the programs they rank. List lengths are broadly similar across gender and academic performance. Priority students submit shorter lists on average (mean $3.70$ vs.\ $4.25$),
  as do applicants with a sibling priority ($3.29$ vs.\ $3.97$) or a school staff priority ($3.21$ vs.\ $3.87$).}
  \label{tab:demo_chile_second} \\
  \hline
  \textbf{Variable} & \textbf{Q1} & \textbf{Median} & \textbf{Q3} & \textbf{Mean} & \textbf{SD} \\
  \hline
  \endfirsthead
  \hline
  \textbf{Variable} & \textbf{Q1} & \textbf{Median} & \textbf{Q3} & \textbf{Mean} & \textbf{SD} \\
  \hline
  \endhead
  \hline
  \endfoot
    Gender &  &  &  &  &   \\
    \quad... Male & 2.00 & 3.00 & 5.00 & 3.82 & 2.05 \\
    \quad... Female & 3.00 & 3.00 & 5.00 & 3.92 & 2.07 \\
    Priority Student &  &  &  &  &   \\
    \quad... No & 3.00 & 4.00 & 5.00 & 4.25 & 2.29 \\
    \quad... Yes & 2.00 & 3.00 & 5.00 & 3.70 & 1.92 \\
    High-Performing Student &  &  &  &  &   \\
    \quad... No & 2.00 & 3.00 & 5.00 & 3.82 & 2.04 \\
    \quad... Yes & 3.00 & 3.00 & 5.00 & 4.03 & 2.12 \\
    Sibling Priority &  &  &  &  &   \\
    \quad... No & 3.00 & 3.00 & 5.00 & 3.97 & 2.11 \\
    \quad... Yes & 2.00 & 3.00 & 4.00 & 3.29 & 1.67 \\
    Already Enrolled Priority &  &  &  &  &   \\
    \quad... No & 2.00 & 3.00 & 5.00 & 3.88 & 2.08 \\
    \quad... Yes & 2.00 & 3.00 & 5.00 & 3.83 & 1.94 \\
    Former Student Priority &  &  &  &  &   \\
    \quad... No & 2.00 & 3.00 & 5.00 & 3.85 & 2.05 \\
    \quad... Yes & 3.00 & 4.00 & 5.00 & 4.16 & 2.22 \\
    School Staff Priority &  &  &  &  &   \\
    \quad... No & 2.00 & 3.00 & 5.00 & 3.87 & 2.06 \\
    \quad... Yes & 2.00 & 3.00 & 4.00 & 3.21 & 1.77 \\
  \end{longtable}

%% file: other/tab_validation_central_ranking_chile.tex
\begin{longtable}{llrr}
\caption{\textbf{Chile:} Applicants who apply to schools in their own region of residence versus in other regions. } \label{validation_central_ranking_Chile} \\

\hline
\textbf{Region Student} & \textbf{Region School} & \textbf{N} & \% \\
\hline
\endfirsthead

\multicolumn{4}{c}{\tablename\ \thetable{} -- continued from previous page} \\
\hline
\textbf{Region Student} & \textbf{Region School} & \textbf{N} & \% \\
\hline
\endhead

\hline
\multicolumn{4}{r}{\textit{Continued on next page}} \\
\endfoot

\hline
\endlastfoot

Región Metropolitana de Santiago & Región Metropolitana de Santiago & 38925 & 99.03 \\
 & Other & 381 & 0.97 \\ \hline
Región de Antofagasta & Región de Antofagasta & 5394 & 99.28 \\
 & Other & 39 & 0.72 \\ \hline
Región de Arica y Parinacota & Región de Arica y Parinacota & 1975 & 99.20 \\
 & Other & 16 & 0.80 \\ \hline
Región de Atacama & Región de Atacama & 3352 & 99.44 \\
 & Other & 19 & 0.56 \\ \hline
Región de Aysén del Gral.Ibañez del Campo & Región de Aysén del Gral.Ibañez del Campo & 431 & 97.51 \\
 & Other & 11 & 2.49 \\ \hline
Región de Coquimbo & Región de Coquimbo & 5653 & 98.38 \\
 & Other & 93 & 1.62 \\ \hline
Región de La Araucanía & Región de La Araucanía & 9245 & 98.36 \\
 & Other & 154 & 1.64 \\ \hline
Región de Los Lagos & Región de Los Lagos & 7170 & 95.77 \\
 & Other & 317 & 4.23 \\ \hline
Región de Los Ríos & Región de Los Ríos & 2958 & 92.93 \\
 & Other & 225 & 7.07 \\ \hline
Región de Magallanes y Antártica Chilena & Región de Magallanes y Antártica Chilena & 1144 & 99.05 \\
 & Other & 11 & 0.95 \\ \hline
Región de Tarapacá & Región de Tarapacá & 2612 & 99.13 \\
 & Other & 23 & 0.87 \\ \hline
Región de Valparaíso & Región de Valparaíso & 11087 & 98.67 \\
 & Other & 149 & 1.33 \\ \hline
Región de Ñuble & Región de Ñuble & 3359 & 97.76 \\
 & Other & 77 & 2.24 \\ \hline
Región del Bío-Bío & Región del Bío-Bío & 11610 & 97.84 \\
 & Other & 256 & 2.16 \\ \hline
Región del Libertador Bernardo O'Higgins & Región del Libertador Bernardo O'Higgins & 8175 & 97.34 \\
 & Other & 223 & 2.66 \\ \hline
Región del Maule & Región del Maule & 9310 & 98.96 \\
 & Other & 98 & 1.04 \\

\end{longtable}

%% file: sections/03_additional_results_chile.tex
\section{Additional Results for Chile}
\label{app:chile_results}

This section reports two additional results for Chile, both based on the observed preference profiles. First, we break down the STB$-$MTB comparison of Figure~\ref{fig:stb_mtb_diff} by gender (Appendix~\ref{app:mtb_stb}). Second, we show how Figure~\ref{fig:chilestbmtblottery} changes under alternative ways of sorting students into lottery deciles under MTB (Appendix~\ref{app:lottery_summary}).

\subsection{Tie-Breaking Results by Gender}
\label{app:mtb_stb}

Table~\ref{tab:mtb_stb_summary} reports the top-$p$ match rates behind Figure~\ref{fig:stb_mtb_diff} for all applicants and separately by gender (Section~\ref{sec:eval:rules}).  In both groups, STB matches more students to one of their top three choices ($p=1$-$3$), with the largest gain at the first choice, while MTB is equal or marginally better from $p=4$ onward and leaves slightly fewer students unmatched. The STB$-$MTB differences are nearly identical for female and non-female applicants, so switching the tie-breaking rule does not change outcomes differently by gender.

\begin{table}[ht!]
\caption{\textbf{Chile:} Top-$p$ match rates (\%) under MTB and STB tie-breaking, for all students as well as broken down by gender. Diff columns report STB$-$MTB in percentage points.}
\centering
\begin{tabular}{c rrr rrr rrr}
\toprule
& \multicolumn{3}{c}{\textbf{Overall}} 
& \multicolumn{3}{c}{\textbf{Female}} 
& \multicolumn{3}{c}{\textbf{Non-female}} \\
\cmidrule(lr){2-4} \cmidrule(lr){5-7} \cmidrule(lr){8-10}
$p$ & MTB & STB & Diff & MTB & STB & Diff & MTB & STB & Diff \\
\midrule
1  & 53.9 & 56.1 & $+2.2$ & 52.2 & 54.3 & $+2.1$ & 55.4 & 57.8 & $+2.4$ \\
2  & 70.6 & 71.5 & $+0.9$ & 68.7 & 69.6 & $+0.9$ & 72.3 & 73.3 & $+0.9$ \\
3  & 78.8 & 79.1 & $+0.3$ & 77.4 & 77.7 & $+0.3$ & 80.1 & 80.4 & $+0.3$ \\
4  & 83.4 & 83.4 & $-0.1$ & 82.4 & 82.4 & $0.0$ & 84.4 & 84.3 & $-0.1$ \\
5  & 85.7 & 85.5 & $-0.2$ & 85.1 & 84.9 & $-0.2$ & 86.4 & 86.2 & $-0.2$ \\
6  & 87.0 & 86.7 & $-0.3$ & 86.5 & 86.2 & $-0.3$ & 87.5 & 87.3 & $-0.3$ \\
7  & 87.7 & 87.3 & $-0.3$ & 87.2 & 86.9 & $-0.3$ & 88.1 & 87.8 & $-0.3$ \\
8  & 88.0 & 87.6 & $-0.3$ & 87.5 & 87.2 & $-0.3$ & 88.4 & 88.1 & $-0.4$ \\
9  & 88.1 & 87.8 & $-0.3$ & 87.7 & 87.4 & $-0.3$ & 88.5 & 88.2 & $-0.3$ \\
10 & 88.2 & 87.9 & $-0.3$ & 87.8 & 87.5 & $-0.3$ & 88.6 & 88.3 & $-0.3$ \\
11 & 88.3 & 87.9 & $-0.3$ & 87.9 & 87.5 & $-0.3$ & 88.7 & 88.3 & $-0.3$ \\
12 & 88.3 & 88.0 & $-0.3$ & 87.9 & 87.5 & $-0.3$ & 88.7 & 88.3 & $-0.3$ \\
Unmatched & 11.7 & 12.0 & $+0.3$ & 12.1 & 12.4 & $+0.3$ & 11.3 & 11.6 & $+0.3$ \\
\bottomrule
\end{tabular}
\label{tab:mtb_stb_summary}
\end{table}

\FloatBarrier
\subsection{Choice of Lottery Summary under MTB}
\label{app:lottery_summary}

\begin{figure}[ht]
    \centering
    \includegraphics[width=\linewidth]{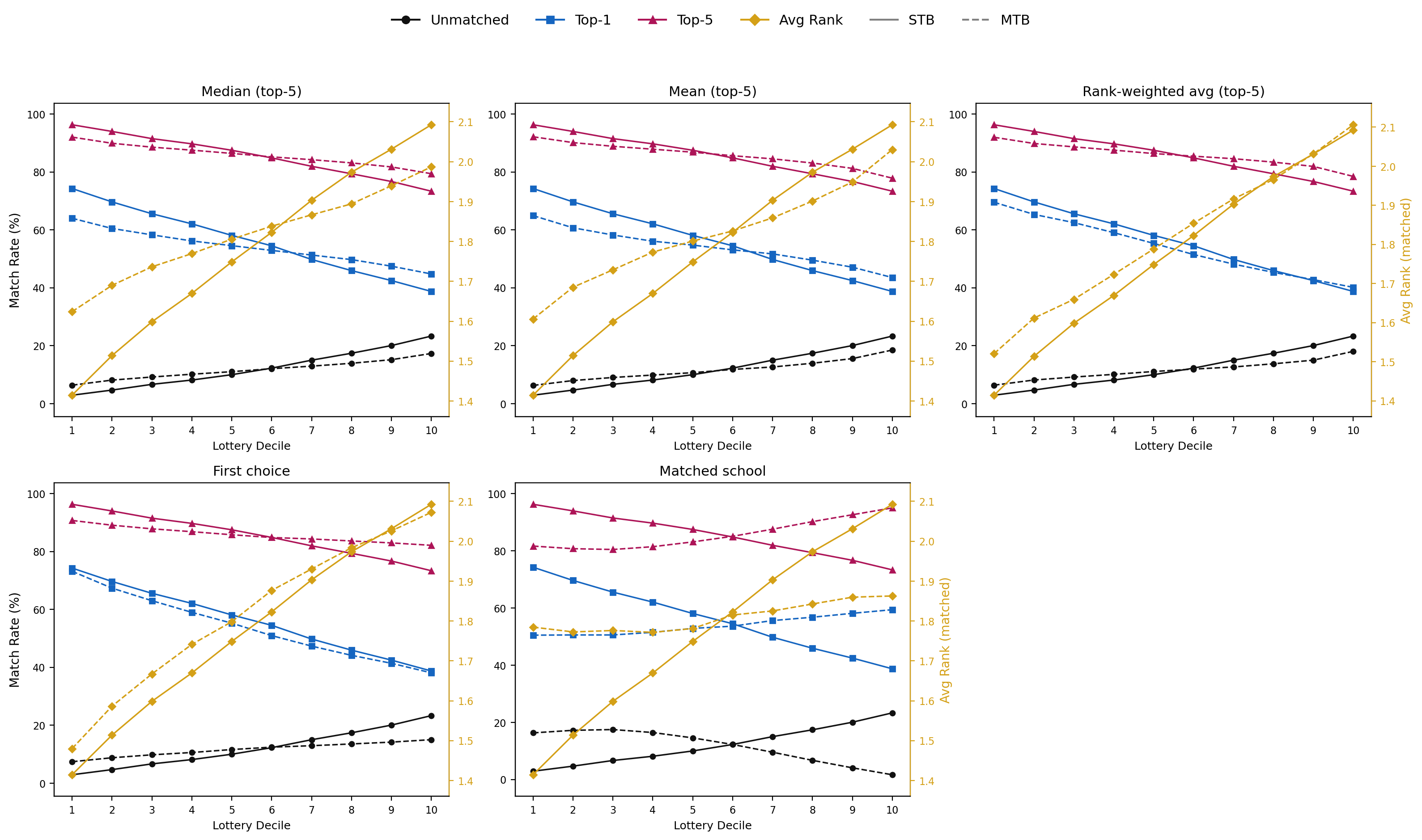}
    \caption{\textbf{Chile:} Figure~\ref{fig:chilestbmtblottery} with alternative summaries of a student's MTB draws for sorting into deciles: median, mean, and rank-weighted mean over the top-5 choices, the first-choice draw, and the matched-program draw. STB curves are identical across panels.}
    \label{fig:fig4_alternates}
\end{figure}

Under STB, each student has one lottery draw $\lambda_i$, so students can be sorted into deciles directly. Under MTB, each student has one draw $\lambda_{ij}$ per program, so the draws must be summarized into a single number. The main text uses the best (lowest) draw among the student's top-5 choices. Figure~\ref{fig:fig4_alternates} shows five alternatives: the median, the mean, and the rank-weighted mean of the top-5 draws, the draw at the first choice, and the draw at the matched program.

The median and the mean give the same picture as the best draw. With the first-choice draw or the rank-weighted mean, $W_1$ under MTB falls almost as steeply as under STB, because the draw at the first choice largely decides whether a student gets it. For all four of these summaries, $W_5$ and $W_{\mathrm{unmatch}}$ still vary more across deciles under STB than under MTB. The matched-program draw reverses the MTB curves, but it depends on the assignment itself and therefore cannot be used to sort students; we show it only for completeness.

%% file: sections/03_additional_validation_results_NYC_and_Chile.tex
  \section{Validation}
\label{app:validation}

This section extends the validation in Section~\ref{sec:eval:val}. We report observed and simulated welfare for every NYC district and Chilean region (Appendix~\ref{app:obs_vs_sim}). For Chile, we also calibrate \sys on subsets of the aggregate statistics (Appendix~\ref{app:held_out}) and rerun the counterfactuals on synthetic profiles (Appendix~\ref{app:fidelity}).

\subsection{Observed vs. Simulated }
\label{app:obs_vs_sim}

\begin{figure}[!htbp]
\centering
\includegraphics[width=\linewidth,height=0.24\textheight,keepaspectratio]{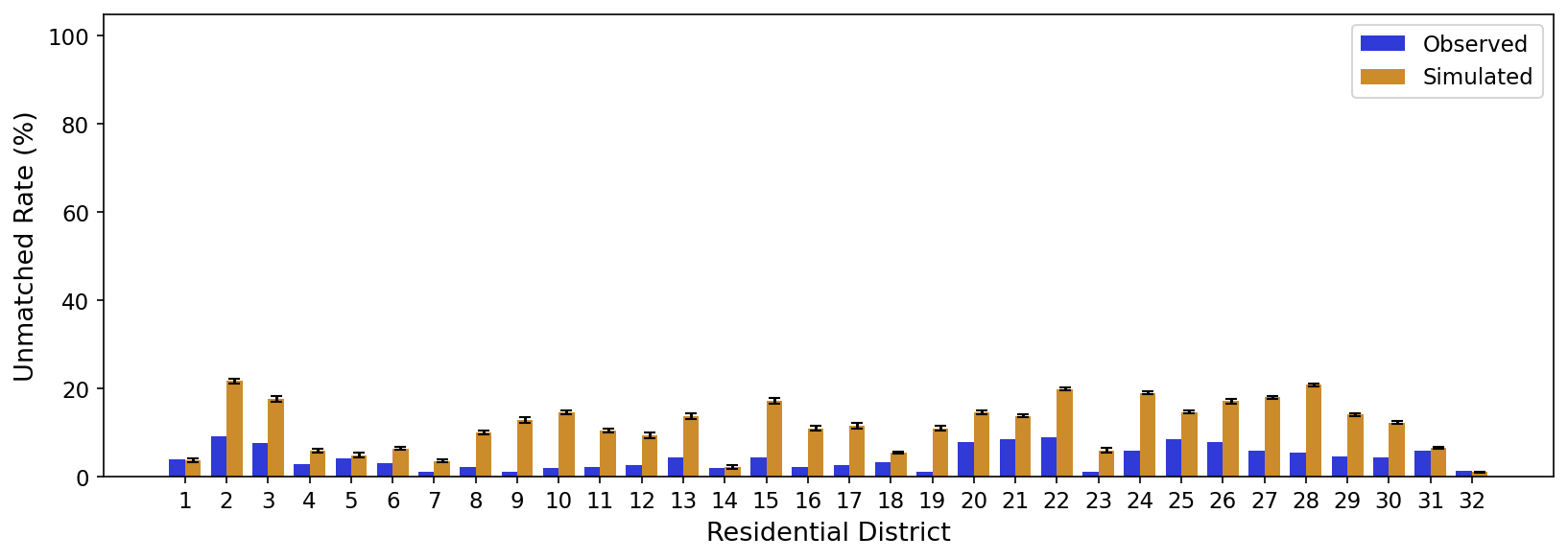}
\caption{\textbf{NYC:} Observed vs.\ simulated unmatched rate $W_{\mathrm{unmatch}}(\mu \mid d)$ by residential district. Simulated values are means over sampled profiles; error bars show 95\% confidence intervals.}
\label{fig:unmatchedvalidation}

\vspace{1em}

\includegraphics[width=\linewidth,height=0.24\textheight,keepaspectratio]{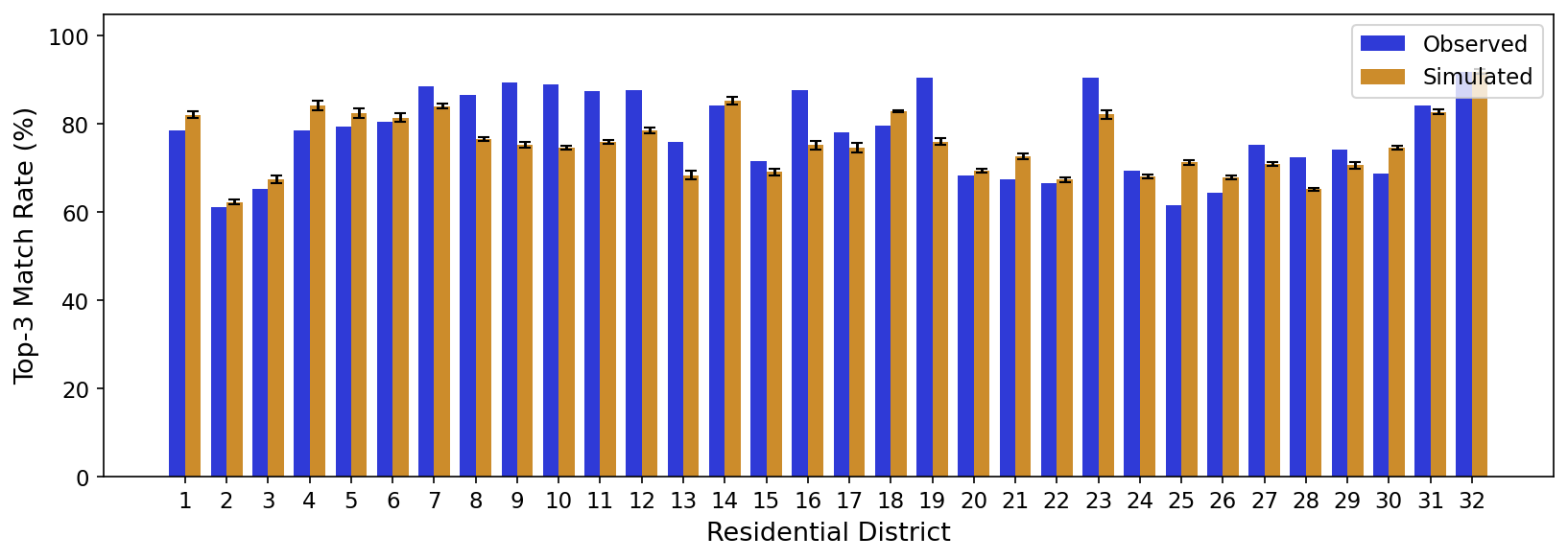}
\caption{\textbf{NYC:} Observed vs.\ simulated top-3 match rate $W_3(\mu \mid d)$ by residential district. Simulated values are means over sampled profiles; error bars show 95\% confidence intervals.}
\label{fig:nyc_val_top3}
\end{figure}

\begin{figure}[!htbp]
\centering
\includegraphics[width=\linewidth,height=0.24\textheight,keepaspectratio]{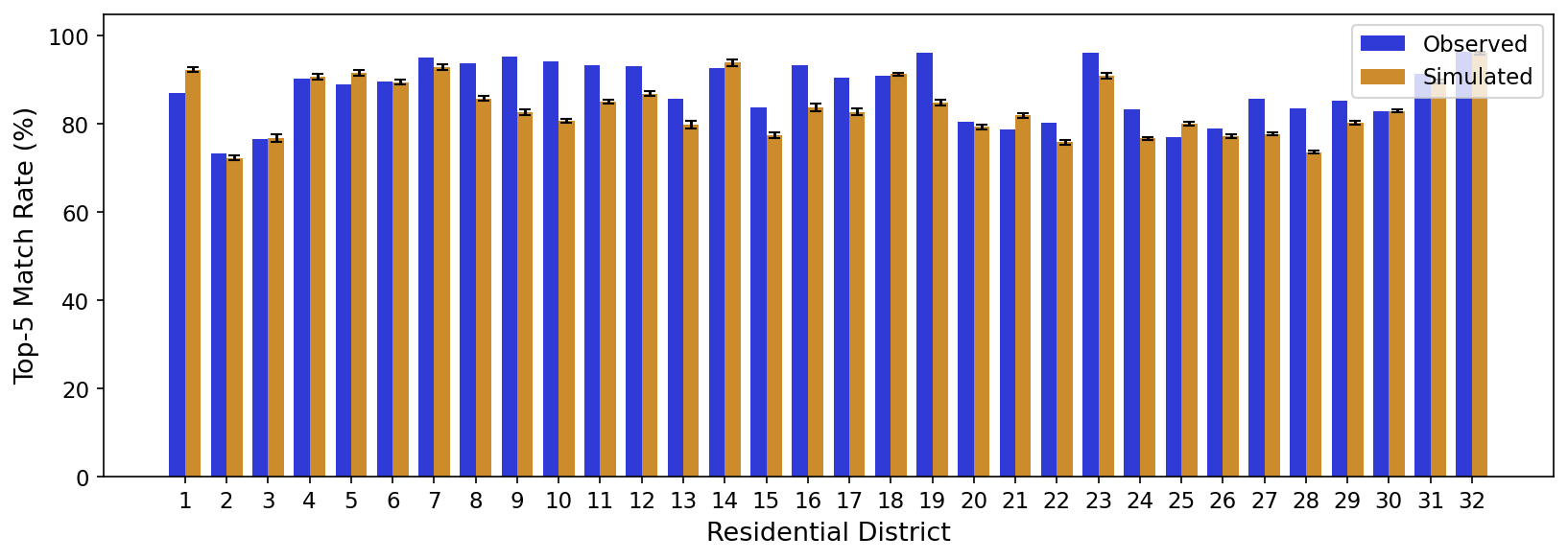}
\caption{\textbf{NYC:} Observed vs.\ simulated top-5 match rate $W_5(\mu \mid d)$ by residential district. Simulated values are means over sampled profiles; error bars show 95\% confidence intervals.}
\label{fig:top5validation}

\vspace{1em}

\includegraphics[width=\linewidth,height=0.24\textheight,keepaspectratio]{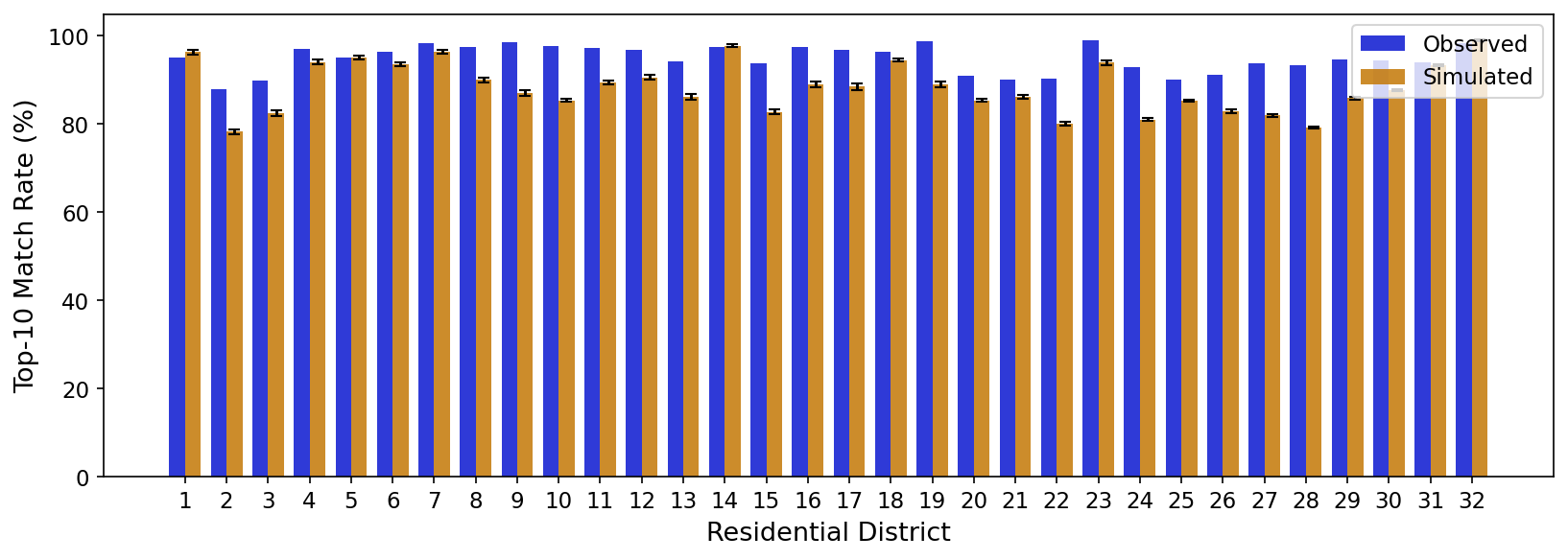}
\caption{\textbf{NYC:} Observed vs.\ simulated top-10 match rate $W_{10}(\mu \mid d)$ by residential district. Simulated values are means over sampled profiles; error bars show 95\% confidence intervals.}
\label{fig:top10validation}

\vspace{1em}

\includegraphics[width=\linewidth,height=0.24\textheight,keepaspectratio]{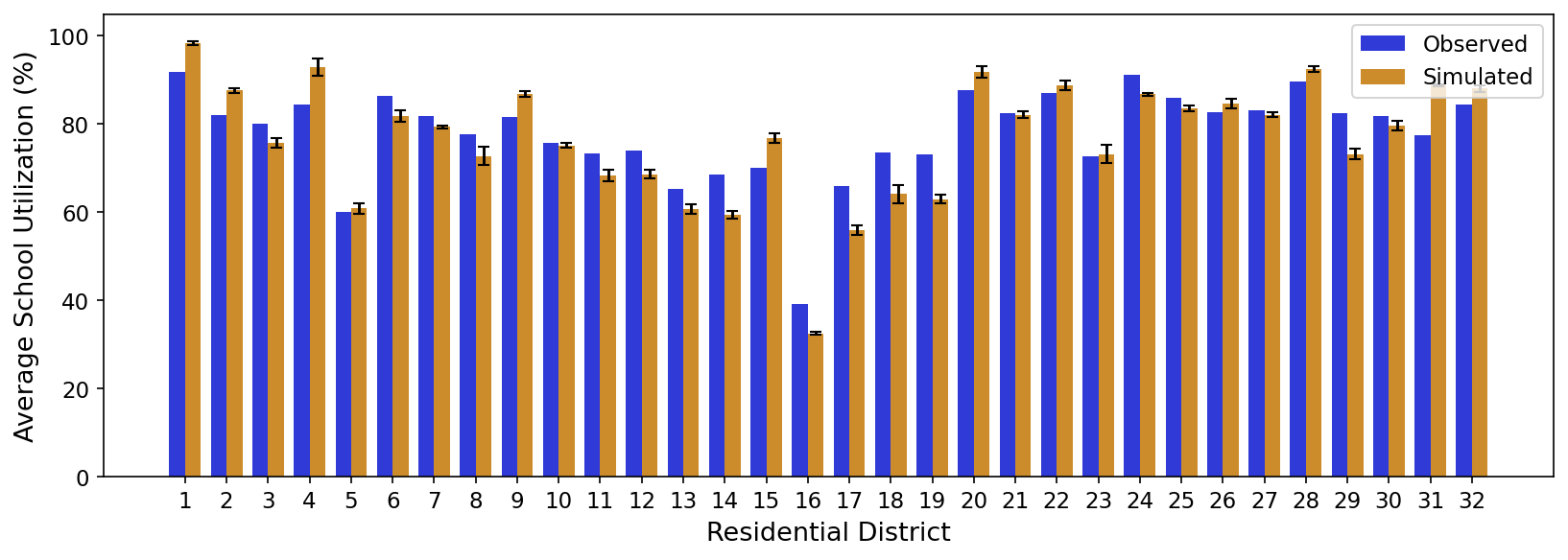}
\caption{\textbf{NYC:} Observed vs.\ simulated program utilization $\bar u_d(\mu)$ by residential district. Simulated values are means over sampled profiles; error bars show 95\% confidence intervals.}
\label{fig:nyc_val_util}
\end{figure}

\begin{figure}[!htbp]
\centering
\includegraphics[width=\linewidth,height=0.24\textheight,keepaspectratio]{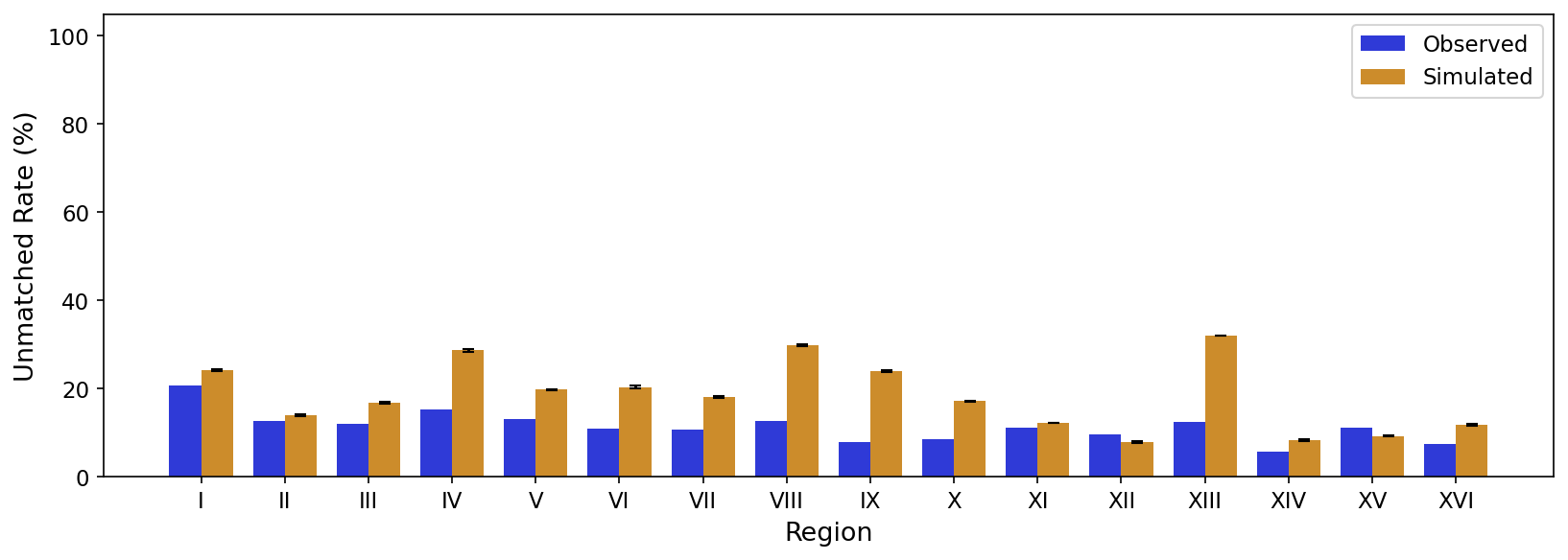}
\caption{\textbf{Chile:} Observed vs.\ simulated unmatched rate $W_{\mathrm{unmatch}}(\mu \mid d)$ by region. Simulated values are means over sampled profiles; error bars show 95\% confidence intervals.}
\label{fig:chileunmatchedvalidation}

\vspace{1em}

\includegraphics[width=\linewidth,height=0.24\textheight,keepaspectratio]{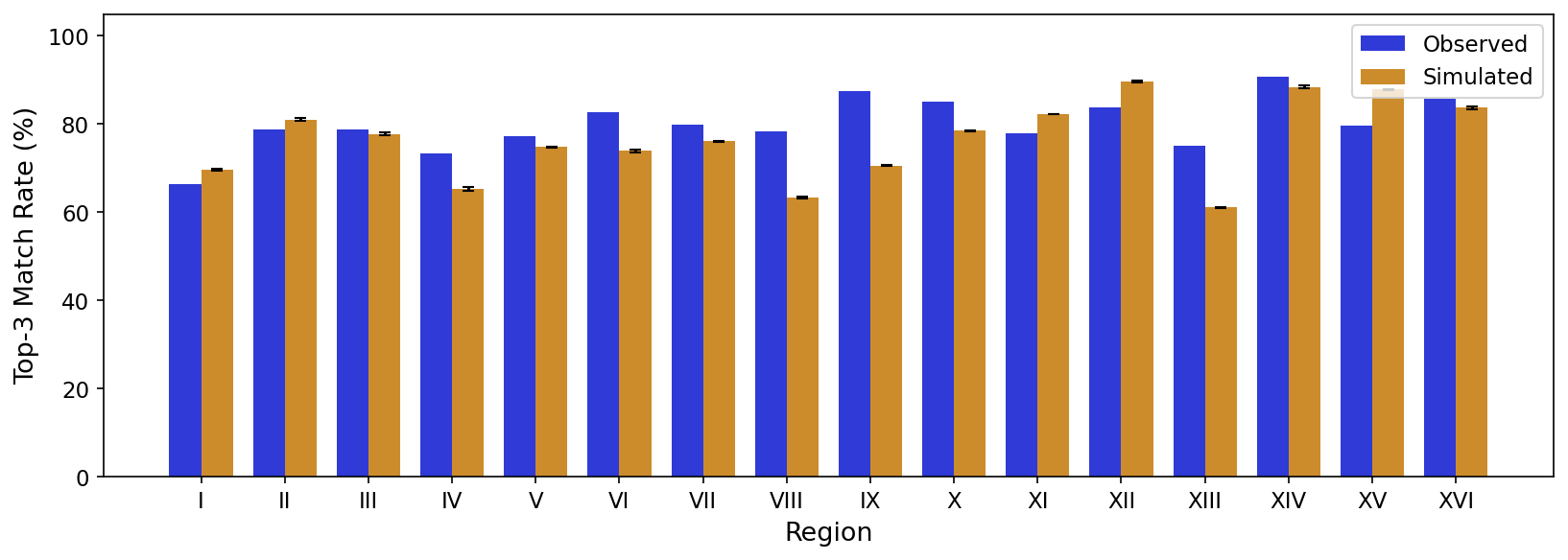}
\caption{\textbf{Chile:} Observed vs.\ simulated top-3 match rate $W_3(\mu \mid d)$ by region. Simulated values are means over sampled profiles; error bars show 95\% confidence intervals.}
\label{fig:chile_val_top3}

\vspace{1em}

\includegraphics[width=\linewidth,height=0.24\textheight,keepaspectratio]{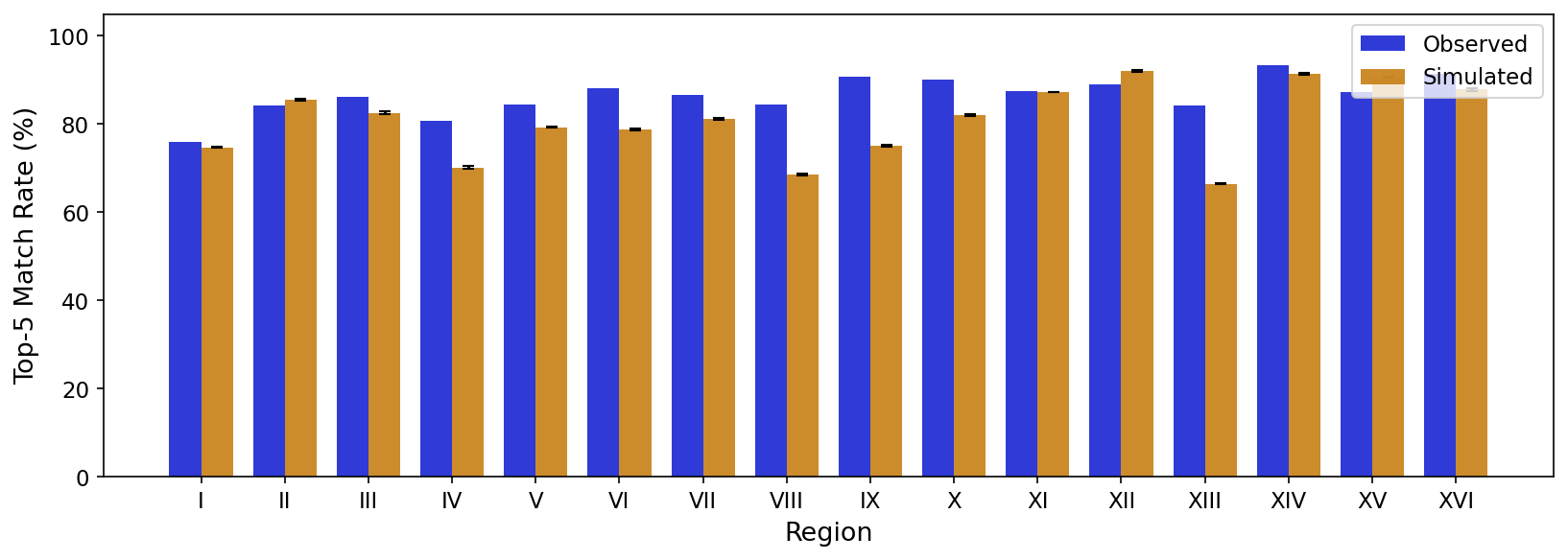}
\caption{\textbf{Chile:} Observed vs.\ simulated top-5 match rate $W_5(\mu \mid d)$ by region. Simulated values are means over sampled profiles; error bars show 95\% confidence intervals.}
\label{fig:chiletop5validation}
\end{figure}

\begin{figure}[!htbp]
\centering
\includegraphics[width=\linewidth,height=0.24\textheight,keepaspectratio]{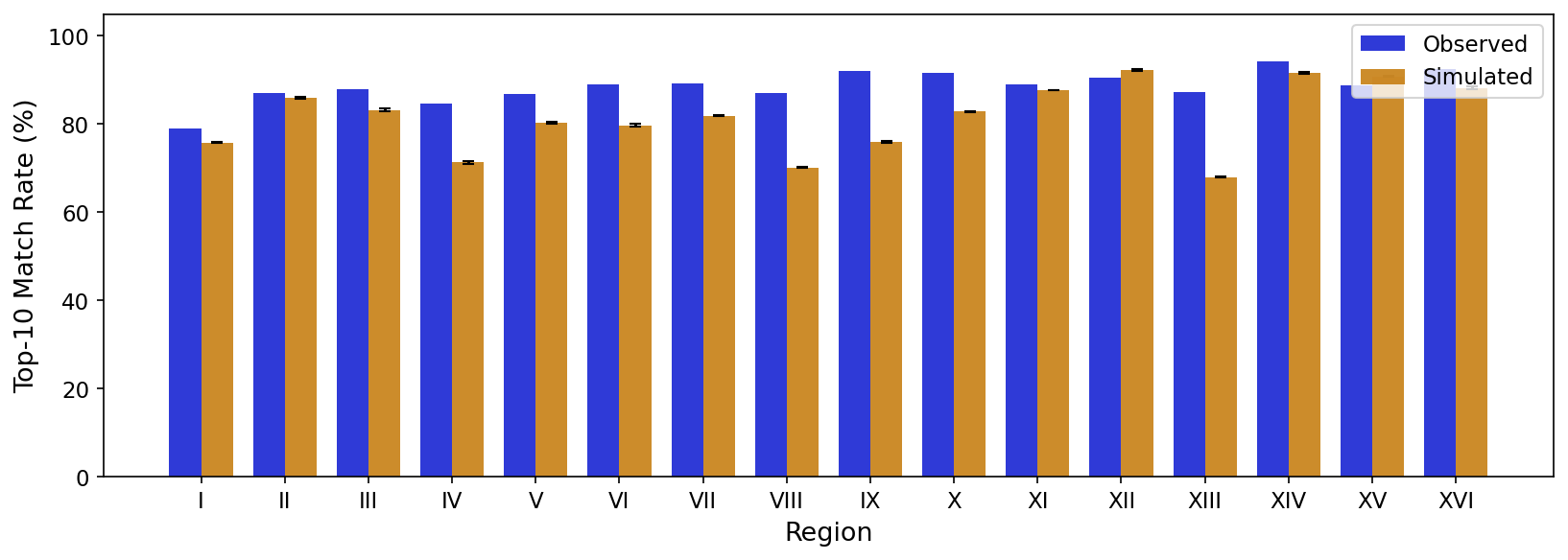}
\caption{\textbf{Chile:} Observed vs.\ simulated top-10 match rate $W_{10}(\mu \mid d)$ by region. Simulated values are means over sampled profiles; error bars show 95\% confidence intervals.}
\label{fig:chiletop10validation}

\vspace{1em}

\includegraphics[width=\linewidth,height=0.24\textheight,keepaspectratio]{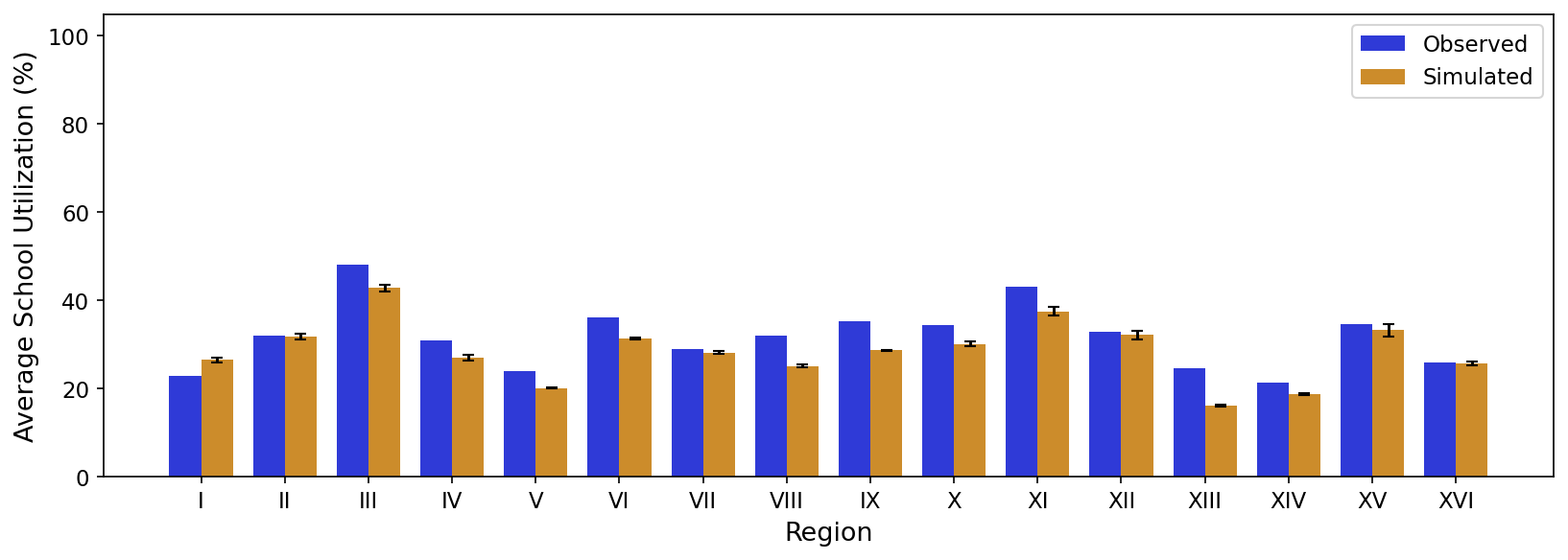}
\caption{\textbf{Chile:} Observed vs.\ simulated program utilization $\bar u_d(\mu)$ by region. Simulated values are means over sampled profiles; error bars show 95\% confidence intervals.}
\label{fig:chile_val_util}
\end{figure}

The figures below provide a more detailed version of Figure~\ref{fig:val_scatter}. In addition to $W_3(\mu \mid d)$ and $\bar u_d(\mu)$, we report $W_{\mathrm{unmatch}}(\mu \mid d)$, $W_5(\mu \mid d)$, and $W_{10}(\mu \mid d)$. For each subdivision, we compare the observed value with the mean over profiles sampled from the calibrated parameters. NYC results are by residential district (Figures~\ref{fig:unmatchedvalidation}-\ref{fig:nyc_val_util}). For Chile, as in the main text, \sys is calibrated by province and the results are aggregated to the 16 regions (Figures~\ref{fig:chileunmatchedvalidation}-\ref{fig:chile_val_util}).

\FloatBarrier
\subsection{Held-out Statistics (Chile)}
\label{app:held_out}

\begin{figure}[!htbp]
    \centering
    \includegraphics[width=0.8\textwidth]{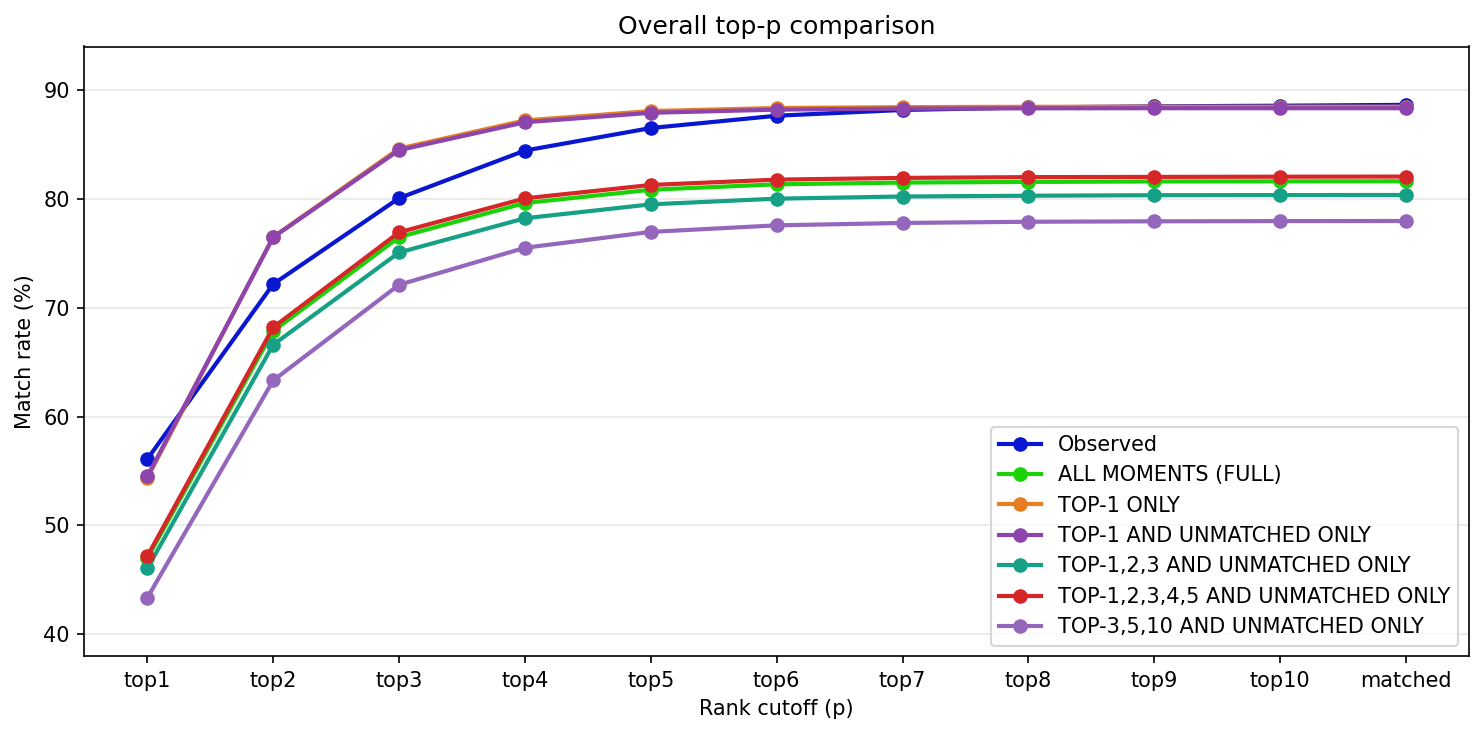}
    \caption{\textbf{Chile:} Observed (dark blue) and simulated top-$p$ match rates $W_p$ for all applicants, under calibrations that use different subsets of the aggregate statistics as targets. ``Matched'' is the overall match rate. Target sets: full (green), (i) orange, (ii) dark purple, (iii) teal, (iv) red, and (v) light purple; (v) is the set of statistics available for NYC.}
    \label{fig:heldout_overall}
\end{figure}

\begin{figure}[!htbp]
    \centering
    \begin{subfigure}[b]{\textwidth}
        \centering
        \includegraphics[width=\textwidth,height=0.26\textheight,keepaspectratio]{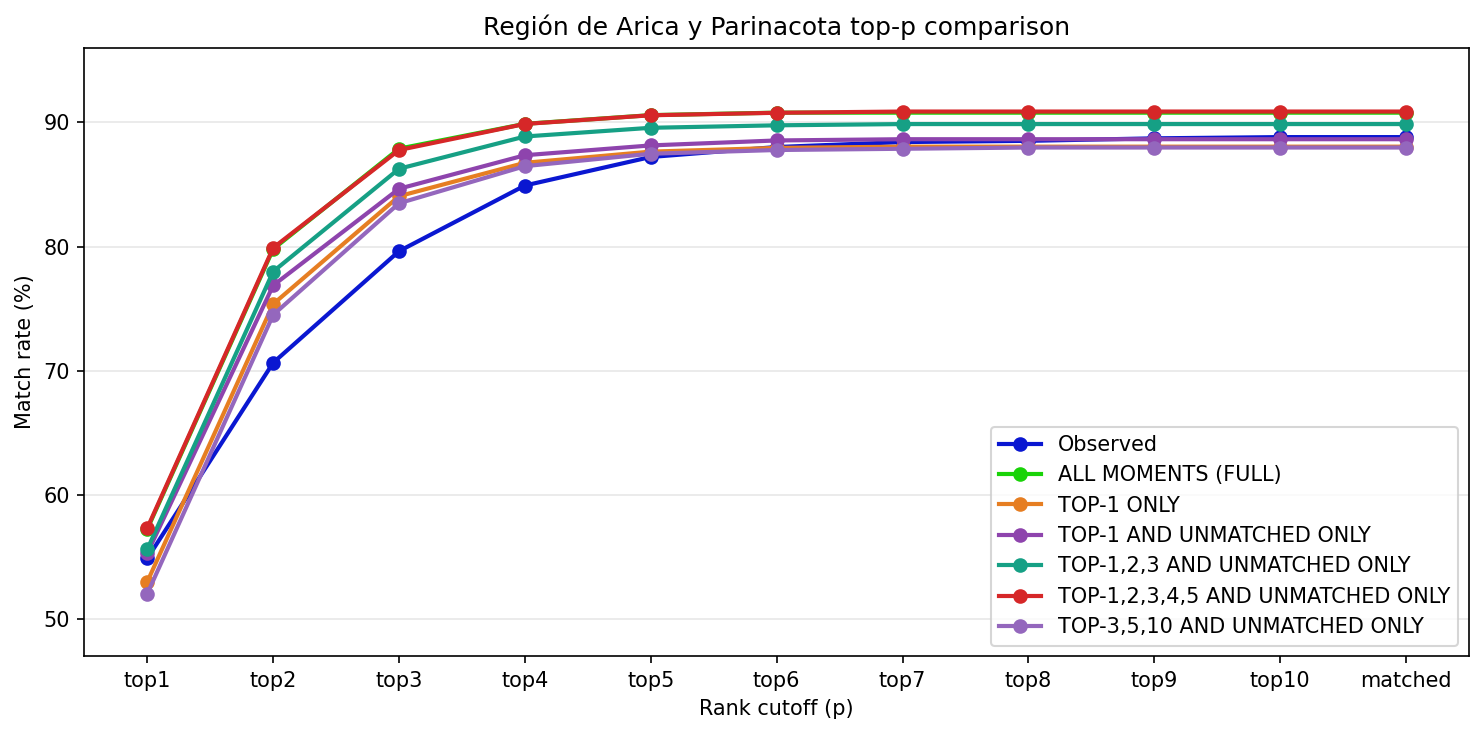}
        \caption{Arica y Parinacota (XV)}
        \label{fig:heldout_arica}
    \end{subfigure}

    \vspace{0.5em}

    \begin{subfigure}[b]{\textwidth}
        \centering
        \includegraphics[width=\textwidth,height=0.26\textheight,keepaspectratio]{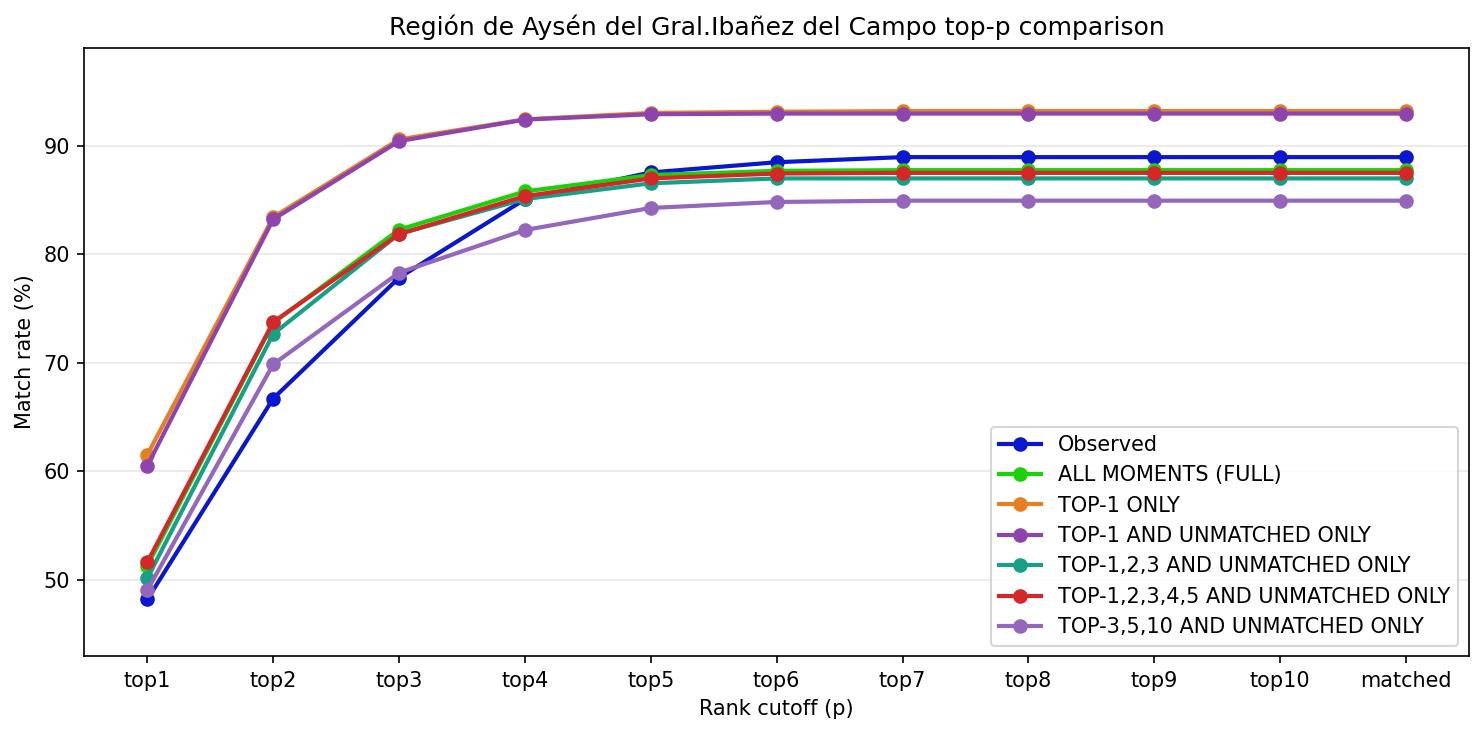}
        \caption{Aysén (XI)}
        \label{fig:heldout_aysen}
    \end{subfigure}

    \vspace{0.5em}

    \begin{subfigure}[b]{\textwidth}
        \centering
        \includegraphics[width=\textwidth,height=0.26\textheight,keepaspectratio]{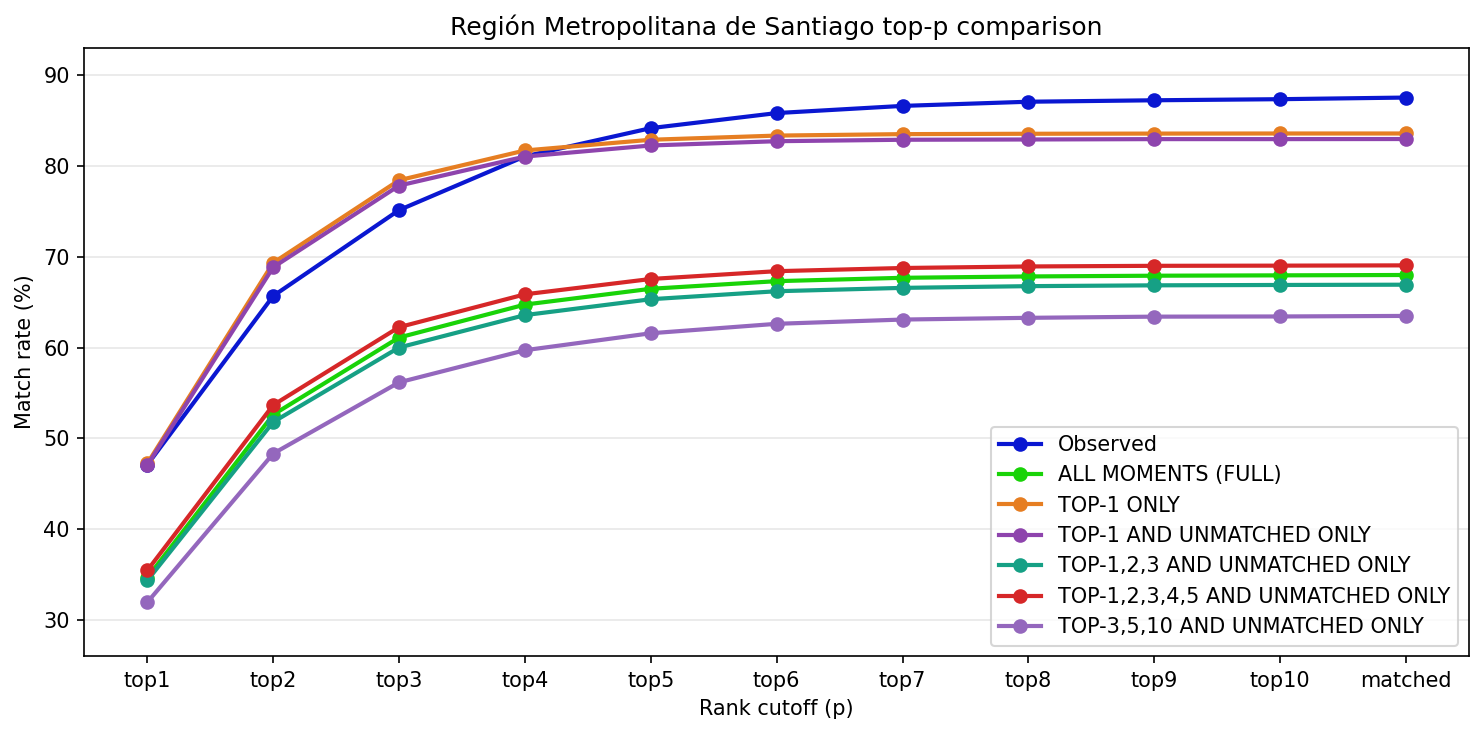}
        \caption{Santiago Metropolitan Region (XIII)}
        \label{fig:heldout_santiago}
    \end{subfigure}

    \caption{\textbf{Chile:} Held-out comparison of Figure~\ref{fig:heldout_overall} for three regions. Which target set comes closest to the observed values differs across regions.}
    \label{fig:heldout_regions}
\end{figure}

\paragraph{\textbf{Setup.}} Can \sys also reproduce statistics that were not used for calibration? We investigate this for Chile, where observed preference profiles allow us to compute all statistics from  $W_1, \dots, W_{10}$ and $W_{\mathrm{unmatch}}$. We calibrate \sys on the full set and on five subsets (below), and compare simulated and observed values of all statistics. Colors refer to Figures~\ref{fig:heldout_overall} and~\ref{fig:heldout_regions}; observed values are dark blue.

\begin{itemize}[leftmargin=2.5em, itemsep=0pt]
    \item[Full] $W_1, \dots, W_{10}$ and $W_{\mathrm{unmatch}}$ \hfill (green, ``All moments (full)'')
    \item[(i)] $W_1$ only \hfill (orange, ``Top-1 only'')
    \item[(ii)] $W_1$ and $W_{\mathrm{unmatch}}$ \hfill (dark purple, ``Top-1 and unmatched only'')
    \item[(iii)] $W_1, W_2, W_3$ and $W_{\mathrm{unmatch}}$ \hfill (teal, ``Top-1,2,3 and unmatched only'')
    \item[(iv)] $W_1, \dots, W_5$ and $W_{\mathrm{unmatch}}$ \hfill (red, ``Top-1,2,3,4,5 and unmatched only'')
    \item[(v)] $W_3, W_5, W_{10}$ and $W_{\mathrm{unmatch}}$ \hfill (light purple, ``Top-3,5,10 and unmatched only'')
\end{itemize}
Subset~(v) is exactly the set of statistics available for NYC, so it shows how well \sys does with NYC-style data in a setting where the full profiles are known. Because the $W_p$ are cumulative, they are strongly correlated, and fewer targets mean less, but not independent, information.

\paragraph{\textbf{Results.}} Figure~\ref{fig:heldout_overall} shows the results for all applicants, and Figure~\ref{fig:heldout_regions} for three regions. Three findings stand out.

First, every target set reproduces the shape of the observed curve: a steep rise up to approx. $p = 4$, then a plateau. This shape drives the counterfactual conclusions in Section~\ref{sec:eval}.

Second, the level depends on the target set, and more targets do not always help. Nationally, the two sets, (i) and (ii), come closest and recover the overall match rate ($88.3\%$ vs.\ $88.6\%$ observed). The richer sets (full, (iii), (iv)) are about 7pp too low, and the NYC-style set~(v) is furthest off, about 11pp too low. This pattern is driven mainly by the Santiago Metropolitan Region (XIII), which has about a third of all applicants (Figure~\ref{fig:heldout_regions}c).

Third, no target set is best everywhere. In Aysén (XI), for example, the order reverses. Sets (i) and (ii), the closest nationally, are now the furthest off, while set~(v) is the closest for $p \leq 3$ (Figure~\ref{fig:heldout_regions}b).

\paragraph{\textbf{Implications.}} \sys generalizes beyond the statistics it is calibrated on. Even with a single target ($W_1$), it reproduces the full top-$p$ curve, and the sparse sets recover the observed overall match rate to within 0.3pp. With NYC-style data (set~(v)), \sys is among the closest fits in some regions, such as Arica y Parinacota and Aysén. The counterfactual conclusions in Section~\ref{sec:eval} rest on this shape and on differences between policies, and Appendix~\ref{app:fidelity} shows that synthetic profiles preserve both. We therefore report NYC results as patterns and policy differences, and read absolute levels with a margin of several percentage points.

\FloatBarrier
\subsection{Fidelity of Counterfactual Conclusions (Chile)}
\label{app:fidelity}

\begin{figure}[ht]
    \centering
    \includegraphics[width=0.6\linewidth]{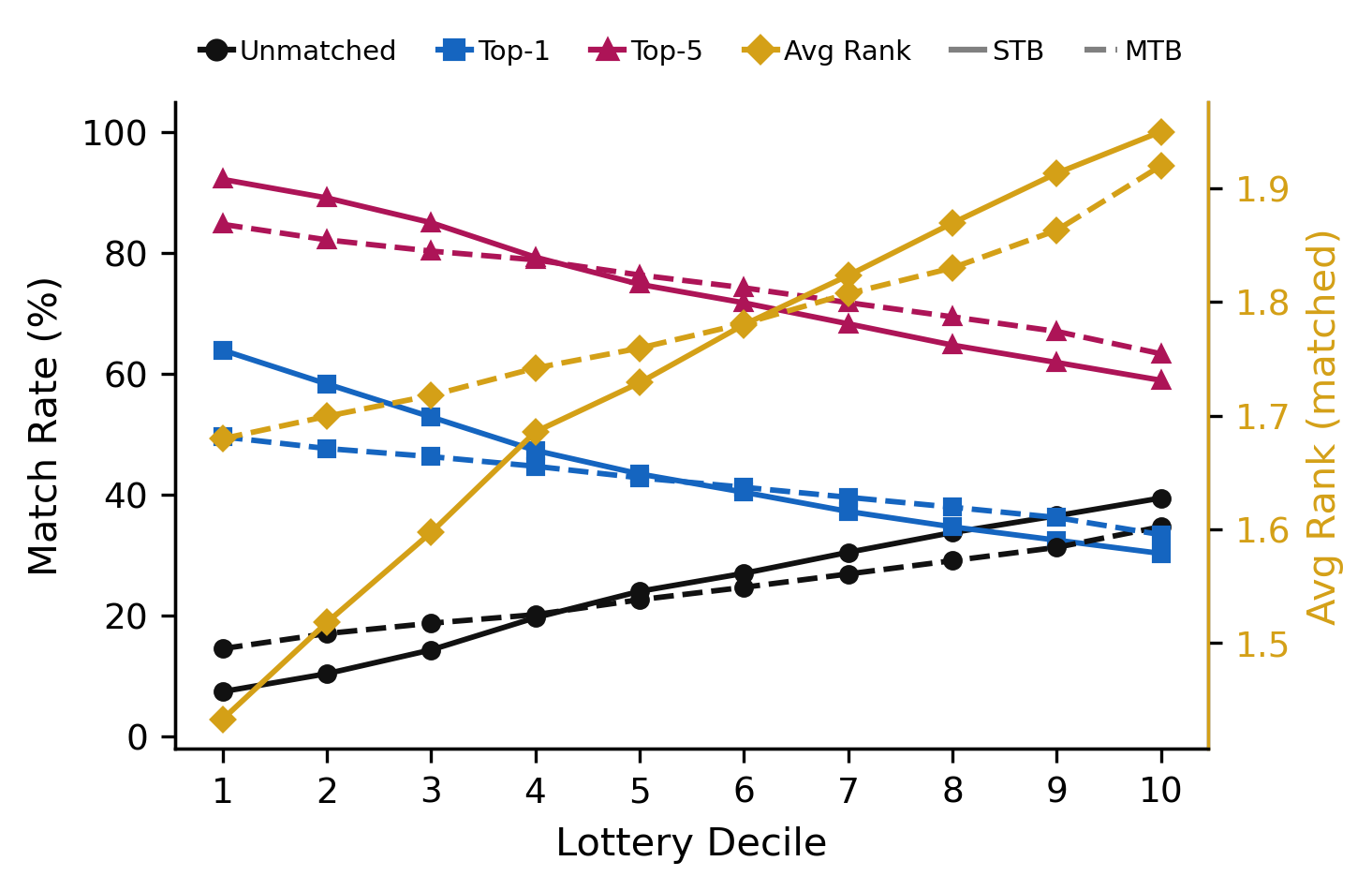}
    \caption{\textbf{Chile, synthetic profiles:} $W_1$, $W_5$, $W_{\mathrm{unmatch}}$, and $\bar r$ by lottery decile under STB and MTB; recreation of Figure~\ref{fig:chilestbmtblottery}. MTB deciles use each student's best (minimum) draw among their top-5 choices.}
    \label{fig:fig4validation}
\end{figure}

\begin{figure}[ht]
    \centering
    \includegraphics[width=0.6\linewidth]{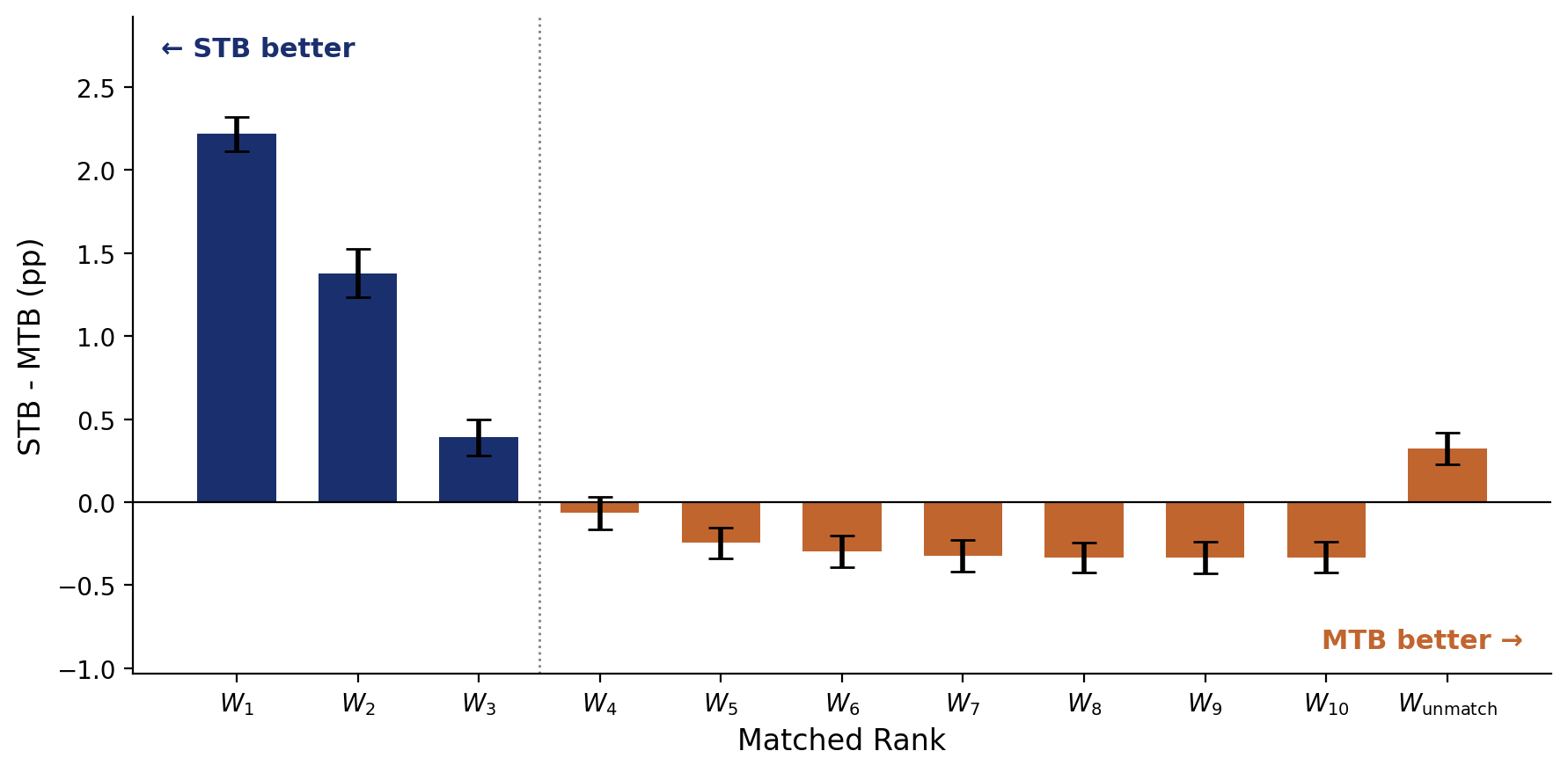}
    \caption{\textbf{Chile, synthetic profiles:} STB$-$MTB difference in $W_p$ and $W_{\mathrm{unmatch}}$ (pp), with 95\% confidence intervals; recreation of Figure~\ref{fig:stb_mtb_diff}.}
    \label{fig:fig6validation}
\end{figure}

\begin{figure}[ht]
    \centering
    \includegraphics[width=0.6\linewidth]{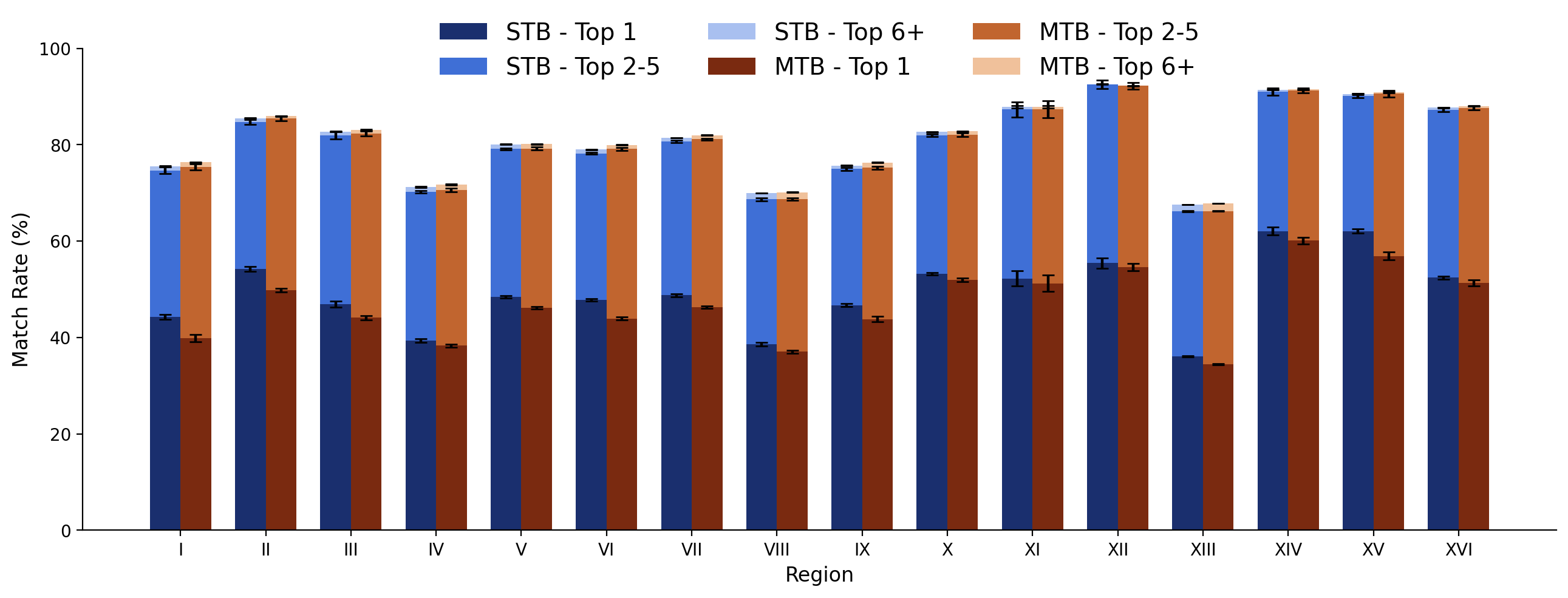}
    \caption{\textbf{Chile, synthetic profiles:} Share of students matched to their first choice, to a top-2--5 choice, and to a lower choice under STB and MTB, by region; recreation of Figure~\ref{fig:chilestbmtbregion}.}
    \label{fig:fig8validation}
\end{figure}

\paragraph{\textbf{Setup.}} Many preference profiles produce the same aggregate statistics, and \sys does not try to recover the true one. Instead, it should produce profiles that lead to the same policy conclusions as the true profile. Chile lets us test this directly. We take the synthetic profiles calibrated on Chilean aggregates only (Section~\ref{sec:eval:val}), rerun the three counterfactual analyses of Section~\ref{sec:eval} on them, and compare the results with those obtained from the observed profiles: (1) welfare by lottery decile (Figure~\ref{fig:chilestbmtblottery} vs.\ Figure~\ref{fig:fig4validation}); (2) the STB$-$MTB difference by rank (Figure~\ref{fig:stb_mtb_diff} vs.\ Figure~\ref{fig:fig6validation}); and (3) outcomes by region (Figure~\ref{fig:chilestbmtbregion} vs.\ Figure~\ref{fig:fig8validation}).

\paragraph{\textbf{Results.}} The synthetic profiles lead to the same three policy conclusions as the observed profiles.
\begin{itemize}
    \item \emph{Lottery deciles.} The lottery draw matters more under STB than under MTB. From the best to the worst decile, the shares of students matched to their first choice ($W_1$) and to a top-5 choice ($W_5$) fall more steeply under STB, and the unmatched rate ($W_{\mathrm{unmatch}}$) and the average matched rank ($\bar r$) rise more steeply. Students with good draws are therefore better off under STB, and students with bad draws under MTB. The STB and MTB curves cross around decile 5 (observed: around decile 5).
    \item \emph{STB vs.\ MTB by rank.} STB matches more students to one of their first few choices, MTB matches more students within a broader top-$p$ cutoff, and STB leaves slightly more students unmatched. STB leads through $p=3$, matching the observed data.
    \item \emph{Regions.} In every region, STB matches more students to their first choice, whereas MTB matches more students to a choice ranked 2--5. The total share of matched students ($W_{\mathrm{match}}$) is nearly the same under both rules.
\end{itemize}

\begin{figure}[ht]
    \centering
    \includegraphics[width=\linewidth]{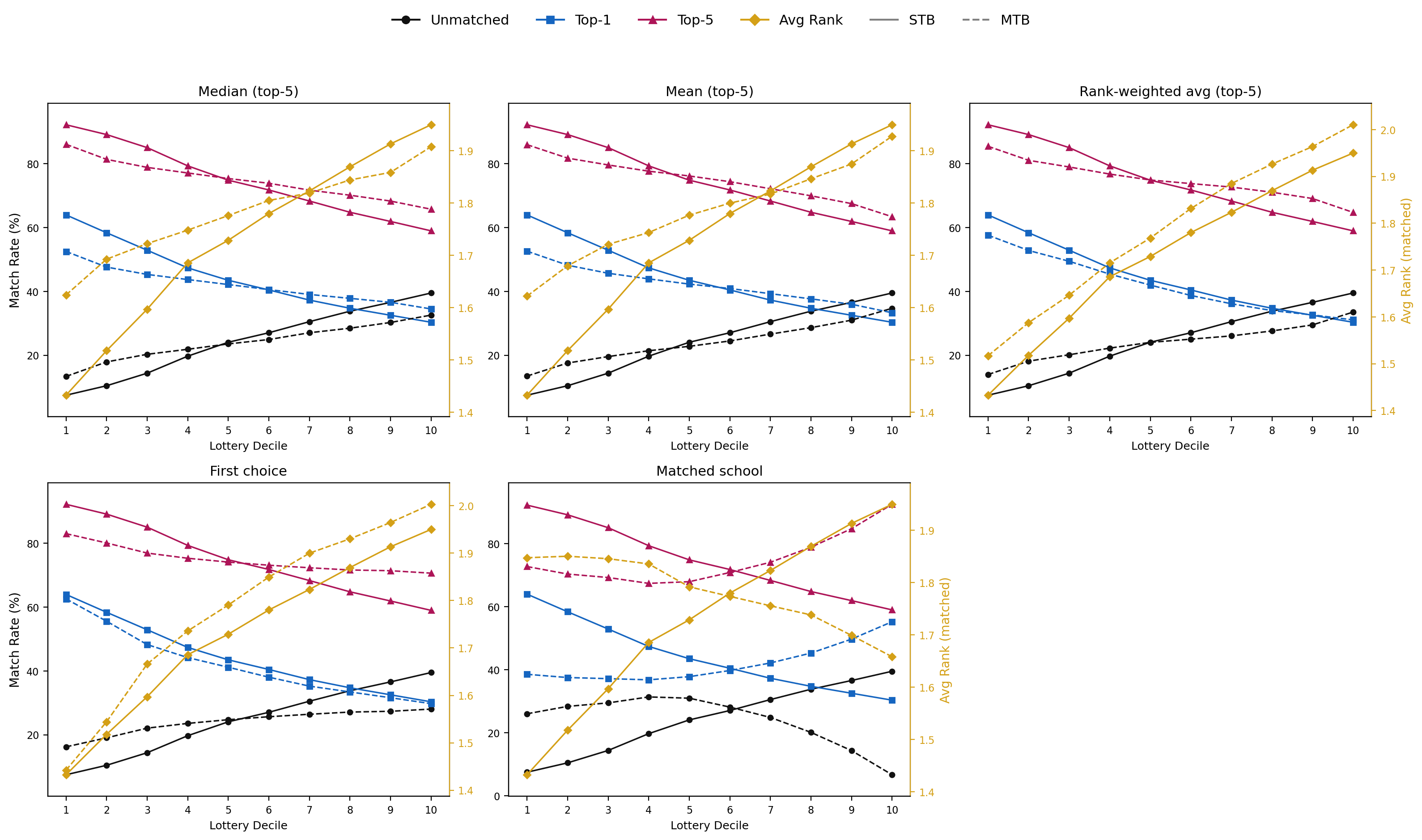}
    \caption{\textbf{Chile, synthetic profiles:} Recreation of Figure~\ref{fig:fig4_alternates} with the alternative MTB summaries.}
    \label{fig:fig4validation_alternates}
\end{figure}

\paragraph{\textbf{Choice of lottery summary under MTB.}} Appendix~\ref{app:lottery_summary} compares alternative ways of summarizing a student's MTB draws for sorting students into lottery deciles (Figure~\ref{fig:fig4_alternates}). Figure~\ref{fig:fig4validation_alternates} repeats this comparison with the synthetic profiles. Observed and synthetic profiles show similar patterns for every summary.

%% file: sections/04_additional_detail_demo.tex
\FloatBarrier
\section{Theoretical Mechanisms Behind Selected Patterns}
\label{app:theory_selected_figures}

This appendix relates selected welfare patterns to two theoretical benchmarks:
a stability implication for applicants with identical preferences and
non-lottery priorities, and the large-market characterization of
\citet{arnostiLotteryDesignSchool2023}.
We retain notation from the main text: applicants $i\in I$, programs
$j\in\mathcal{J}$, submitted lists $R_i$, preference profile $P$,
capacities $c_j$, priority scores $\rho_{ij}$, and matching
$\mu=M(P,c,\rho)$. Lower scores indicate higher priority:
\begin{equation*}
i\succ_j i'
\quad\Longleftrightarrow\quad
\rho_{ij}<\rho_{i'j}.
\end{equation*}
The realized rank is $r_i(\mu)$, with $r_i(\mu)=+\infty$ if applicant
$i$ is unmatched. All rank comparisons refer to submitted preferences.
Auxiliary large-market quantities are distinguished from the finite-market
welfare metrics $W_p$, $W_{\mathrm{match}}$, and $\bar r$.

\subsection{Lottery-decile gradients under STB}
\label{app:lottery_decile_theory}

Figures~\ref{fig:lotto-dist} and the STB series in Figure~\ref{fig:chilestbmtblottery} compare welfare across lottery
deciles. The following comparison holds within a fixed matching under the
capacity and priority model of Section 3.1. It compares two applicants of
the same type; it does not compare two counterfactual matchings obtained
by changing one applicant's draw.

\paragraph{Within-type rank ordering under STB}
Under STB, $\lambda_{ij}=\lambda_i\in[0,1)$ at every program, so
\begin{equation*}
\rho_{ij}=\alpha b_{ij}+\beta t_{ij}+\lambda_i.
\end{equation*}

\begin{lemma}[Within-type rank ordering under STB]
\label{thm:pathwise_monotonicity}
Consider two applicants $i,i'\in I$ in the same market who submit the
same list $R_i=R_{i'}=(j_1,\ldots,j_L)$ and have identical non-lottery
priority components:
\begin{equation*}
b_{ij}=b_{i'j},\qquad t_{ij}=t_{i'j}
\qquad\forall j\in\mathcal{J}.
\end{equation*}
If $\lambda_i<\lambda_{i'}$, the matching produced by student-proposing
DA satisfies $r_i(\mu)\leq r_{i'}(\mu)$.
\end{lemma}

\begin{proof}
The assumptions imply $i\succ_j i'$ at every program. If
$r_i(\mu)>r_{i'}(\mu)$, then $i'$ is matched to a program
$j=\mu(i')$ that $i$ strictly prefers to $\mu(i)$, including the case
where $i$ is unmatched. Program $j$ admits $i'$ although $i$ has higher
priority, so $(i,j)$ is a blocking pair. This contradicts stability.
\end{proof}

\paragraph{First-order stochastic dominance across lottery deciles}
Let $\mathcal{Q}_\ell$ denote lottery decile $\ell$, ordered from best
to worst draws, and define the non-lottery type and type-decile cells by
\begin{equation*}
\xi_i=\bigl(R_i,(b_{ij},t_{ij})_{j\in\mathcal{J}}\bigr),
\qquad
I_{\xi,\ell}
=\{i\in I:\xi_i=\xi,\ \lambda_i\in\mathcal{Q}_\ell\}.
\end{equation*}
With distinct lottery draws, every draw in an earlier decile is smaller
than every draw in a later one.

\begin{proposition}[Within-type first-order stochastic dominance]
\label{thm:fosd}
Fix a realization of the market under STB. For every type $\xi$, every
$\ell<\ell'$ with both $I_{\xi,\ell}$ and $I_{\xi,\ell'}$ nonempty,
and every integer rank cutoff $p\geq1$,
\begin{equation*}
W_p(\mu\mid\xi,\mathcal{Q}_\ell)
\geq W_p(\mu\mid\xi,\mathcal{Q}_{\ell'}),
\qquad
W_p(\mu\mid\xi,\mathcal{Q}_\ell)
=\frac{1}{|I_{\xi,\ell}|}
\sum_{i\in I_{\xi,\ell}}\mathbf{1}\{r_i(\mu)\leq p\}.
\end{equation*}
Thus realized ranks in the better decile are first-order stochastically
smaller, and hence weakly better, conditional on type.
\end{proposition}

\begin{proof}
For every $i\in I_{\xi,\ell}$ and $i'\in I_{\xi,\ell'}$,
Lemma~\ref{thm:pathwise_monotonicity} gives $r_i(\mu)\leq r_{i'}(\mu)$.
Consequently,
$\mathbf{1}\{r_i(\mu)\leq p\}\geq
\mathbf{1}\{r_{i'}(\mu)\leq p\}$.
Averaging over all pairs $(i,i')$ in the two cells yields the claim.
\end{proof}

\paragraph{Implications and limits}
Within type, unmatched rates therefore satisfy
\begin{equation*}
W_{\mathrm{unmatch}}(\mu\mid\xi,\mathcal{Q}_\ell)
\leq W_{\mathrm{unmatch}}(\mu\mid\xi,\mathcal{Q}_{\ell'}).
\end{equation*}
The rank-ordering lemma also orders \emph{average matched ranks within
type}, whenever both cells contain matched applicants:
\begin{equation*}
\bar r(\mu\mid\xi,\mathcal{Q}_\ell)
\leq \bar r(\mu\mid\xi,\mathcal{Q}_{\ell'}).
\end{equation*}
Indeed, every applicant in the better cell has a rank weakly below that
of every matched applicant in the worse cell, and is therefore matched.

These statements require no independence between lottery draws and
applicant characteristics. They do not imply exact ordering of the
\emph{aggregate} decile metrics in every finite realization: deciles can
have different type compositions. Independent, identically distributed
lottery draws make their type compositions comparable in expectation,
but do not impose identical compositions in each realization. Moreover,
aggregate average matched rank,
\begin{equation*}
\bar r(\mu)
=\frac{\sum_{i:\mu(i)\ne\varnothing}r_i(\mu)}
{|\{i:\mu(i)\ne\varnothing\}|},
\end{equation*}
weights types by their representation among matched applicants. Those
weights can vary across deciles, even if unconditional type proportions
are the same. The lemma therefore supports the mechanism behind the
STB gradients in Figures~\ref{fig:lotto-dist} and \ref{fig:chilestbmtblottery} without determining their aggregate
magnitude or guaranteeing aggregate monotonicity.

\subsection{Minimum list length under STB}
\label{app:minlist_theory}

We use the characterization in Theorem~1 of
\citet{arnostiLotteryDesignSchool2023}, including its heterogeneous-program
extension in Section~3.3, to interpret Figure~\ref{fig:nyc_min_impact_lotto}. This benchmark assumes
independent applicants whose lists are sampled without replacement from
a common program-popularity distribution, list lengths drawn independently
of these rankings and lotteries, and priorities determined solely by
lotteries. The numbers of applicants and programs grow at a fixed ratio,
while list lengths, capacities, and normalized program popularities remain
bounded. These assumptions differ from the calibrated preferences and
institutional priorities in our simulations.

Let $\mathcal{L}$ denote the list-length distribution,
$\mathcal{L}(>h)=\Pr(L_i>h)$, and
$\varrho=|I|/|\mathcal{J}|=n/m$ the applicant-to-program ratio, distinct
from the priority scores $\rho_{ij}$. Write $z=1-\lambda$ for lottery
priority, with larger $z$ indicating better priority. Translating
\citet{arnostiLotteryDesignSchool2023}'s Equations~(3), (6), and (7) into our notation gives
\begin{align*}
m_{\mathcal{L}}(q)
  &=\sum_{h=0}^{\infty}\mathcal{L}(>h)(1-q)^h,\\
\Lambda_{\mathcal{L}}(1)&=0,
\qquad
\Lambda_{\mathcal{L}}'(z)
  =-\varrho\,m_{\mathcal{L}}\!\left(V(\Lambda_{\mathcal{L}}(z))\right),\\
F_{\mathrm{STB},\mathcal{L}}(p)
  &=1-\int_0^1
  \left(1-V(\Lambda_{\mathcal{L}}(z))\right)^p\,dz.
\end{align*}
Here $m_{\mathcal{L}}(q)$ is the expected number of proposals when each
successive choice independently accepts with probability $q$.
$\Lambda_{\mathcal{L}}(z)$ is the average higher-priority proposal
intensity per program. The nonincreasing function $V$ is the
\emph{proposal-weighted} vacancy probability, incorporating capacity and
popularity as in \citet{arnostiLotteryDesignSchool2023}'s Equation~(8). For equally popular programs of
common capacity $c_0$, for example,
$V(x)=\Pr\{\operatorname{Poisson}(x)<c_0\}$.
Theorem~1 identifies $F_{\mathrm{STB},\mathcal{L}}(p)$ as the limiting
top-$p$ probability for each list length $\ell\geq p$ in the support of
$\mathcal{L}$, and therefore also conditional on listing at least $p$
programs whenever $\mathcal{L}(\geq p)>0$.

Consider an increase in the minimum list length implemented by appending
programs below existing choices, within this same sampling model. Keep
$\varrho$, program capacities, and popularity fixed. The resulting
distribution $\mathcal{L}'$ satisfies
$\mathcal{L}'(>h)\geq\mathcal{L}(>h)$ for every $h$, so
$m_{\mathcal{L}'}(q)\geq m_{\mathcal{L}}(q)$. Scalar ODE comparison,
after setting $s=1-z$, yields
$\Lambda_{\mathcal{L}'}(z)\geq\Lambda_{\mathcal{L}}(z)$.
Because $V$ is nonincreasing, the displayed characterization implies
\begin{equation*}
F_{\mathrm{STB},\mathcal{L}'}(p)
\leq F_{\mathrm{STB},\mathcal{L}}(p).
\end{equation*}
This comparative-static implication is obtained from Arnosti's
characterization; it is not a separate theorem quoted from his paper.

His total-match identity (Appendix~A.1, Lemma~5) gives the limiting
matched fraction
\begin{equation*}
W_{\mathrm{match},\mathcal{L}}^{\mathrm{STB},\infty}
=\sum_{\ell\geq1}\mathcal{L}(\ell)
 F_{\mathrm{STB},\mathcal{L}}(\ell)
=\frac{1}{\varrho}
 \int_0^{\Lambda_{\mathcal{L}}(0)}V(u)\,du.
\end{equation*}
Thus $\Lambda_{\mathcal{L}'}(0)\geq\Lambda_{\mathcal{L}}(0)$ also implies
\begin{equation*}
W_{\mathrm{match},\mathcal{L}'}^{\mathrm{STB},\infty}
\geq W_{\mathrm{match},\mathcal{L}}^{\mathrm{STB},\infty}.
\end{equation*}
These two inequalities explain a coverage--quality tension in the
benchmark. They do not generally imply that aggregate $W_p$ falls:
applicants whose original lists have fewer than $p$ entries may become
matched within the top $p$ after extending their lists. If all original
lists already contain at least $p$ programs, the conditional and
aggregate top-$p$ quantities coincide in the limit. With heterogeneous
short lists, $F_{\mathrm{STB},\mathcal{L}}(p)$ divided by the total
matched fraction is not the top-$p$ rate among matched applicants.

Figure~\ref{fig:nyc_min_impact_lotto} shows a related pattern in our NYC simulations: comparing the
upper and lower ends of the displayed range, coverage is higher and
aggregate $W_3$ is lower. As specified in its caption, its average-rank
series is restricted to applicants matched at baseline, and its reported
value is higher at the upper endpoint. This restriction excludes newly
matched applicants; a constant-cohort interpretation additionally requires
specifying how baseline-matched applicants who later become unmatched
are treated. These finite-market comparisons, and any intermediate
nonmonotonicities, are simulation findings rather than direct predictions
of \citet{arnostiLotteryDesignSchool2023}'s benchmark.

\subsection{STB-MTB crossover in top-$p$ outcomes}
\label{app:stb_mtb_crossover_theory}

For $B\in\{\mathrm{STB},\mathrm{MTB}\}$, retain the main-text metrics
$W_p^B=W_p(\mu^B)$, where $\mu^B=M(P,c,\rho^B)$, and define
$\Delta_p=W_p^{\mathrm{STB}}-W_p^{\mathrm{MTB}}$.
To interpret Figure~\ref{fig:stb_mtb_diff} and Table~\ref{tab:mtb_stb_summary}, fix
$\mathcal{L}$ and write $F_B(p)=F_{B,\mathcal{L}}(p)$ for Arnosti's
conditional large-market probabilities. For
$\mathcal{L}(>1)>0$, his Theorem~2 gives a cutoff
$p^\star_{\mathrm A}\geq1$, with
$\mathcal{L}(>p^\star_{\mathrm A})>0$, such that
\begin{equation*}
\begin{aligned}
F_{\mathrm{STB}}(p)&>F_{\mathrm{MTB}}(p)
&&\text{for }1\leq p\leq p^\star_{\mathrm A},\\
F_{\mathrm{STB}}(p)&\leq F_{\mathrm{MTB}}(p)
&&\text{for }p>p^\star_{\mathrm A}.
\end{aligned}
\end{equation*}
If everyone lists exactly one program, the two lottery rules instead
give the same assignment probabilities for that sole choice.
Under STB, a favorable draw applies at every program; under MTB, later
applications provide new independent lottery draws. This explains why
the benchmark favors STB for early choices and MTB for the match
probability of applicants submitting the longest lists.

The distinction between conditional and aggregate outcomes matters.
Under \citet{arnostiLotteryDesignSchool2023}'s assumptions, Theorem~1 implies
\begin{equation*}
W_p^B\xrightarrow{\mathbb P}W_p^{B,\infty}
=\sum_{\ell=1}^{p-1}\mathcal{L}(\ell)F_B(\ell)
 +\mathcal{L}(\geq p)F_B(p),
\end{equation*}
where a term with zero weight is omitted. An applicant with a list
shorter than $p$ contributes to aggregate top-$p$ welfare whenever
matched anywhere on that list. The conditional single-crossing result
therefore does not, by itself, establish an aggregate crossing at the
same cutoff, or guarantee the existence and uniqueness of such a crossing.

\citet{arnostiLotteryDesignSchool2023} Theorem~3 supplies sufficient conditions for the total-match
comparison: convexity of $f(q)=1/m_{\mathcal{L}}(q)$ weakly favors MTB;
concavity weakly favors STB. In particular, in the nondegenerate case
above and under convexity, the limiting aggregate difference satisfies
\begin{equation*}
\Delta_1^\infty>0,
\qquad
\Delta_p^\infty
=W_{\mathrm{match}}^{\mathrm{STB},\infty}
 -W_{\mathrm{match}}^{\mathrm{MTB},\infty}
\leq0
\quad\text{for }p\geq L_{\max},
\end{equation*}
where $L_{\max}$ bounds the support of $\mathcal{L}$ and
$\Delta_p^\infty=W_p^{\mathrm{STB},\infty}-W_p^{\mathrm{MTB},\infty}$.
Thus at least one transition from positive to weakly nonpositive
aggregate differences must occur in this benchmark. If MTB's total-match
advantage is strict, sufficiently broad top-$p$ comparisons strictly
favor MTB. This endpoint argument determines neither the location nor
the uniqueness of an aggregate crossover.

There is also an institutional distinction: Arnosti explicitly notes
in Section~4.1.2 that Theorem~2 does not directly cover multiple
priority classes; he investigates some such configurations by simulation.
Our Chilean results likewise concern a market with program-specific
priority tiers. Overall, STB increases $W_1$ 
and leads through $p=3$, while MTB has a small advantage for the
reported ranks $p=4,\ldots,10$. The unmatched rate is  higher under STB, which also favors MTB. The observed crossover is consistent with \citet{arnostiLotteryDesignSchool2023}'s mechanism, but the crossover point is identified empirically.